\RequirePackage{ifpdf}
\documentclass[28pt, letterpaper, hyper]{JHEP3}

\pdfoutput=1

\usepackage[utf8]{inputenc}

\usepackage{epic,eepic}
\usepackage{amsmath,amssymb,amsfonts}
\usepackage{makeidx}

\usepackage{dsfont}
\usepackage{graphicx}
\usepackage{csquotes}
\usepackage{url}

\usepackage{float}
\newlength \savewidth

	\def\bea{\begin{eqnarray}}
	\def\eea{\end{eqnarray}}
	\def\nn{\nonumber\\}
	
	\def\i{\item} 
	\def\mm{\mathcal}
	
	\def\pa{\partial}

	\def\a{\alpha} \def\b{\beta} \def\g{\gamma} \def\d{\delta} \def\e{\epsilon}   \def\n{\nu} \def\m{\mu} \def\s{\sigma} \def\r{\rho} \def\t{\theta}  \def\l{\lambda}  
	     \def\L{\Lambda}  \def\S{\Sigma} \def\O{\Omega}

\newcommand{\ben}{\begin{equation}}
\newcommand{\een}{\end{equation}}
\newcommand{\be}{\begin{equation}}
\newcommand{\ee}{\end{equation}}
\newcommand{\ba}{\begin{eqnarray}}
\newcommand{\ea}{\end{eqnarray}}

\newcommand{\beq}{\begin{equation}}
\newcommand{\eeq}{\end{equation}}
\newcommand{\beqa}{\begin{eqnarray}}
\newcommand{\eeqa}{\end{eqnarray}}
\newcommand{\beqar}{\begin{eqnarray*}}
	\newcommand{\eeqar}{\end{eqnarray*}}
\newcommand{\reef}[1]{(\ref{#1})}
\newcommand{\ssc}{\scriptscriptstyle}
\newcommand{\eg}{{\it e.g.,}\ }
\newcommand{\ie}{{\it i.e.,}\ }
\newcommand{\comment}[1]{{\bf [[[#1]]]}}

\newcommand{\cO}{{\cal O}}

\newcommand{\labell}[1]{\label{#1}} %{\quad\mt{#1}\label{#1}} % 
\def\pa {\partial}

\def\t6 {T_\mt{D6}}

\newcommand{\mt}[1]{\textrm{\tiny #1}}

\def\calb         {{\cal B}}
\def\calc         {{\cal C}}
\def\cald         {{\cal D}}
\def\cale         {{\cal E}}

\def\calm         {{\cal M}}

\def\calo         {{\cal O}}

\def\calv         {{\cal V}}

\def\ee           {{\rm e}}

\def\sqr#1#2{{\vcenter{\vbox{\hrule height.#2pt
				\hbox{\vrule width.#2pt height#1pt \kern#1pt
					\vrule width.#2pt}\hrule height.#2pt}}}}

\def\a{\alpha}
\def\b{\beta}

\def\r{\rho}

\def\e{\epsilon}

\def\g{\gamma}
\def\ee{\cale}

\def\aa1{\phi}
\def\cc1{\psi}

\def\eps{\epsilon}
\def\s{\sigma}

\newcommand{\nt}[1]{{\bf t}_#1}
\newcommand{\nk}[1]{{\bf k}_#1}
\newcommand{\hn}[1]{{\bf \hat n}_#1}
\newcommand{\hnt}[1]{{\bf \hat t}_#1}
\newcommand{\hnk}[1]{{\bf \hat k}_#1}
\newcommand{\nnv}{{\bf n}}

\newcommand{\nnn}{{\bf n}}

\newcommand{\nkk}{{\bf k}}

\newcommand{\cv}{{\cal C}_\mt{V}}
\newcommand{\ca}{{\cal C}_\mt{A}}
\newcommand{\bn}{{\bf n}}
\newcommand{\bt}{{\bf t}}
\newcommand{\bs}{{\bf s}}
\newcommand{\tth}{{\theta'}}

\newcommand{\nlj}[1]{{\bf n}_#1}

\title{\textbf{{\Large   Comments on Holographic Complexity}}}
\author{Dean Carmi,${}^{1,2}$  Robert C. Myers${}^1$ and Pratik Rath${}^{1,3}$
	\\
	${}^1$ \textit{Perimeter Institute for Theoretical Physics\\
\ $\,$	31 Caroline Street North, Waterloo, ON N2L 2Y5, Canada}
	\\
	${}^2$ \textit{Raymond and Beverly Sackler Faculty of Exact Sciences \\
	\ $\,$	School of Physics and Astronomy, Tel-Aviv University, Ramat-Aviv 69978, Israel}
\\
${}^3$ \textit{Center for Theoretical Physics and Department of Physics\\ 
\ $\,$	University of California, Berkeley, CA 94720, USA}
	
}

\abstract{We study two recent conjectures for holographic complexity: the complexity=action conjecture and the complexity=volume conjecture. In particular, we examine the structure of the UV divergences appearing in these quantities, and show that the coefficients can be written as local integrals of geometric quantities in the boundary. We also consider extending these conjectures to evaluate the complexity of the mixed state produced by reducing the pure global state to a specific subregion of the boundary time slice. The UV divergences in this subregion complexity have a similar geometric structure, but there are also new divergences associated with the geometry of the surface enclosing the boundary region of interest. We discuss possible implications arising from the geometric nature of these UV divergences.
}	

\begin{document}

\tableofcontents	
\setcounter{footnote}{0}

\allowdisplaybreaks

% % % % % % % % % % % % % % % % % % % % % % % % % % % % % % % % % % % % % % % % % % % % % % % % % % % % % % % % % % % % % % % % % % % % % % % % % % % % % % % % % % % % % % % % % % % % %

\section{Introduction}

Concepts and perspectives from quantum information science are having a rapidly growing influence in investigations of quantum field theory and quantum gravity.  Quantum complexity is one such concept which has recently begun to be discussed. Loosely speaking, the complexity of a particular state corresponds to the minimum number of simple (universal) gates needed to build a quantum circuit which prepares this state from a particular reference state, \eg see \cite{watrous2009quantum,2014arXiv1401.3916G,0034-4885-75-2-022001}. 
In the context of  the AdS/CFT correspondence, discussions have focused on understanding the growth of the Einstein-Rosen bridge for AdS black holes in terms of quantum complexity in the dual boundary CFT \cite{Susskind:2014rva,2014arXiv1403.5695S,Stanford:2014jda,Susskind:2014jwa,Susskind:2014moa,Brown:2015bva,Brown:2015lvg}.

There are two independent proposals to evaluate the complexity of a holographic boundary state, which we will refer to as the complexity=volume (CV) conjecture \cite{Susskind:2014rva,2014arXiv1403.5695S,Stanford:2014jda,Susskind:2014jwa,Susskind:2014moa} and the complexity=action (CA) conjecture \cite{Brown:2015bva,Brown:2015lvg}. The first of the proposals states  that the complexity of the boundary state is dual to the volume of the extremal codimension-one bulk hypersurface which meets the asymptotic boundary on the desired time slice.\footnote{An alternative proposal related to complexity=volume was recently put forward in \cite{Couch:2016exn}. We also note that similar extremal volumes in the interior of asymptotically flat black holes were studied in \cite{Christodoulou:2014yia,Christodoulou:2016tuu}.} More precisely, the CV duality states that the complexity of the state on a time slice $\S$ is given by:
\begin{equation}
\cv(\S) =\ \mathrel{\mathop {\rm
max}_{\scriptscriptstyle{\S=\partial \mathcal{B}}} {}\!\!}\left[\frac{\mathcal{V(B)}}{G_N \, \ell}\right] \, ,\labell{defineCV}
\end{equation}
where $\mathcal B$ is the corresponding bulk surface and $\ell$ is some length scale associated with the bulk geometry, \eg the AdS curvature scale or the horizon radius of a black hole. The ambiguity in choosing the latter scale is an unappealing feature of CV duality and provided some motivation for developing CA duality \cite{Brown:2015bva,Brown:2015lvg}. This second conjecture equates the complexity with the gravitational action evaluated on a particular bulk region, now known as  the Wheeler-DeWitt (WDW) patch:
\bea
\ca(\S) =  \frac{I_\mt{WDW}}{\pi\, \hbar}\,. \labell{none}
\eea
The WDW patch can be defined as the domain of dependence of any Cauchy surface in the bulk which asymptotically approaches the time slice $\Sigma$ on the boundary.

The complexity evaluated with either the CV or CA duality satisfies a number of expected properties, \eg they continue to grow (linearly with time) after the boundary theory reaches thermal equilibrium. However, the second conjecture has certain advantages. In particular, as noted above, CV duality requires choosing an additional length scale, while there are no free parameters in eq.~\reef{none} for the CA duality. However, the latter faced the obstacle that when the conjecture was originally proposed, there was no rigorous method for evaluating the gravitational action on spacetime regions with null boundaries. This problem was recently overcome with a careful analysis of the boundary terms which must be added to the gravitational action for null boundary surfaces and for joints where such null boundaries intersect with other boundary surfaces \cite{Lehner:2016vdi}.

On the gravity side, either of these dualities deals with a geometric entity which  extends to the asymptotic AdS boundary and as a result, the holographic complexity is divergent.
To understand these divergences, it is natural to draw upon lessons from holographic entanglement entropy \cite{Ryu:2006bv,Ryu2}. In particular, for both the CV duality and holographic entanglement entropy, the bulk calculations evaluate the volume of an extremal surface extending to the asymptotic boundary. Now UV divergences are found in calculating holographic entanglement entropy, \eg \cite{Hung:2011xb,Hung:2011ta}, and these divergences are related to the existence of correlations down to arbitrarily short scales in the boundary CFT. The leading divergence gives rise to the famous `area law' term \cite{Bombelli:1986rw,Srednicki:1993im} and the subleading divergent terms involve integrals of curvature invariants, both extrinsic and intrinsic, over the entangling surface in the boundary. In this paper, we will examine the divergent contributions appearing in holographic complexity and show that the structure of these UV divergent terms in the holographic complexity have a similar geometric interpretation in the boundary CFT, \ie the coefficients in these divergent terms are given by local geometric integrals over the time slice of interest. One might have anticipated that the UV divergences would appear in the complexity from the necessity of establishing correlations down to the cut-off scale in the boundary CFT. As we comment in the discussion section, the geometric structure of these divergences leads to some unusual behaviour for the complexity.

We also consider the structure of divergences for the holographic complexity of subregions, \ie evaluating the complexity of the mixed state produced by reducing the global boundary state to specific subregion on the time slice. The latter idea was discussed previously for time-independent geometries in, \eg \cite{Alishahiha:2015rta,Ben-Ami:2016qex}. We will begin by proposing a covariant extension of eqs.~\reef{defineCV} and \reef{none} which aims to evaluate subregion complexity. Our proposals are motivated by the suggestion that the mixed state on the boundary is encoded in the corresponding entanglement wedge in the bulk \cite{ewedge1,Headrick:2014cta}. We then find a similar geometric structure for the UV divergences in this subregion complexity, but there are also new divergences associated with the entangling surface which encloses the boundary region.

The remainder of the paper is organized as follows: Section~\ref{sec:extremal} considers the CV conjecture and we investigate the structure of UV divergences appearing in $\cv(\S)$. The coefficients of the divergences are given in terms of extrinsic and intrinsic curvatures integrated over the boundary time slice. In section~\ref{sec:action}, we study the analogous divergences arising in the CA conjecture. In particular, $\ca(\S)$ contains an additional class of divergences involving the logarithm of the cutoff scale, which are produced where the null boundaries reach the asymptotic AdS boundary. Next we extend our studies to consider the complexity  of subregions on the boundary. In section~\ref{sec:subregions}, we propose an extension of CV duality with a covariant definition of extremal surface whose volume defines the subregion complexity, and we study the divergence structure of this quantity. In section~\ref{sec:subaction}, we propose an extension of CA duality to evaluate subregion complexity and we examine the corresponding UV divergences. In section~\ref{discuss}, we conclude with a brief discussion of our results, and we consider some directions for future research. In appendix~\ref{sec:manuel}, we review the prescription introduced by \cite{Lehner:2016vdi} for computing the gravitational action in the presence of null boundary surfaces. In appendix~\ref{sec:extremal3}, we apply the CV duality to a simple example and compare the results to the general geometric expressions found in section \ref{sec:extremal}. In appendix~\ref{newA}, we apply the CA duality to the simple example of global AdS and compare the results to the general geometric expressions found in section \ref{sec:action}. We also show the coefficients of the logarithmic divergences agree for two different schemes to regulate the bulk divergences. In appendix~\ref{sec:actionXXX}, we provide some geometric details which are needed for our CA duality calculations in section \ref{sec:action}.

% % % % % % % % % % % % % % % % % % % % % % % % % % % % % % % % % % % % % % % % % % % % % % % % % % % % % % % % % % % % % % % % % % % % % % % % % % % % % % % % % % % % % % % % % % % %
% % % % % % % % % % % % % % % % % % % % % % % % % % % % % % % % % % % % % % % % % % % % % % % % % % % % % % % % % % % % % % % % % % % % % % % % % % % % % % % % % % % % % % % % % % % %

\section{Complexity Equals Volume Conjecture \labell{sec:extremal}}
\label{sec:polkijh}

In this section, we examine the structure of UV divergences appearing in holographic complexity for the complexity=volume (CV) conjecture \cite{Susskind:2014rva,2014arXiv1403.5695S,Stanford:2014jda,Susskind:2014jwa,Susskind:2014moa}. 
Recall that the CV duality is captured by eq.~\reef{defineCV} and the corresponding construction is illustrated in Figure \ref{fig:CV}. Of course, the asymptotic AdS metric diverges at the boundary and so this prescription would yield a divergent volume for the extremal bulk surface, and hence a divergent complexity. As usual, we regulate the calculation by introducing a cut-off surface at some large radius, which will be related to a short-distance cut-off in the boundary theory --- see, \eg \cite{Emparan:1999pm,sken1,sken2}. Given this framework, we want to study the structure of the UV divergences appearing in the complexity.
\begin{figure}[t]
\centerline{\includegraphics[scale=0.4]{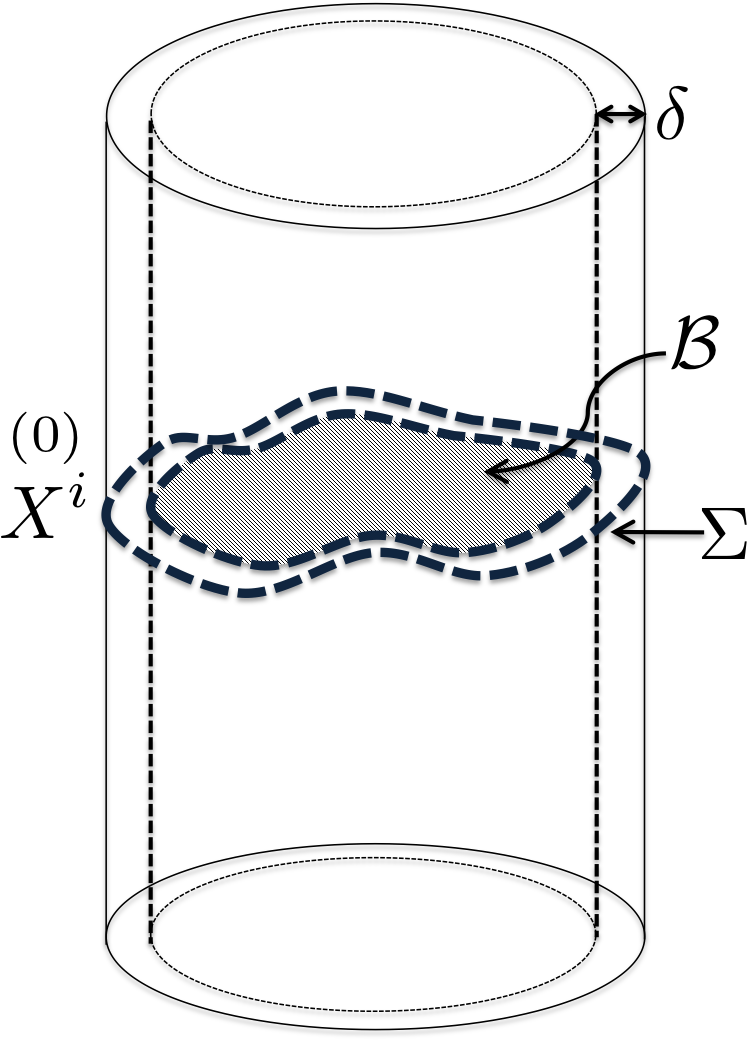}}
		\caption{Showing the extremal volume construction for the CV conjecture for $AdS_3$. We regulate the volume by introducing a cutoff surface at $z=\d$, where $\d$ is the short-distance cutoff in the boundary theory. 
		\labell{fig:CV} }
\end{figure}

In the following, we are borrowing results from \cite{Hung:2011xb,Hung:2011ta} --- see also \cite{sken1,sken2,Schwimmer:2008yh}.
According to the Fefferman-Graham (FG) construction \cite{Fefferman:2007rka,Graham:1999pm}, any asymptotically AdS geometry can be described with the following metric:
\bea
 %ds^2=\frac{L^2}{4}\,\frac{d\r^2}{\r^2} + \frac{1}{\r}\,g_{ij}(x,\r)dx^idx^j
ds^2=\frac{L^2}{z^2}\left(d z^2 + \,g_{ij}(x,z)\,dx^i dx^j\right) \,. 
 \labell{FGform}
\eea 
where $L$ is the AdS radius, $x^i$ denote the boundary directions,\footnote{Our notation in the following will be: Greek indices $\mu$ denote tensors in the bulk spacetime and run from $0$ to $d$; Latin indices $i$ from the middle of the alphabet denote tensors in the boundary spacetime, running from $0$ to $d-1$; and Latin indices $a$ from the beginning of the alphabet denote tensors in the boundary time slice, running from $1$ to $d-1$. We note that often the metric \reef{FGform} is expressed in terms of the dimensionless radial coordinate $\rho=z^2/L^2$ --- see \cite{sken1,sken2}.} and $z$ is the radial coordinate in the bulk. For $d$ boundary dimensions,  $g_{ij}(x,z)$ admits a Taylor series expansion in $z^2$ near the asymptotic boundary, \ie as $z\to0$, 
\bea
g_{ij} (x,\rho ) = \overset{(0)}{g_{ij}}( x^i ) + z^2\, \overset{(1)}{g_{ij}}\left( x^i \right) + z^4\, \overset{(2)}{g_{ij}}\left( x^i \right) +\cdots + z^{d} \overset{(d/2)}{g_{ij}}\left( x^i \right) + z^d\, \log (z/L) \, f_{ij}(x^i)+\cdots
\labell{eq:poli}
\eea 
where $\overset{(0)}{g_{ij}}$ is the boundary metric. We note that the logarithmic term arises only for even $d$. We see that this Taylor series breaks down at $\calo(z^d)$ where the logarithmic terms above start appearing for even $d$ ($ f_{ij}(x^i)$ is determined by $\overset{(0)}{g_{ij}}( x^i )$), or where odd powers, \eg $z^d$, start appearing for odd $d$. All of the expansion coefficients $\overset{(n)}{g_{ij}}$ for $n<{d/2}$ are determined in terms of the boundary metric $\overset{(0)}{g_{ij}}$ via the Einstein equations, \eg see \cite{Hung:2011xb,sken1,sken2}. For example,
\begin{equation}
\overset{(1)}{g_{ij}}\!\left( x^i \right) = -\frac{1}{d-2}\left( \mathcal{R}_{ij}[ \overset{(0)}{g}] - \frac{\overset{(0)}{g_{ij}}}{2(d-1)} \mathcal{R}[\overset{(0)}{g}] \right)\labell{eq:g1}
\end{equation}
where $\mathcal{R}_{ij}[ \overset{(0)}{g}]$ and $\mathcal{R}[\overset{(0)}{g}]$ are the Ricci tensor and Ricci scalar, respectively, calculated using the boundary metric $\overset{(0)}{g_{ij}}$. At order $\calo(z^d$), an independent solution (starting from $\overset{(d/2)}{g_{ij}}\left( x^i \right)$) appears which cannot be fixed by the boundary metric alone and it contains information about the expectation value of the boundary stress tensor, \eg \cite{sken1,sken2}.

As described earlier, we pick a time slice $\Sigma$ on the boundary and then look for the extremal codimension-one bulk surface $\mm{B}$ which approaches this slice at the boundary. We describe this submanifold embedded in the $d+1$ dimensional bulk using coordinates $X^{\mu}=X^{\mu}\left(\tau,\sigma^a \right)$, where $X^{\mu}=\left\lbrace z,x^i\right\rbrace$ and $\lbrace\tau,\sigma^a\rbrace$ are coordinates intrinsic to the submanifold $\mm{B}$. The induced metric on the bulk surface is:
\bea
\label{eq:ldfdfs}
h_{\alpha \beta}=\partial_{\alpha}X^{\mu} \partial_{\beta} X^{\nu}\, G_{\mu \nu}[X]
\eea
For simplicity, we make the gauge choice
\begin{equation}
\tau=z\,, \qquad	\quad h_{a \tau}=0\,.
\end{equation}
Then from eq.~\reef{defineCV}, the complexity is:
\begin{equation}
\cv = \frac{1}{G_NL}\,\int_\mm{B} d^{d-1}\sigma \, d\tau\ \sqrt{h}
\labell{eq:lkopj}
\end{equation}
where for simplicity, we have chosen $\ell=L$, the AdS scale.

Extremizing the volume above gives the following equation of motion:
\bea
\frac{1}{\sqrt{h}}\partial_{\alpha}\left(\sqrt{h}h^{\alpha \beta} \partial_{\beta} X^{\mu}\right) +h^{\alpha \beta}\Gamma^{\mu}_{\nu \sigma}\partial_{\alpha}X^\nu \partial_{\beta}X^\sigma = 0 \,.\labell{eq:eom}
\eea
where $\Gamma^{\mu}_{\nu \sigma}$ are the Christoffel symbols associated with the bulk metric $G_{\m \n}$.
This equation can be solved order by order near the boundary with a series solution for $X^i(\tau,\sigma^a)$ \cite{Graham:1999pm}:
\begin{equation}
X^i(\tau,\sigma^a)= \overset{(0)}{X^i}(\sigma^a) + z^2\,\overset{(1)}{X^i}(\sigma^a)  +z^4\, \overset{(2)}{X^i}(\sigma^a) + \dotsb \,.
\end{equation}
For  $n<\frac{d+1}{2}$ the term $\overset{(n)}{X^i}$ is determined by $\overset{(0)}{X^i}$.
For example, solving for $\overset{(1)}{X^i}$ yields:
\beq
\overset{(1)}{X^i}=\frac{L^2}{2(d-1)}\left(\nabla_{a}\partial^a \overset{(0)}{X^i} + \partial^a \overset{(0)}{X^j}\partial_a \overset{(0)}{X^k} \Gamma^i_{jk}\right)=\frac{L^2}{2(d-1)} \,K\,\bn^i \,,
\labell{lefft}
\eeq
where $\bn^i$ is the (future pointing) timelike unit normal  to the time slice $\S$ (in the boundary), and $K$ is the trace of the corresponding extrinsic curvature.

Using eqs.~\reef{eq:g1} and \reef{lefft}, one can begin to write the induced metric \reef{eq:ldfdfs} in a near boundary series expansion as well, \eg
\bea
\label{eq:posdf1}
h_{zz} = \frac{L^2}{z^2} \Big(1+ z^2\,\overset{(1)}{h_{zz}}+ \dotsb \Big) \ , \qquad  h_{ab}= \frac{L^2}{z^2}\Big( \overset{(0)}{{h}_{ab}} +  z^2\,\overset{(1)}{{h}_{ab}}+\dotsb \Big)
\eea 
where $\overset{(0)}{{h}_{ab}}$ represents the induced metric on the boundary time slice and $\overset{(1)}{{h}_{\a\b}}$ are the first order corrections. In particular, we find 
\beqa
\overset{(1)}{h_{zz}}&=&\frac{4 \overset{(1)}{X^i} \overset{(1)}{X^j}\overset{(0)}{g_{ij}}}{L^2} = - \frac{1}{(d-1)^2}\,K^2\,,
\label{eq:posdf2}\\
\overset{(1)}{{h}_{ab}}&=&\left(\partial_{a} \overset{(1)}{X^i}\partial_{b} \overset{(0)}{X^j}+\partial_{a} \overset{(0)}{X^i}\partial_{b} \overset{(1)}{X^j}\right)\overset{(0)}{{g}_{ij}}+ \partial_{a} \overset{(0)}{X^i}\partial_{b} \overset{(0)}{X^j}\overset{(1)}{{g}_{ij}} 
\nn
&=& -\frac{1}{d-1}\left( \frac{d-1}{d-2}\,\mathcal{R}_{ab} - \frac{\mathcal{R}}{2(d-2)}\,\overset{(0)}{h_{ab}} -K K_{a b}\right)\,,
 \nonumber
\eeqa
where  $\mathcal{R}_{ab}$ is the projection of the boundary Ricci tensor into the time slice $\Sigma$, \ie $\mathcal{R}_{ab}=e^i_a\,e^j_b\,\mathcal{R}_{ij}$ with $e^i_a= \frac{\pa X^i}{\pa \s^a}$. Further note that in both of these expressions, we have implicitly used $\bn^i \bn^j \overset{(0)}{g_{ij}}=-1$.

Given the bulk metric \reef{FGform}, we introduce a regulator surface at $\rho=\delta^2/L^2$, where $\delta$ then plays the role of a short distance cut-off in the boundary CFT. From eq.~\reef{eq:lkopj}, we can now extract the leading divergences in the complexity
\bea
\label{eq:volume}
 \cv &=&\frac{1}{G_NL}\int d^{d-1}\sigma \int_{z = \d} dz \frac{L^d}{z^d}\sqrt{\overset{(0)}{h}}\Bigg(1+\frac{z^2}{2}\,\bigg(\overset{(1)}{h_{zz}}+\overset{(0)}{h^{a b}}\overset{(1)}{h_{a b}}\bigg)+\dotsb \Bigg)
\nn
&=&\frac{L^{d-1}}{G_N}\int d^{d-1}\sigma \sqrt{\overset{(0)}{h}} \Bigg(\frac{1}{(d-1)\d^{d-1}}+\frac{1}{2(d-3)\d^{d-3}}\left(\overset{(1)}{h_{\tau \tau}}+\overset{(0)}{h^{a b}}\overset{(1)}{h_{a b}}\right)
+\dotsb \Bigg)\,.
\eea
Substituting eq.~\reef{eq:posdf2}, we find
\bea
\cv=\frac{L^{d-1}}{(d-1)G_N} \int d^{d-1}\sigma \sqrt{{h}} \left[\frac{1}{\d^{d-1}}-\frac{(d-1)}{2(d-2)(d-3)\d^{d-3}}\left(\mathcal{R}^{a}_{a}-\frac{1}{2}\,\mathcal{R}-\frac{(d-2)^2}{(d-1)^2}\,K^2\right)+\dotsb \right]
\labell{eq:divergence}
\eea
where to simplify the notation, we denote the induced metric on the boundary time slice as simply: ${h}_{ab}=\overset{(0)}{h}{}_{ab}$. In the above, we also use
$\mathcal{R}^{a}_{a}={h}^{ab}\,\mathcal{R}_{ab}$. The power law divergent terms here are regulator dependent, but we see that their coefficients have a geometric interpretation, \eg the leading divergence scales as the volume of $\Sigma$ while the sub-leading terms involve integrals of curvature invariants over this time slice. Of course, this is very similar in nature to the divergence structure found in holographic entanglement entropy. 
If we consider the special case of $d=3$ dimensions, the first sub-leading divergence in eq.~\reef{eq:divergence} is replaced by a logarithmic term
\begin{equation}
\cv^\mt{(universal)}= \log\left(\frac{\d}{L}\right)\frac{L^2}{8G_N}\int d^2\sigma \sqrt{{h}}\left(4\,\mathcal{R}^a_a - 2\,\mathcal{R}- K^2\right)\,.
\labell{roundx}
\end{equation}

Similar logarithmic divergences will generally appear whenever the boundary dimension is odd, \ie when the extremal surface is even dimensional. Of course, they are related to the submanifold conformal anomalies studied in \cite{GraWit}.
The dimensionless coefficient(s) of these logarithmic divergences will be regulator-independent parameters characterizing the underlying boundary theory,%\footnote{Sergey Solodukhin suggested that these coefficients should be related to boundary terms in the conformal anomaly of the boundary CFT, as studied in \cite{Solo15}.}
As expected in eq.~\reef{roundx}, this parameter is proportional to $L^2/G_N\sim L^2/\ell_{\ssc Planck}^2$. This ratio is well known to characterize the number of degrees of freedom in the boundary CFT dual to (four-dimensional) Einstein gravity. However, $L^2/G_N$ is the only dimensionless parameter intrinsic to the bulk theory and so the same ratio appears in any physical quantity involving some count of degrees of freedom, \eg the entropy density of a thermal bath. One approach to distinguish the various parameters appearing in different physical quantities in holographic boundary theories is to consider higher curvature theories for the bulk gravity, \eg see \cite{Hung:2011xb,gb2,Myers:2010jv,corner}. The challenge in the present case would be developing the extension of the CV duality \reef{defineCV} for higher curvature bulk theories.

Let us provide a few geometric comments on the above result: Using the Gauss-Codazzi relations, we could replace $\mathcal{R}^{a}_{a}$ in eqs.~\reef{eq:divergence} or \reef{roundx} in favour of the intrinsic Ricci scalar on the time slice $\Sigma$, as well as a term proportional to $K^{a b} K_{a b}$. 
Note that in the case of entanglement entropy, the first subleading contribution, \eg  the universal contribution for $d=4$ contains a term with the Weyl curvature $C_{ijkl}$ of the boundary metric \cite{Solodukhin:2008dh}, however, we see that this tensor does not appear in eq.~\reef{eq:divergence}. The key difference is that for holographic entanglement, one is considering a codimension-two surface and hence there are two normal vectors which can be contracted with $C_{ijkl}$. On the other hand, in evaluating holographic complexity, one considers a codimension-one surface in the boundary and hence there is one normal vector. Then given the symmetries and traceless property of the Weyl curvature, there are not enough geometric structures to construct a scalar which is linear in $C_{ijkl}$. However, the Weyl tensor might appear in higher order contributions to the complexity with a scalar  such as $C_{ijkl}C^{ijkl}$. In Appendix~\ref{sec:extremal3}, we study the divergence structure in a specific example of a CFT living on a sphere.

\subsection*{General divergence structure}

While the calculations above are somewhat preliminary, our experience with analogous calculations for holographic entanglement entropy \eg \cite{Hung:2011xb,Hung:2011ta,ajay}, suggests the following framework:
With $d$ boundary dimensions, the general structure of divergences appearing in the CV duality is:
\bea
\cv(\S) =\frac{1}{\delta^{d-1}}\,\int_\Sigma d^{d-1}\sigma\,\sqrt{{h}}\ v(\mathcal{R},K)\qquad{\rm where}\ \ \ v(\mathcal{R},K)=\sum_{n=0}^{\lfloor{\frac{d-1}{2}}\rfloor}\sum_{i}\ \, c_{i,n}(d)\,\delta^{2n}\,[\mathcal{R},K]^{2n}_{i} \,.
\labell{eq:e1}
\eea
That is, there can be a number of power law divergences beginning with $1/\delta^{d-1}$ where the coefficient is proportional to the volume of the time slice $\Sigma$. The power of the subsequent divergences is reduced by two at each step and the coefficients of these terms are fixed purely in terms of a local integral over $\Sigma$ of various curvature invariants on the boundary. The schematic expression $[\mathcal{R},K]^{2n}_{i}$ indicates invariant combinations of boundary curvatures (represented by $\mathcal{R}$) and the extrinsic curvature of the time slice (represented by $K$), with a mass dimension of $2n$, so that the combination $\delta^{2n}\,[\mathcal{R},K]^{2n}_{i}$ is dimensionless. Of course, for odd (even) $d$, there are only even (odd) power divergences. Further, in odd dimensions with the special case that $2n=d-1$, logarithmic divergences appear which provide universal parameters characterizing the underlying CFT, as discussed below eq.~\reef{roundx}. 

Again in eq.~\reef{eq:e1} and in the preceding example \reef{eq:divergence}, we observe that there are only even (odd) power divergences for a boundary CFT in odd (even) $d$. At first sight, one may have thought the first subleading divergence would be proportional to $\int_\Sigma d^{d-1}\sigma\sqrt{{h}}\, K /\delta^{d-2}$, which would have disrupted this pattern. However, the simple reason that this term can not appear is that it depends on the orientation of time, \ie the orientation of $\bn^i$, while the bulk volume in eq.~\reef{defineCV} does not. We should add that in the boundary theory, the natural notions of complexity are intrinsic to a given state, and are independent of the time evolution of the state under some Hamiltonian. That is, the invariance of $\cv(\S)$ under reversing the time orientation can be counted as a success of the definition in eq.~\reef{defineCV}.

% % % % % % % % % % % % % % % % % % % % % % % % % % % % % % % % % % % % % % % % % % % % % % % % % % % % % % % % % % % % % % % % % % % % % % % % % % % % % % % % % % % % % % % % % % % %
% % % % % % % % % % % % % % % % % % % % % % % % % % % % % % % % % % % % % % % % % % % % % % % % % % % % % % % % % % % % % % % % % % % % % % % % % % % % % % % % % % % % % % % % % % % %
% % % % % % % % % % % % % % % % % % % % % % % % % % % % % % % % % % % % % % % % % % % % % % % % % % % % % % % % % % % % % % % % % % % % % % % % % % % % % % % % % % % % % % % % % % % %

\section{Complexity Equals Action Conjecture
	\labell{sec:action}}

In this section, we examine the divergence structure emerging in the CA duality \cite{Brown:2015bva,Brown:2015lvg}. The procedure to evaluate the gravitational action for the Wheeler-DeWitt patch was carefully examined in \cite{Lehner:2016vdi}. 
In particular, the WDW patch has null boundary surfaces and ref.~\cite{Lehner:2016vdi} constructed the boundary terms which must be added to the gravitational action for these null boundaries and for the joints where such null boundaries intersect with other boundary surfaces. We review these results in appendix \ref{sec:manuel}.%\footnote{However, we present the boundary terms for the gravitational action with a slightly different conventions than in \cite{Lehner:2016vdi}.}

Again, the holographic complexity $\ca$ diverges because the WDW patch extends to the asymptotic AdS boundary and the focus here is to examine the structure of the resulting UV divergences. As in the previous section, we adopt the usual approach to regulating our calculations of introducing a cut-off surface at some large radius, \eg see \cite{Emparan:1999pm,sken1,sken2}. However, given this framework, we can propose two different approaches to regulating the WDW action, as illustrated in  figure~\ref{fig:space}. In particular, in figure~\ref{fig:space}a, we discard the portion of  the WDW patch extending beyond the regulator surface, \ie we only integrate the bulk action out to this maximum radius. In this case, the regulated WDW region has a new timelike boundary segment and two null joints where the regulator surface intersects with the null sheets defining the past and future boundaries of the WDW patch, both of which contribute to $I_\mt{WDW}$. In figure~\ref{fig:space}b, we instead regulate the action by simply shifting the edge of the WDW patch inwards to the regulator surface and hence we only have a single null joint at this time slice. In Appendix~\ref{newA}, we will show that the structure of the UV divergences in the corresponding complexity $\ca$ is the same for both regularization procedures with a simple example. For simplicity, in the following general discussion, we adopt the second regulator which is shown in figure~\ref{fig:space}b.
\begin{figure}[h]
	\centering
	\includegraphics[scale=0.34]{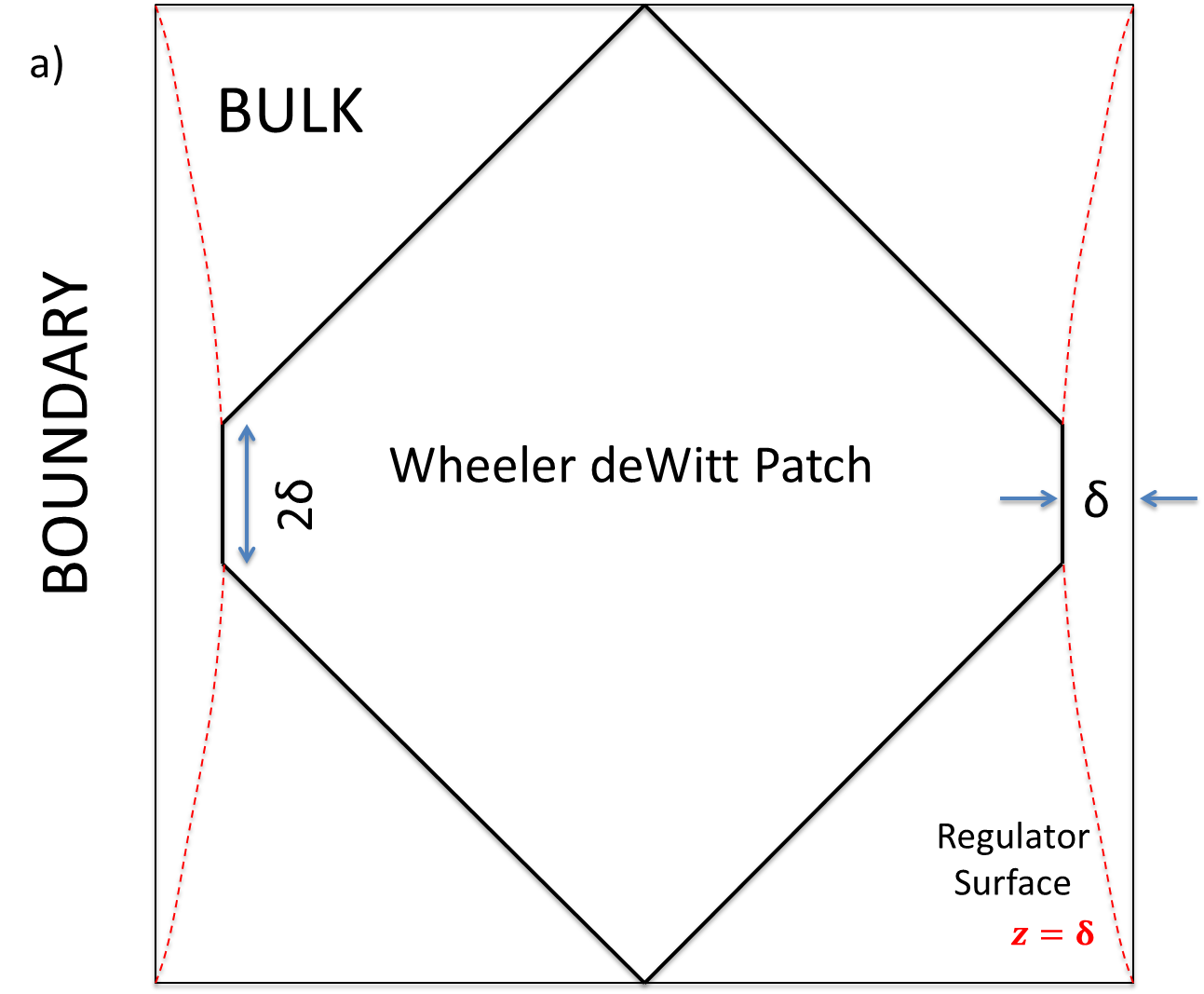}
	\includegraphics[scale=0.34]{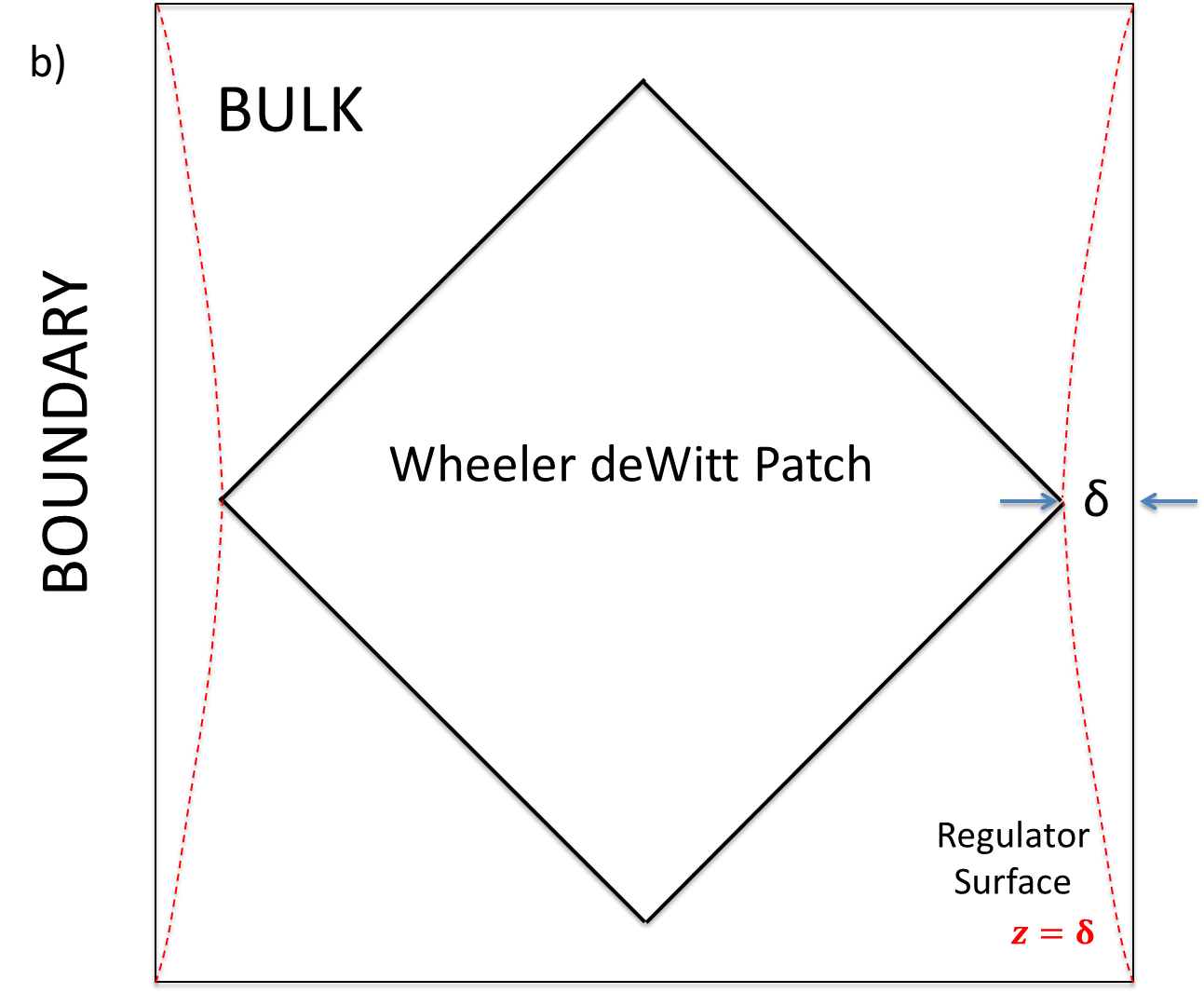}
	\caption{Wheeler-DeWitt patch with two different regularizations. In both cases, the WDW patch terminates at the regulator surface: (a) The edge of the WDW patch is the time slice on the asymptotic boundary. The action contains a GHY surface term and two joint terms from the new boundary at $z=\d$. (b) The edge of the WDW patch is the time slice in the regulator surface.  The action contains null joint term from the edge at $z=\d$. 
		\labell{fig:space}}
\end{figure}

We begin again with the bulk metric in FG gauge, as  in eq.~\reef{FGform}.
%\bea
%ds^2=\frac{L^2}{z^2}\left(d z^2 + \,g_{ij}(x,z)\,dx^i dx^j\right) \,. 
%\labell{eq:polinn}
%\eea
For simplicity, we restrict the boundary metric to take the following form
\bea
\overset{(0)}{g_{ij}}( x^i ) 
\,dx^i\, dx^j = -dt^2 + h_{ab}(t,\s)\,d\s^a d\s^b\,.
\labell{eqLpolikmnb}
\eea
In particular, the ($ta$)-components of the boundary metric are fixed to be zero and the ($tt$)-component is simply --1. However, we should note that this form is not  preserved in the full tensor $g_{ij}(x,z)$ in eq.~\reef{FGform}. For example, the first correction \reef{eq:g1} appearing in the Taylor expansion around $z=0$, \ie at order $z^2$, will generally introduce a nonvanishing $g_{ta}$ and nontrivial dependence on $x^i=(t,\s^a)$ in $g_{tt}$. We made this choice for the boundary metric \reef{eqLpolikmnb} as it greatly simplifies the analysis of the WDW action below, but it is still general enough that most of curvature invariants appearing in the power law divergences are still nontrivial. In the following, we will compute the leading divergences of the WDW action, working to second order in the near boundary expansion.

We begin by determining the equations defining the null boundaries of the WDW patch near the asymptotic boundary, \ie $z=0$.\footnote{We would like to thank Run-Qiu Yang for pointing out an error in an earlier version of this discussion.} We will also set the time slice $\S$ to be $t=0$ and hence in our calculations, we will expand both for small $z$ and for small $t$. For simplicity, we focus on the future null boundary for most of the discussion. 
Given the boundary metric \reef{eqLpolikmnb}, this null surface can be described as $t=z-\delta+\cdots$ to leading order and the corresponding normal would be $\nk1=\a_1(dt-dz+\cdots)$, where $\a_1$ is some (positive) normalization constant.\footnote{We have adopted the convention that $\nk1$ points outward from the region of interest --- see appendix \ref{sec:manuel}.} Now we wish to extend the former equation to
\beq
S^+\ :\quad t= t_+(z,\s^a)=\ f_+(z,\s^a)-f_+(\delta,\s^a)
\labell{surfx1}
\eeq
for $t\ge0$\footnote{In general, we could have a different functional dependence on $\left(\delta,\sigma^a\right)$, but this works to the required order since we simply integrate using eq.~\reef{noraX}}. In the vicinity of the boundary, $f_+(z,\s^a)$ has an expansion in powers of $z$, which we write as
\beq
f_+(z,\s^a)=z+\frac{z^2}2\,f^{(2)}(\s^a)+\frac{z^3}6\,f^{(3)}(\s^a)+\cdots\,,
\labell{surfx2}
\eeq
where the leading term was fixed in the above discussion. The form of the second expression in eq.~\reef{surfx1} was chosen to ensure that $t=0$ at $z=\delta$ (for all $\s^a$) order by order in this $z$ expansion. Now, in fact, to the order that we will be interested in here, the coefficients in eq.~\reef{surfx2} can be fixed by demanding that the normal to $S^+$ is null. That is, we determine $f^{(2)}(\s^a)$ and $f^{(3)}(\s^a)$ by imposing that $\nk1\cdot\nk1=0$  with the one-form $\nk1$ given by the exterior derivative of the function determining the boundary surface (up to an overall normalization factor), \ie $\nk1=\a_1\,d[t-t_+(z,\s^a)]$. The result of this calculation is that $f^{(2)}(\s^a)=0$ while
\beq
f^{(3)}(\s^a)=n^i n^j\,\overset{(1)}{g_{ij}}\!\left( \s^a,t=0 \right)  =  -\frac{1}{d-2}\left.\left( \mathcal{R}^{a}_{a} - \frac{2d-3}{2(d-1)} \mathcal{R} \right) \right|_{t=0} 
\labell{surfx3}
\eeq
where $n_i=\delta^t_i$ is the unit normal to the boundary surface $t=0$. The second expression written in terms of the boundary curvature follows from eq.~\reef{eq:g1}, as well as making the replacement that $n^i n^j\mathcal{R}_{ij}=\mathcal{R}^a_a-\mathcal{R}$. Hence to order $z^3$, we can write the null boundaries of the WDW patch as
\bea
S^+\ :&\quad t=t_+(z,\s^a)=\ \ (z-\d) + \frac{ f^{(3)}(\s^a)}{6}  (z^3-\d^3) + \cdots
& \quad{\rm for}\ \ t\ge 0\,, 
\labell{eq:fjksss}\\
S^-\ :&\quad t=t_-(z,\s^a)= -(z-\d) - \frac{f^{(3)}(\s^a)}{6}  (z^3-\d^3) + \cdots
& \quad{\rm for}\ \ t\le 0\,. 
\nonumber
\eea
In the second line above, the result for the past null boundary $S^-$ is found with the same analysis as that given for $S^+$ above.

At this point, we are ready to evaluate the WDW action with eq.~\reef{actall}, and we begin with the bulk integral of the Einstein-Hilbert term and the cosmological constant. Using the Einstein equations for the bulk, we may substitute ${R}=-d(d+1)/L^2$, and this contribution simplifies to evaluating the spacetime volume of the WDW patch
\bea
I_\mt{bulk} =-\frac{d}{8\pi G_NL^2}\,{\cal V}({\mm{W}}) \,.
\labell{racoon}
\eea  
Now recall $g_{at}\sim \calo(z^2)$ and thus we find to leading order 
\beq
\sqrt{-g(x,z)} =\sqrt{-g_{tt}(x,z)}\,\sqrt{{\rm det} [g_{ab}(x,z)]} + \calo(z^4)
\labell{rocky}
\eeq
in the measure of the above bulk integral, \ie  the cross terms with $g_{at}$ only appear at order $\calo(z^4)$.
Hence it is useful to write a double expansion for $\sqrt{\g} \equiv \sqrt{{\rm det} [g_{ab}(x,z)]}$:
\bea
\sqrt{\g}=\sqrt{h(\s)}\Big([1+q_0^{(2)}(\s^a)z^2+\cdots ]+[q_1^{(0)}(\s^a)+q_1^{(2)}(\s^a)z^2 +\cdots]t+ [q_2^{(0)}(\s^a)
%+q_2^{(2)}(\s^a)z^2
 +\cdots]t^2  +\dotsb \Big)
\labell{pout}
\eea
where $\sqrt{h (\s)}\equiv\sqrt{{\rm det}[h_{ab}(\s^a,t=0)]}$ using the boundary metric in eq.~\reef{eqLpolikmnb}. Hence using these expressions, as well as eq.~\reef{surfx3} to substitute for $\overset{(1)}{g_{tt}}$, we identify the leading contributions near the asymptotic boundary in eq.~\reef{racoon} as  
\bea
I_\mt{bulk}&=&-\frac{d}{8\pi G_NL^2}\,\int d^{d-1}\s \int_{\delta}dz\int_{t_-(z,\s^a)}^{t_+(z,\s^a)}\!\!\!dt\ \frac{L^{d+1}}{z^{d+1}}\,\sqrt{-g(x,z)}
\nn
&=& -\frac{d\,L^{d-1}}{8\pi G_N}\,\int d^{d-1}\s\sqrt{h (\s)} \int_{\delta}\frac{dz}{z^{d+1}}\int_{t_-(z,\s^a)}^{t_+(z,\s^a)}\!\!\!dt\,   \bigg[1 - \frac{1}{2} f^{(3)}(\s^a) z^2  +  q_0^{(2)}(\s^a)z^2  + q_1^{(0)}(\s^a)t + q_2^{(0)}(\s^a)t^2 + \cdots \bigg] 
\nn
&=&-\frac{d\,L^{d-1}}{8\pi G_N}\,\int d^{d-1}\s\sqrt{h (\s)} \int_{\delta}\frac{dz}{z^{d+1}}\bigg[\big(1 - \frac{1}{2} f^{(3)}(\s^a) z^2  +  q_0^{(2)}(\s^a)z^2\big)\big[t_+(z,\s^a)-t_-(z,\s^a)\big]  
\labell{bullw1}\\
&&\qquad\qquad\qquad\qquad\qquad\qquad
+ \frac{q_1^{(0)}(\s^a)}{2}\big[t^2_+(z,\s^a)-t^2_-(z,\s^a)\big] + \frac{q_2^{(0)}(\s^a)}{3} \big[t^3_+(z,\s^a)-t^3_-(z,\s^a)\big]  + \cdots \bigg] \,.
\nonumber
\eea 
Now substituting $t_+(z,\s^a)-t_-(z,\s^a)=2(z-\d)+\cdots$ into eq.\reef{eq:fjksss}, the leading divergence in the above expression becomes 
\bea
-\frac{dL^{d-1}}{4\pi G_N }\int d^{d-1}\s \sqrt{h} \int_{\delta}  \frac{dz}{z^{d+1}}  (z-\d) = -\frac{L^{d-1}}{4\pi G_N (d-1) }\frac{1}{\d^{d-1}} \int d^{d-1}\s  \sqrt{h}  \,. \labell{eq:polikm1}
\eea
That is, the leading divergence in the bulk integral is proportional to $\calv(\S)/\d^{d-1}$. We note, however, that this leading term is {\it negative} --- see further comments below. 

Next, $t^2_+(z,\s^a)-t^2_-(z,\s^a)$ vanishes to the order that we are calculating and hence the first subleading divergence in the above expression becomes
\bea
&&-\frac{dL^{d-1}}{8\pi G_N }\int d^{d-1}\s \sqrt{h}  \int_{\delta} \frac{dz}{z^{d+1}} \bigg[(z-\d)z^2 \Big(2q_0^{(2)}(\s^a) - f^{(3)}(\s^a)\Big)  + \frac{f^{(3)}(\s^a)(z^3-\d^3)}{3}  + \frac{2q^{(0)}_2(\s^a)(z-\d)^3}{3}  \bigg] 
\nn
&&\qquad=-\frac{dL^{d-1}}{4\pi G_N } \frac{1}{\d^{d-3}} \int d^{d-1}\s \sqrt{h} \,   \bigg[ -\frac{f^{(3)}(\s^a)}{d(d-2)(d-3)} +\frac{q^{(2)}_0(\s^a)}{(d-2)(d-3)} + \frac{2q^{(0)}_2(\s^a)}{d(d-1)(d-2)(d-3)}   \bigg]
\nn
&&\qquad=-\frac{L^{d-1}}{16\pi G_N } \frac{1}{\d^{d-3}} \int  \frac{d^{d-1}\s\, \sqrt{h}}{(d-1)(d-2)(d-3)}\,  \bigg[4K^2+ 4K_{ab}K^{ab}
+(d-7)\,\mathcal{R}-2 (d-3)\,\mathcal{R}^a_a
\bigg]
\labell{big22}
\eea
We used a number of identities to produce the geometric expression in the last line above. In particular,
we substitute the result in eq.~\reef{surfx2} and further, in appendix \ref{sec:actionXXX}, we derive:
\bea
	q^{(2)}_0(\s^a) &=&	-\frac{1}{2(d-2)}\left.\left(\mathcal{R}^{a}_{a}-\frac{1}{2}\,\mathcal{R}\right)\right|_{t=0} \,,  
\labell{rumpt}\\
	q_2^{(0)}(\s^a)    &=& \frac{1}{2}\left. \Big(  K^2 +   K_{ab}K^{ab}+ \mathcal{R}^{a}_{a}- \,\mathcal{R}  \Big)\right|_{t=0} \,.\nonumber
\eea 

In addition to the bulk term above, we must include the contribution from the joint where the past and future null sheets \reef{eq:fjksss} intersect, \ie $(z,t)=(\d,0)$ --- see figure \ref{fig:space}b. Given eq.~\reef{eq:fjksss}, we may write the null normals to order $\calo(z^2)$ as
\bea 
S^+\ :&&\quad\nk1= \a_1\big(dt-dt_+(z,\s^a)\big)\simeq \a_1\left(dt-dz-\frac{z^2}2\,f^{(3)}\,dz
+\cdots\right)\,,
\labell{noraX}\\
S^-\ :&&\quad\nk2= \a_2\big(-dt+dt_-(z,\s^a)\big) \simeq \a_2\left(-dt-dz-\frac{z^2}2\,f^{(3)}\,dz
+\cdots\right)\,.
\nonumber
\eea 
Hence their inner product yields
\bea 
\nk1 \cdot \nk2 \simeq \a_1 \a_2 \frac{z^2}{L^2} \left(- g^{tt}  + \left(1+ \frac{z^2}2\,f^{(3)}\right)^2 +\calo(z^4)\right)  = 2 \a_1 \a_2 \frac{z^2}{L^2}\left(1+{z^2}\,f^{(3)}\right)+ \calo(z^4)\,,
\eea 
where we have used $g^{tt}=-1- \overset{(1)}{g_{tt}}z^2+ \calo(z^4)$, as well as substituting eq.~\reef{surfx3} for $\overset{(1)}{g_{tt}}$.
Now using the prescription in appendix \ref{sec:manuel}, as well as eq.~\reef{pout}, the leading contributions from the joint term are 
\bea
I_\mt{jnt}&=&-\frac{L^{d-1}}{8\pi G_N \d^{d-1}}\int d^{d-1}\s \sqrt{\g} \,\left. \log \Big( \frac{\nk1\cdot\nk2}{2}\Big)\right|_{(z,t)=(\d,0)} \nn
&\simeq&-\frac{L^{d-1}}{4\pi G_N \d^{d-1}}\log\left(\frac{\sqrt{\a_1 \a_2}\,\d}{L}\right)\int d^{d-1}\s \sqrt{\g}
-\frac{L^{d-1}}{8\pi G_N \d^{d-3}}\int d^{d-1}\s \sqrt{\g}\, f^{(3)}(\s^a)
\nn
&\simeq&\frac{L^{d-1}}{4\pi G_N}\log\left(\frac{L}{\sqrt{\a_1 \a_2}\,\d}\right) \int d^{d-1}\s \sqrt{h}\left[\frac{1}{\d^{d-1}}-\frac{2\mathcal{R}^a_a-\mathcal{R}}{4(d-2)\,\d^{d-3}} \right]
\labell{pole}\\
&&\qquad\qquad+\frac{L^{d-1}}{8\pi G_N \d^{d-3}}\int d^{d-1}\s \sqrt{h}\,\frac{1}{d-2}\left[ \mathcal{R}^{a}_{a} - \frac{2d-3}{2(d-1)} \mathcal{R} \right] 
\,.\nonumber
\eea
The leading term here is proportional to $\log\left({L}/{\d}\right)\calv(\S)/\d^{d-1}$ and hence this contribution from the asymptotic joint $S^+\cap S^-$ becomes the leading divergence in the WDW action. This joint divergence is always positive in contrast to the leading divergence in the bulk action \reef{eq:polikm1}, which guarantees the positivity of the corresponding complexity in the boundary theory.

Combining the contributions in eqs.~\reef{eq:polikm1}, \reef{big22} and \reef{pole}, we find the leading divergences in the holographic complexity \reef{none},
\bea
\ca(\S)&=&  \frac{1}\pi\left(I_\mt{bulk}+I_{jnt} \right)
\simeq-\frac{L^{d-1}}{4\pi^2 G_N } \int d^{d-1}\s \sqrt{h }\, \Bigg[ \frac{1}{d-1}\,\frac{1}{\d^{d-1}}   
\labell{trumpet2}\\
&&\qquad  + \frac{1}{\d^{d-3}}   \frac{1}{2(d-1)(d-2)(d-3)}\,  \bigg(2K^2+ 2K_{ab}K^{ab}
+(d^2-4d+1)\,\mathcal{R}-d (d-3)\,\mathcal{R}^a_a    \bigg)
\Bigg]
\nn
&&\qquad + \frac{L^{d-1}}{4\pi^2 G_N}\log\left(\frac{L}{\sqrt{\a_1 \a_2}\,\d}\right) \int d^{d-1}\s \sqrt{h }\,\Big[\frac{1}{\d^{d-1}}-\frac{1}{4(d-2)\d^{d-3}} (2\mathcal{R}^a_a-\mathcal{R}) \Big]\,.
\nonumber
\eea 
In Appendix \ref{newA}, we compare these results with an explicit example in global AdS, and we find that the leading divergences match, as expected. We also examine the alternate regularization in figure \ref{fig:space}a applied to this example.

\subsection*{General divergence structure}

From the insights coming from the above calculation, we expect the general structure of the divergences in the CA duality to be:
\bea
\label{eq:ksdfj}
\ca(\S) &=&\frac{1}{\delta^{d-1}}\,\int_\Sigma d^{d-1}\sigma\,\sqrt{{h}}\,\left[ v_1(\mathcal{R},K)+\log\left(\frac{L}{\alpha\,\d}\right)\ v_2(\mathcal{R},K)\right]
\labell{cute1}\\
&&\quad{\rm with}\qquad
\ v_k(\mathcal{R},K)=\,\sum_{n=0}^{\lfloor{\frac{d-1}{2}}\rfloor}\,\sum_{i}\  c^{[k]}_{i,n}(d)\ \d^{2n}\ [\mathcal{R},K]^{2n}_{i} \,,
\labell{cute2}
\eea
for $d$ boundary dimensions. As in eq.~\reef{eq:e1}, the schematic expressions $[\mathcal{R},K]^{2n}_{i}$ appearing in each of the integrands indicate invariant combinations of boundary curvatures (denoted $\mathcal{R}$) and the extrinsic curvature of the time slice (denoted $K$), with a mass dimension of $2n$, so that the combination $\delta^{2n}\,[\mathcal{R},K]^{2n}_{i}$ is dimensionless. 
Hence for the CA duality, there are two sets of divergences: The first (coming from the bulk term in the action) associated with $v_1$ is a series of power law divergences beginning with $1/\delta^{d-1}$ and then lower powers decreasing in steps of two. The second set of divergences (coming from the joint term) identified with $v_2$ involve a $\log\d$ multiplying powers $1/\d^{d-2n-1}$. In both series, only even (odd) power divergences appear for odd (even) $d$. Further, when $d$ is odd, the final term in $v_1$ with $2n=d-1$ yields an extra $\log(L/\d)$. Hence in odd $d$, the universal term (proportional to $\log\d$) has contributions coming from both $v_1$ and $v_2$, while in even $d$ there is no $\log\d$ term.
Again all of the coefficients in the two integrands are determined by local integrals on $\S$ of various curvature invariants on the boundary. We note that there is some ambiguity in these expressions related the logarithmic factor in eq.~\reef{cute1}, and in particular, because of the coefficient $\a$ in the argument there. We will return to discuss this point in section \ref{discuss}.

We should comment again on the appearance of only even (odd) power divergences for odd (even) $d$. As discussed at the end of section \ref{sec:extremal}, this indicates that are no contributions in $v_{1,2}$ which are proportional to an odd power of the extrinsic curvature. However, this is actually a requirement for a holographic definition of the complexity since the latter should be independent of the orientation of time on the boundary. The gravitational action, as described in appendix \ref{sec:manuel}, is independent of the orientation of time\footnote{We have adopted slightly different conventions in appendix \ref{sec:manuel} than originally presented in \cite{Lehner:2016vdi}. Our prescription is entirely equivalent to that given in \cite{Lehner:2016vdi}, but the latter has the disadvantage that it explicitly refers to the time orientation. Since this is not the case in appendix \ref{sec:manuel}, the above statement becomes manifest with the present prescription.} and so the definition \reef{none} of CA duality satisfies this requirement.

% % % % % % % % % % % % % % % % % % % % % % % % % % % % % % % % % % % % % % % % % % % % % % % % % % % % % % % % % % % % % % % % % % % % % % % % % % % % % % % % % % % % % % % % % % % %
% % % % % % % % % % % % % % % % % % % % % % % % % % % % % % % % % % % % % % % % % % % % % % % % % % % % % % % % % % % % % % % % % % % % % % % % % % % % % % % % % % % % % % % % % % % %
% % % % % % % % % % % % % % % % % % % % % % % % % % % % % % % % % % % % % % % % % % % % % % % % % % % % % % % % % % % % % % % % % % % % % % % % % % % % % % % % % % % % % % % % % % % %

\section{Subregion Complexity: CV Duality
\labell{sec:subregions}}

It is also interesting to extend holographic complexity to subregions. That is, one would evaluate the complexity of the mixed state produced by reducing the boundary state to a specific subregion of the boundary time slice. Given the proposal that in holography, this mixed state is encoded in the corresponding entanglement wedge in the bulk \cite{ewedge1,Headrick:2014cta}, it is natural that the holographic prescription for the  complexity of this state should involve the entanglement wedge. These ideas were first considered for time-independent geometries by \cite{Alishahiha:2015rta} in the context of CV duality --- see also \cite{Ben-Ami:2016qex}. Below, we propose a covariant definition of the appropriate volume, which can be applied in a time-dependent bulk and reduces to \cite{Alishahiha:2015rta} for static geometries. We then examine the structure of the UV divergences for this subregion complexity.

For a static bulk geometry, the CV duality for subregions \cite{Alishahiha:2015rta} evaluates the volume of the extremal codimension-one surface in the bulk which is bounded by the subregion on the asymptotic boundary and the Ryu-Takayanagi (RT) surface \cite{Ryu:2006bv,Ryu2} for this subregion.  A natural extension of this prescription to a time-dependent bulk spacetime refers to the Hubeny-Rangamani-Takayanagi (HRT) prescription for holographic entanglement entropy \cite{Hubeny:2007xt} (see also \cite{ha}) as follows:\footnote{There is some ambiguity in producing a covariant definition of the CV duality for subregions, just as there was for holographic entanglement entropy \cite{Hubeny:2007xt}. However, we think this proposal is the most natural as it connects the complexity directly to the entanglement wedge \cite{Headrick:2014cta}.}
\begin{itemize}
\i Beginning with a subregion $A$ on a given boundary time slice $\Sigma$, one constructs $\cale_{ A}$, the corresponding extremal HRT surface in the bulk --- this defines the inner edge of the entanglement wedge \cite{Headrick:2014cta}. Then consider the codimension-one bulk surfaces $\mm{R}_A$ which are bounded by this HRT surface $\cale_{A}$ and the boundary subregion $A$. The subregion complexity is then conjectured to be given by maximizing the volume $\mathcal{V(R}_A)$ over this class of surfaces:
\begin{equation}
\mathcal{C}_{\text{V}}(A) =\ \mathrel{\mathop {\rm
max}_{\scriptscriptstyle{A\,\cup\, \cale_{A}\,=\,\partial \mm{R}_A}} {}\!\!}\left[\frac{\mathcal{V(R}_A)}{G_N \, \ell}\right] \, ,
\labell{defineCVx}
\end{equation}
where as in eq.~\reef{defineCV}, $\ell$ is some length scale associated with the bulk geometry, \eg the AdS radius.
\end{itemize}
We note that in defining the entanglement wedge \cite{Headrick:2014cta}, reference was made to `homology surfaces,'  which had precisely the definition of $\mm{R}_A$ above. Hence, our proposal for CV duality for subregions assigns a special role to the homology surface with maximal volume. 

\subsection*{Divergence Structure}	

Now we make some general comments on the divergence structure of the complexity of a subregion $A$ that would arise from the above proposal \reef{defineCVx}.
If the subregion is extended to the full time slice on the boundary, \ie $A=\S$, we will reproduce the divergence structure found in section \ref{sec:extremal}. In particular, the coefficients of the various power law divergences are determined by local integrals of geometric invariants over $A=\S$, as in eq.~\reef{eq:e1}.  When the subregion $A$ is a proper subregion of $\S$, we will still have the same `volume' contributions $v(\mathcal{R},K)$ now integrated only over the subregion. However, there is the additional possibility that new divergences may arise associated with
the boundary of the subregion $\pa A$, which we will refer to as the entangling surface following the discussions of entanglement entropy. Thus, we expect that the full divergence structure of the subregion complexity has the following general form:
\bea
\cv(A)=  \frac{1}{\delta^{d-1}}\int_{ A}d^{d-1}\s \,\sqrt{h}\ v(\mathcal{R},K) + \frac{1}{\delta^{d-2}}\int_{\pa A} d^{d-2}\tilde{\s}\,\sqrt{\tilde{h}}\ b({\mm{R}},\tilde{K};{\bs},\bt)\,.
\labell{eq:jdkkks}
\eea 
Again, the first term would be identical to that found in eq.~\reef{eq:e1} except that the integration is restricted to the subregion $A$. In the second term, we have a local integral over the entangling surface, and $\tilde{h}_{ab}$ is the induced metric on $\pa A$. Now the integrand $b({\mm{R}},\tilde{K};{\bs},\bt)$ is again a dimensionless quantity constructed from the cut-off $\delta$ and various geometric curvatures including  $\mathcal{R}$, the background curvatures of the AdS boundary, and $\tilde{K}^i_{ab}$, the extrinsic curvatures of the codimension-two entangling surface, \ie
\beq
b({\mm{R}},\tilde{K};{\bs},\bt)=\sum_{n=0}^{d-1}\sum_{i}\ \, \tilde c_{i,n}(d)\,\delta^{n}\,[\mathcal{R},\tilde{K};{\bs},\bt ]^{n}_{i} \,.
\labell{purple}
\eeq
However, we have also introduced an explicit dependence on a particular basis of vectors in transverse space. The entangling surface $\pa A$ is a codimension-two surface and so the transverse space is spanned by a basis of two unit vectors. In  discussions of entanglement entropy in (relativistic) theories, there is nothing to distinguish one such basis from another. However, in the present discussion, we have defined a preferred time slice $A$ where the state resides for which we are evaluating the complexity. Hence there is a preferred basis in the space transverse to $\pa A$: ${\bs}^i$, the spacelike unit vector which is in the tangent space of $A$, points outward from $A$, and  is orthogonal to $\pa A$; and ${\bf t}^i$,  the timelike unit vector which is points to the future from $A$ (or $\S$), and  is orthogonal to both ${\bf s}^i$ and $\pa A$.\footnote{Note that $\bt^i$ actually coincides with the timelike normal $\bn^i$ to $\S$ when the latter is evaluated on  $\pa A$.}

We now wish to constrain the function $b({\mm{R}},\tilde{K};{\bs},\bt)$ with some general considerations. First, as discussed at the end of section \ref{sec:extremal}, the complexity should be invariant if the time orientation is reversed. This invariance should also apply for the subregion complexity $\cv(A)$ and in fact, it follows because the bulk volume in eq.~\reef{defineCVx} does not depend on the orientation of time. Therefore, we must have
\beq
b({\mm{R}},\tilde{K};{\bs},\bt)=b({\mm{R}},\tilde{K};{\bs},-\bt)\,,
\labell{goal}
\eeq
\ie this functional only contains terms that are even in the timelike normal ${\bf t}^i$. Note that this restriction was enough to eliminate the possibility of any odd powers of $\delta$ appearing in $v(\mathcal{R},K)$, the integrand in the integral over $A$ in eqs.~\reef{eq:e1} and \reef{eq:jdkkks}. However, the integrand on the entangling surface may still contain odd powers of ${\bf s}_i\tilde{K}^i_{ab}$, which would produce odd powers of $\d$.

Next, let us consider a pure global state on the time slice $\S = A+ \bar A$ dual to a time-symmetric bulk geometry. Now if we choose $\Sigma$ to be the time-symmetric time slice in the boundary, the extremal volume surface yielding the complexity of any of these regions, \ie $\S$, $A$ or $\bar A$, will lie in the special Cauchy slice running through the moment of time symmetry in the bulk and hence from eq.~\reef{defineCVx}, we will find
\bea
{\rm time\ symmetry:}\quad\cv(\S) = \cv(A)+\cv(\bar{A})  \,.
\labell{eq:jdkkks33}
\eea
Now as a result of the time symmetry, various extrinsic curvatures must vanish, \ie the extrinsic curvature of the time slice vanishes and on the entangling surface, ${\bf t}_i\tilde{K}^i_{ab}=0$. Now combining eqs.~\reef{eq:jdkkks} and \reef{eq:jdkkks33}, we find
\bea
{\rm time\ symmetry:}\quad
b({\mm{R}},\tilde{K};{\bs},\bt)\big|_{\pa A} + b({\mm{R}},\tilde{K};{\bs},\bt)\big|_{\pa \bar A}=0\,,
\labell{eq:sdnsd}
\eea 
where since this cancellation is a general result, we have assumed that the integrands must cancel point by point.
Now the only geometric quantity in this expression that distinguishes $\pa A$ from $\pa \bar A$ is the spacelike normal ${\bs}^i$, which points outward from the corresponding subregion, \ie ${\bf s}^i|_{\pa A}=-{\bf s}^i|_{\pa \bar A}$. Therefore we can write eq.~\reef{eq:sdnsd} as 
\bea
{\rm time\ symmetry:}\quad
\left[\,b({\mm{R}},\tilde{K};{\bs},\bt)+ b({\mm{R}},\tilde{K};-\bs,\bt)\,\right]_{\pa A} =0\,.
\labell{simp}
\eea 
That is, this geometric functional on the entangling surface only contains terms with odd powers of the normal vector $\bs^i$.

In particular then, eq.~\reef{simp} rules out the possibility that $b$ contains a constant term, \ie the coefficient $\tilde c_{1,0}=0$ in eq.~\reef{purple} and there will {\it not} be a contribution proportional to ${\cal V}(\pa A)/\d^{d-2}$ in eq.~\reef{eq:jdkkks}. Hence given the constraint \reef{goal}, there is only one possible term which can appear at the next order, namely, $b = \tilde c_{1,1}\,\d\,\bs_i\tilde{K}^i+{\cal O}(\d^2)$ where $\tilde{K}^i$ is the trace of the extrinsic curvature on $\pa A$. More generally, it is not hard to show that in the time-symmetric situation, all of the terms in $b$ will involve odd powers of $\delta$, \ie all of the even $n$ coefficients in eq.~\reef{purple} vanish. Therefore the subregion complexity only contains divergences with odd (even) powers of $\delta$ in even (odd) $d$ in this case. However, it is unclear whether this property extends to cases without time symmetry. For example, one can imagine a term of the form $\delta^2 \,({\bf t}_i\tilde{K}^i)^2$  appearing, which would lead to a divergence of $\cO(1/\delta^{d-4})$. It would be interesting to examine explicit examples for the appearance of such divergences.

\subsection*{Example: Ball-shaped Region}

As an explicit example, consider a ball-shaped region $B$ in a flat background. Hence we consider AdS space in Poincar\'e coordinates,
\bea
ds^2 =\frac{L^2}{z^2}\big[dz^2-dt^2 +dx_i^2\big]
\labell{poink}
\eea 
and we take $B$ to be the region defined by  $\sum_i x_i^2 \leq R^2$ on some constant time slice. With the bulk volume computed in \cite{Alishahiha:2015rta}, the subregion complexity \reef{defineCVx} becomes
\bea 
\cv(B)= \frac{\O_{d-2}L^{d-1}}{G_N (d-1)}\Big(\frac{1}{d-1}\frac{R^{d-1}}{\d^{d-1}} -\frac{d-1}{2(d-3)} \frac{R^{d-3}}{\d^{d-3}}+ \frac{(d-1)(d-3)}{8(d-5)} \frac{R^{d-5}}{\d^{d-5}} + \dots \Big)\,.
\labell{eq:alisha}
\eea 
Since this is a time-symmetric configuration, there are only odd or even powers of $\d$ appearing above, as expected from the discussion above. Now we can recognize a factor of the volume of $B$ in the first term, \ie $\mm{V}(B)=\Omega_{d-2}R^{d-1}/(d-1)$, and so this contribution is simply the first term in $v$ in eq.~\reef{eq:jdkkks}. Now the background curvature vanishes since we are considering flat space and the extrinsic curvature of $B$ also vanishes since it was chosen to live on a constant time slice. Hence all of the subleading terms (\ie $n\ge1$) in $v$ must vanish and hence the remaining contributions in eq.~\reef{eq:alisha} must be associated with boundary divergences. Further note we are considering a time-symmetric situation and so $\cv(B)$ only contains odd (even) powers of $\d$ for even (odd) dimensions, as follows from eq.~\reef{simp}. 
From the same equation, we also argued  that the leading term in $b$ in must be proportional to
\bea
\bs_i\tilde{K}^i=\bs_i\,\tilde{K}^i_{ab}\,\tilde{h}^{ab} =\frac{d-2}{R}\,.
\eea
Hence comparing to eq.~\reef{eq:jdkkks}, we can write the above result as 
\bea
\cv(B)= \frac{L^{d-1}}{G_N}\bigg[\frac{{\cal V}(B)}{(d-1)\,\d^{d-1}} - \frac{1}{2(d-2)(d-3)\,\d^{d-3}} \int_{\partial A}d^{d-2}\tilde{\sigma}
\sqrt{\tilde h}\,\bs_i\tilde{K}^i + \cdots \bigg]\,. \labell{track}
\eea
The term proportional to $(R/\d)^{d-5}$ in eq.~\reef{eq:alisha} is also a boundary divergence and the coefficient is given by some linear combination of terms in $b$ proportional to $(\bs_i\tilde{K}^i)^3$ and $\bs_i\tilde{K}^i\,(\bs_j\tilde{K}^j_{ab})^2$.

% % % % % % % % % % % % % % % % % % % % % % % % % % % % % % % % % % % % % % % % % % % % % % % % % % % % % % % % % % % % % % % % % % % % % % % % % % % % % % % % % % % % % % % % % %

\section{Subregion Complexity: CA duality
\labell{sec:subaction}}

In this section, we consider generalizing the CA duality to subregions, and study the resulting divergence structure. We re-iterate that given the proposal that the mixed state associated with a subregion in the boundary theory is encoded in the corresponding entanglement wedge in the bulk \cite{ewedge1,Headrick:2014cta}, it is natural that the holographic prescription for the  complexity of this state should involve this bulk region. This was the motivation for the approach taken in the previous section with the CV duality and it motivates the following proposal here for the CA duality:

\begin{itemize}
\i Beginning with a subregion $A$ on a given boundary time slice $\Sigma$, we construct the corresponding entanglement wedge $\mm{W}_{\cal E}[A]$  \cite{Headrick:2014cta}, as well as the Wheeler-DeWitt patch $\mm{W}_\mt{WDW}[\S]$. Next we define the bulk region $\widetilde{\mm{W}}$ as the intersection of these two bulk regions: $\widetilde{\mm{W}} = \mm{W}_{\cal E}[A] \cap \mm{W}_\mt{WDW}[\S]$ --- see figure \ref{Wedge2}. The subregion complexity is then conjectured to be given by the gravitational action evaluated on $\widetilde{\mm{W}}$:
\bea
\ca(A) = \frac{I_\mt{WDW}(\widetilde{\mm{W}})}{\pi \hbar} 
\labell{trump}
\eea 
In the limit when the subregion $A$ is the entire time slice $\S$, we have $\widetilde{\mm{W}} = \mm{W}_\mt{WDW}[\S]$ and we recover eq.~\reef{none} for the original CA duality.
\end{itemize}
\begin{figure}[!h]
	\centering
	\includegraphics[scale=0.5]{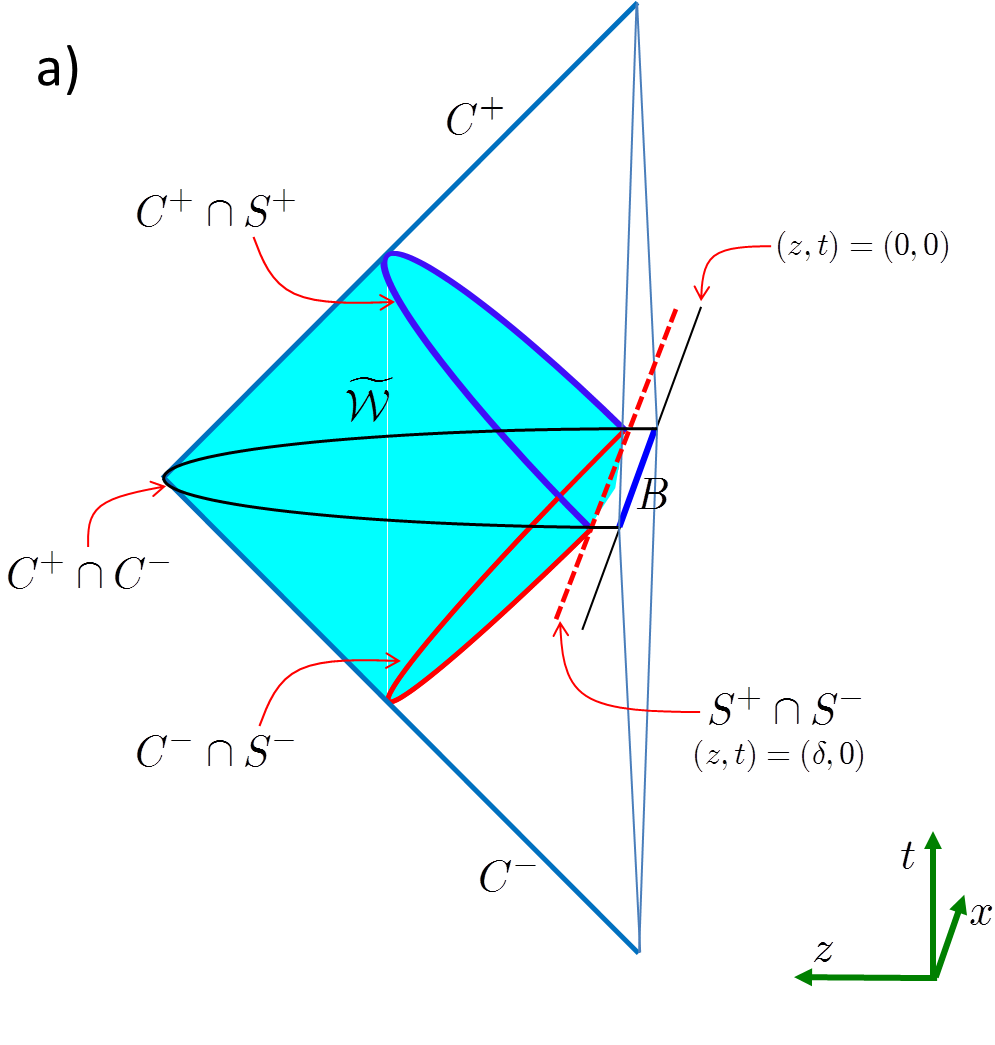}	
	$\qquad$
	\includegraphics[scale=0.5]{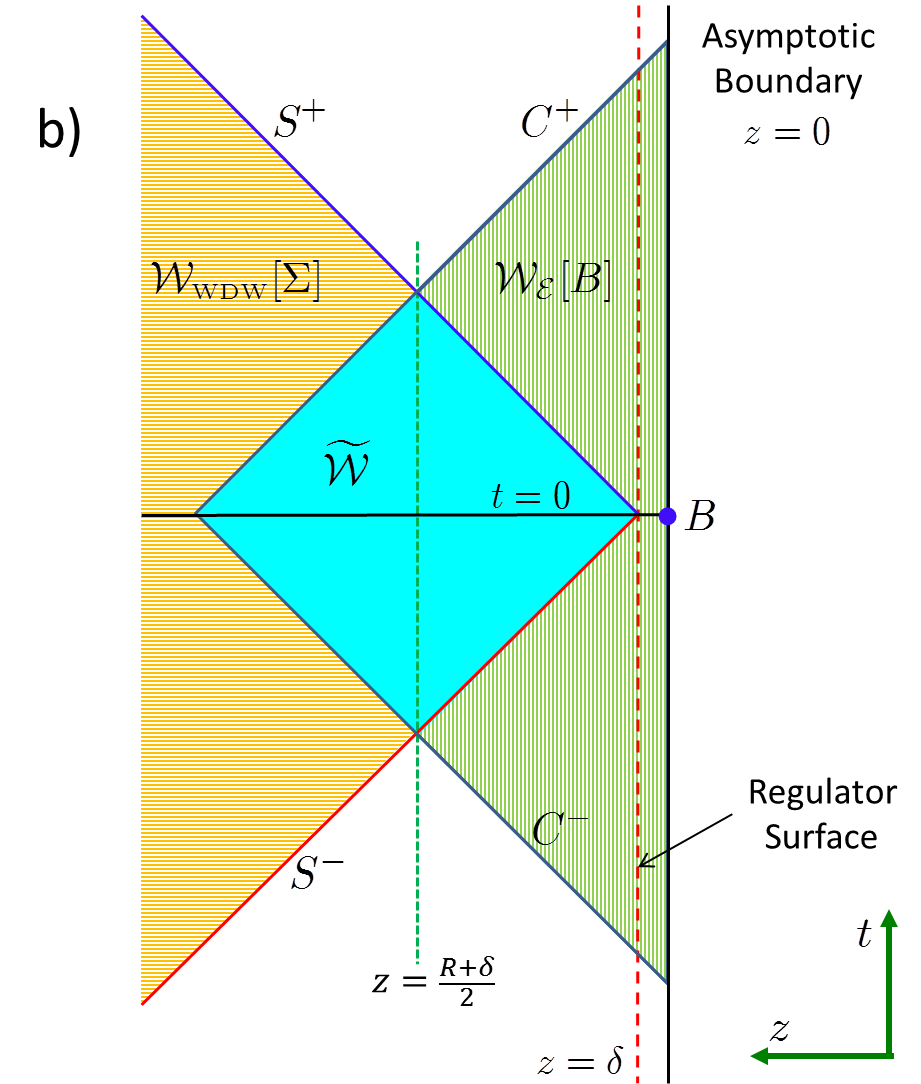}
	\caption{For a ball-shaped boundary region $B$, the bulk region $\widetilde{\mm{W}}$ is the intersection of the entanglement wedge $\mm{W}_{\cal E}[B]$ and the WDW patch $\mm{W}_\mt{WDW}[\S]$. (a)  Showing details of the null joints appearing in the boundary of $\widetilde{\mm{W}}$. (b) Showing a cross-section of $\widetilde{\mm{W}}$ at $r=0$. (See the main text for the notation.) \labell{Wedge2}}
\end{figure}

Consistency of this new holographic definition for subregion complexity would require that the result is independent of the specific choice of the time slice $\S$ used to define the WDW patch. In particular, our definition only fixes the global time slice to coincide with the subregion of interest but leaves the extension of this boundary surface outside of this subregion unspecified. While it is obvious that the bulk region $\widetilde{\mm{W}}$ is independent of the choice of time slice for simply connected subregions, the situation is less clear when the boundary subregion consists of a number of disconnected components. Yet it is straightforward to explicitly verify that this property holds in a number of simple situations, \eg for a number of disconnected intervals on the boundary of pure AdS$_3$ or for a number of parallel strips on the boundary of AdS$_{d+1}$. 

A general proof of the desired time slice independence is as follows:\footnote{We thank Veronika Hubeny for a discussion on this issue.} Denote the boundary subregion as $A$, which may be comprised of any number of disconnected components. Choose a Cauchy surface $\Sigma$ on the boundary which contains $A$ but is otherwise arbitrary, and denote the complement of $A$ on this time slice as $A^c$. We denote the boundary causal development of these two subregions as $\cald(A)$ and $\cald({A^c})$, and the corresponding entanglement wedges, as $\mm{W}_{\cal E}[A]$ and $\mm{W}_{\cal E}[A^c]$. Recall that $\cald(A)$ and $\cald({A^c})$ are the asymptotic boundaries of $\mm{W}_{\cal E}[A]$ and $\mm{W}_{\cal E}[A^c]$ respectively, \ie these boundary regions are where the entanglement wedges meet the asymptotic AdS boundary \cite{Headrick:2014cta}. In passing, we also note that $\cald(A)$ and $\cald({A^c})$ are independent of the particular choice made above for the global time slice $\S$. Now the desired time slice independence follows if we can show that all points in $\cald({A^c})$ are space-like separated from all points in $\widetilde{\mm{W}}$. But here, we simply recall the definition $\widetilde{\mm{W}} = \mm{W}_{\cal E}[A] \cap \mm{W}_\mt{WDW}[\S]$. Therefore $\widetilde{\mm{W}}$ is contained inside the entanglement wedge $\mm{W}_{\cal E}[A]$. However, we know that all points in $\mm{W}_{\cal E}[{A^c}]$ and hence $\cald({A^c})$ are space-like separated from all points in $\mm{W}_{\cal E}[A]$ \cite{Headrick:2014cta}.\footnote{Recall that this result holds as long as the
bulk obeys the null energy condition. Further, this property was required because otherwise the reduced density matrix on $A$ could be affected by operations in $\cald({A^c})$.} Therefore we have proven that the bulk region $\widetilde{\mm{W}}$, and hence the corresponding complexity \reef{trump}, is independent of the choice of the Cauchy surface $\S$ on the boundary outside of the subregion $A$.

In the rest of this section, we examine the above proposal in a specific example where $A$ is a ball-shaped region in a flat background. In particular, we focus on the structure of the divergences and this example allows us to infer general properties of the divergence structure.

\subsection*{Example: Ball-shaped Region}

As in the previous section, let us apply the proposed CA duality to evaluate the subregion complexity for a ball-shaped region $B$ in a flat background. Hence we consider AdS space in Poincar\'e coordinates 
\bea 
ds^2 =\frac{L^2}{z^2}\big[dz^2-dt^2 +dr^2+r^2\,d\Omega^2_{d-2}\big]\,
\labell{poinc}
\eea 
where we use polar coordinates in the spatial boundary directions. For simplicity, we then take $B$ to be the region defined by:  $r \leq R$ and $t=0$. The extremal bulk surface on which we would evaluate the holographic entanglement entropy is a hemisphere \cite{Ryu:2006bv,Ryu2}: $R^2= r^2+z^2$. The entanglement wedge $\mm{W}_{\cal E}[B]$ is the bulk region enclosed by the two null cones:\footnote{The boundary of the entanglement wedge has no caustics in this simple example.}
%\beq
%\sqrt{r^2+z^2}=\left\lbrace\begin{array}{ll}
%		 R-t & \quad{\rm for}\ \ 0\le t\le R\,, \\
%		 R+t & \quad{\rm for}\ \ 0\ge t\ge -R \,.
%	\end{array}\right.
%\eeq
\beqa
C^+\ :&\quad t=\ R-\sqrt{r^2+z^2}\  & \quad{\rm for}\ \ 0\le t\le R\,, 
\labell{ncone}\\
C^-\ :&\quad t=-R+\sqrt{r^2+z^2} & \quad{\rm for}\ \ 0\ge t\ge -R \,.
\nonumber
\eeqa
As in section \ref{sec:action}, we must regulate the WDW patch by introducing a regulator surface at $z=\d$. In particular, we will use the approach illustrated in figure \ref{fig:space}b, where the null boundaries begin at the time slice on this regulator surface, \ie they begin at $(z,t)=(\d,0)$. The boundary of the WDW patch $\mm{W}_\mt{WDW}[t=0]$ is then the two null sheets:
%\beq
%z=\left\lbrace\begin{array}{ll}
%		 \d+t & \quad{\rm for}\ \ t\ge 0\,, \\
%		 \d-t & \quad{\rm for}\ \  t\le 0 \,.
%	\end{array}\right.
%\eeq
\beqa
S^+\ :&\quad t=\ z-\d\ \ \ & \quad{\rm for}\ \ t\ge 0\,, 
\labell{nsheet}\\
S^-\ :&\quad t= -(z-\d) & \quad{\rm for}\ \ t\le 0 \,.
\nonumber
\eeqa
Now following eq.~\reef{trump}, we compute the gravitational action on the intersection of these two bulk regions: $\widetilde{\mm{W}} = \mm{W}_{\cal E}[A] \cap \mm{W}_\mt{WDW}[\S]$, as illustrated in figure~\ref{Wedge2}. 

Following the prescription in appendix \ref{sec:manuel}, there are only two kinds of nonvanishing contributions in eq.~\reef{actall}, which need to be considered here. That is, we must evaluate the Einstein-Hilbert integral and four null joint contributions,
\beq
I({\widetilde{\mm{W}}})= \frac1{16\pi G_N}\int_{\widetilde{\mm{W}}} \!\!d^{d+1}x \sqrt{-g} \left(\mm{R}+\frac{d(d-1)}{L^2}
\right) + \frac1{8\pi G_N}\int_{\Sigma'} d^{d-1}x \sqrt{\sigma}\, a\,.
\labell{actred}
\eeq
No other contributions need to be considered because all of the boundary surfaces for $\widetilde{\mm{W}}$ are null.

Let us begin with the Einstein Hilbert term. Using the Einstein equations, we may substitute $\mm{R}=-d(d+1)/L^2$, and this contribution simplifies to evaluating the spacetime volume of the intersection region
\bea
I_\mt{bulk} =-\frac{d}{8\pi G_NL^2}\,{\cal V}(\widetilde{\mm{W}}) \,.
\labell{eq:corjkllj0}
\eea  
As shown in figure \ref{Wedge2}b, it is straightforward to evaluate this volume by first dividing it into two parts:
\bea
{\cal V}({\widetilde{\mm{W}}}) = {\cal V}_1\left(z>\frac{R+\d}{2}\right)+{\cal V}_2\left(z<\frac{R+\d}{2}\right)\,,
\eea 
where ${\cal V}_1(z>(R+\d)/{2})$ is the volume of the portion of the region bounded above and below entirely by the null cones $C^\pm$, and ${\cal V}_2(z<(R+\d)/{2})$ is the volume of the portion of the region which is also bounded above and below by the null sheets $S^\pm$. We begin with the former
\bea
{\cal V}_1&=&2 L^{d+1}\O_{d-2} \int_0^{\frac{R-\d}{2}} dt \int_{\frac{R+\d}{2}}^{R-t} \frac{dz}{z^{d+1}} \int_{0}^{\sqrt{(R-t)^2-z^2}} dr\, r^{d-2} \nonumber\\
&=&\frac{2L^{d+1} \O_{d-2} }{d-1}  \int_0^{\frac{R-\d}{2}} dt \int_{\frac{R+\d}{2}}^{R-t} dz\, \frac{((R-t)^2-z^2)^{\frac{d-1}{2}}}{z^{d+1}}
\eea
This volume remains finite in the limit $\d\to0$ since the integration does not reach the asymptotic boundary. Hence in an expansion for $\delta/R\ll1$, ${\cal V}_1$ only contains positive powers of $\delta$. Turning to the volume of the region with $z<(R+\d)/{2}$, we find
\bea
{\cal V}_2
%& =&2 L^d\O_{d-2} \int_0^{\frac{R-\d}{2}} dt \int^{\frac{R-\d}{2}}_{t+\d} \frac{dz}{z^{d+1}} \int_{0}^{\sqrt{(R-t)^2-z^2}} dr\, r^{d-2} \nonumber\\
&=&\frac{2L^{d+1} \O_{d-2} }{d-1}  \int_0^{\frac{R-\d}{2}} dt \int^{\frac{R+\d}{2}}_{t+\d} dz\, \frac{((R-t)^2-z^2)^{\frac{d-1}{2}}}{z^{d+1}}
\labell{eq:corjkllj}\\
&=& \frac{2L^{d+1} \O_{d-2} }{d(d-1)}  \left[\frac{R^{d-1}}{(d-1)\d^{d-1}}- \frac{R^{d-2}}{(d-2)\d^{d-2}}+ \frac{(-d^2+3d-4)R^{d-3}}{2(d-2)(d-3)\d^{d-3}} +\cdots \right]\,, 
\eea

where in the second line, we have expanded the integrand for small $z$ to identify the divergent terms arising from the integration near the asymptotic boundary.	
Hence the divergences appearing in the bulk action \reef{eq:corjkllj0} become
\bea
I_\mt{bulk} %=-\frac{d}{8\pi G_NL^2}\,{\cal V}(\widetilde{\mm{W}})  
= - \frac{L^{d-1}}{4\pi G_N}\,\frac{\O_{d-2}}{d-1}  \left[\frac{R^{d-1}}{(d-1)\d^{d-1}}- \frac{R^{d-2}}{(d-2)\d^{d-2}}+ \frac{(-d^2+3d-4)R^{d-3}}{2(d-2)(d-3)\d^{d-3}} +\cdots \right]\,.
\labell{walk}
\eea
Note that there are both even and odd power law divergences in this expression and also that the overall sign is negative.

Now we move on to compute the contributions of the null joint in eq.~\reef{actred}.
The region $\widetilde{\mm{W}}$ has four null joints coming from the intersections of the various null boundaries: $S^+\cap S^-$, $C^+\cap C^-$, $C^+\cap S^+$ and $C^-\cap S^-$ --- see figure \reef{Wedge2}a. Hence we divide the null joint term into the four corresponding contributions
\bea
I_\mt{jnt} = I^{(1)}(S^+\cap S^-) +  I^{(2)}(C^+\cap C^-) + I^{(3)}(C^+\cap S^+)+ I^{(4)}(C^-\cap S^-)
\eea 
and evaluate each in turn using the prescription given in appendix~\ref{sec:manuel}.
Note that the latter requires writing the (outward directed) null normals for each of the corresponding surfaces, which we find using eqs.~\reef{ncone} and \reef{nsheet}:
\bea
&&S^+:\ \nk1=\a\,(\,dt-dz)\,,\qquad\qquad\qquad\ \ 
 S^-:\ \nk2=\a\,(-dt-dz)\,,
\labell{nnorm}\\
&&C^+:\ \nk3=\b\left(dt+\frac{r\,dr+z\,dz}{\sqrt{r^2+z^2}}\right)\,,\qquad C^-:\ \nk4=\b \left(-dt+\frac{r\,dr+z\,dz}{\sqrt{r^2+z^2}}\right)\,,
\nonumber
\eea
where $\a$ and $\b$ are arbitrary (dimensionless) normalization constants for the null normals. For simplicity, we have chosen the same normalization constant on $S^+$ and $S^-$, and on $C^+$ and $C^-$.
	
Beginning with $I^{(1)}(S^+\cap S^-)$, we find $a=-2\log(\a\,\d/L)$ using eq.~\reef{ccddx}. Hence the joint contribution becomes
\bea
I^{(1)} %=- \frac{L^{d-1}}{4\pi G}\log \Big(\frac{\sqrt{\a_1 \a_2} }{L}\d \Big) \frac{\O_{d-2}}{\d^{d-1}}  \int_0^{\sqrt{R^2-\d^2}} dr r^{d-2}   
&=& - \frac{L^{d-1}}{4\pi G_N}\,\frac{\O_{d-2}}{d-1}\log \left(\frac{\a\,\d }{L} \right)\, \frac{(R^2-\d^2)^{\frac{d-1}{2}}}{\d^{d-1}} 
\labell{crown}\\
&=&- \frac{L^{d-1}}{4\pi G_N}\frac{\O_{d-2}}{d-1}\log \left(\frac{\a\,\d }{L} \right) \left[\frac{R^{d-1}}{\d^{d-1}} - \frac{d-1}{2} \frac{R^{d-3}}{\d^{d-3}} + \frac{(d-1)(d-3)}{8} \frac{R^{d-5}}{\d^{d-5}} +\cdots     \right]\,.
\nonumber
\eea
Hence the divergence structure here involves a logarithmic divergence multiplying power law divergences, with only odd (even) powers for even (odd) $d$. Note that the overall sign of this null joint contribution is positive.

Next for $C^+\cap C^-$, we have $a=-2\log(\b z/L)$ and so the joint contribution becomes\footnote{We thank Shira Chapman for pointing out typos in a previous version of this paper.}	
\bea
I^{(2)}&=&  - \frac{L^{d-1}\O_{d-2}}{4\pi G_N}  \int_\d^{R} dz\,R \frac{(R^2-z^2)^{\frac{d-3}{2}}}{z^{d-1}}  \log \left(\frac{\b z}{L} \right) 
\labell{tiara}\\
&=&-\frac{L^{d-1}\O_{d-2}}{4\pi G_N} \log \left(\frac{\b\, \d}{L} \right)  \left[\frac{R^{d-2} }{(d-2)\d^{d-2}} -\frac{(d-3)}{2(d-4)}\frac{R^{d-4} }{\d^{d-4}}
+\frac{(d-3)(d-5)}{8(d-6)}\frac{R^{d-6} }{\d^{d-6}} 
+ \cdots \right] 
\nn 
&&\qquad-\frac{L^{d-1}\O_{d-2}}{4\pi G_N}   \left[\frac{R^{d-2} }{(d-2)^2\d^{d-2}} -\frac{(d-3)}{2(d-4)^2}\frac{R^{d-4} }{\d^{d-4}}
+\frac{(d-3)(d-5)}{8(d-6)^2}\frac{R^{d-6} }{\d^{d-6}} 
+ \cdots \right] 
\nonumber
\eea 
Again, we find that there are only even or odd powers, but not both. There are also terms involving both power law divergences multiplied by a logarithmic divergence.
	
For $C^+\cap S^+$, we have $a=\log\left(\frac{\a\b}{2}\frac{z^2}{L^2}\frac{R+\d}{R+\d-z}\right)$ and the joint contribution is given by	
\bea
I^{(3)}& =&  \frac{L^{d-1}\O_{d-2}}{8\pi G_N} \big(R+\delta\big)^{d-2} \int_\d^{\frac{R+\d}{2}} dz \frac{\Big(1-\frac{2z}{R+\d}\Big)^{\frac{d-3}{2}}}{z^{d-1}}  \log \left(\frac{\a\b}{2}\frac{z^2}{L^2}\frac{R+\d}{R+\d-z}\right)
\labell{stroll}\\
&=&\frac{L^{d-1}\O_{d-2}}{4\pi G_N}   \log \left(\sqrt{\frac{\a\b}{2}}\frac{\d}{L}\right) \left[\frac{R^{d-2} }{(d-2)\d^{d-2}} -\frac{(d-3)}{2(d-4)}\frac{R^{d-4} }{\d^{d-4}}
+\frac{(d-3)(d-5)}{8(d-6)}\frac{R^{d-6} }{\d^{d-6}} 
+ \cdots \right] 
\nn   
&&\quad+   \frac{L^{d-1}\O_{d-2}}{4\pi G_N}    \left[\frac{R^{d-2} }{(d-2)^2\d^{d-2}} +\frac{d-4}{2(d-3)(d-2)}\frac{R^{d-3} }{\d^{d-3}}
-\frac{3d^2-20d+36}{4(d-4)^2(d-2)}\frac{R^{d-4} }{\d^{d-4}}
+ \cdots \right] \,.
\nonumber
\eea 
Note that this joint contributions has both even and odd power divergences, as well as a logarithmic factor in some of the contributions. 
%rcm new sentence
Further, we note in passing that both here and in eq.~\reef{tiara}, the integrals can generate additional logarithms and so we may find divergences of the form $\log^2\delta$. For example, such terms would appear in both eqs.~\reef{tiara} and \reef{stroll} for $d=2$.
Next, we must evaluate the joint contribution coming from $C^-\cap S^-$ but by the symmetry of the present geometry under $t\to-t$, we have $I^{(4)}(C^-\cap S^-)=I^{(3)}(C^+\cap S^+)$.

Up to this point, it seems that we have taken into account all of the contributions to the gravitational action, however, we need to point out that our discussion has overlooked one geometric structure in the boundary of $\widetilde{\mm{W}}$. In particular, there is a codimension-three `corner' where all four null surfaces simultaneously intersect, \ie $S^+\cap S^- \cap C^+\cap C^-$ which is the spherical surface given by $(z,t,r)=(\delta,0,\sqrt{R^2-\d^2})$. As discussed in \cite{Lehner:2016vdi}, the boundary terms that might be required in the gravitational action for such higher codimension corners requires further analysis and remain unknown at the present time. While our intuition is that the contribution from this corner vanishes in our example,\footnote{In part, this intuition is informed by the vanishing contribution of similar singularities in the examples in \cite{shira}.} this provides further motivation for a detailed study of such higher order intersections of boundary surfaces. In any event, in our example, it seems that all possible divergences are already appearing in our final result, and hence even if this extra corner were to make a contribution to the action, it would not add anything conceptually new. 

Finally, the total action $I(\widetilde{\mm{W}})$ combining the results in eqs.~\reef{walk} and (\ref{crown}--\ref{stroll}) and then eq.~\reef{trump} yields the leading divergence structure for the subregion complexity as 
\bea
\ca(B) &=&  \frac{1}\pi\left(I_\mt{bulk}+I^{(1)}+I^{(2)}+2\,I^{(3)}\right)
\labell{trumpet}\\
&=&-\frac{L^{d-1}}{4\pi^2 G_N}\,\frac{\O_{d-2}}{d-1}  \left[\frac{R^{d-1}}{(d-1)\d^{d-1}}- \left(\frac{2d-3}{(d-2)^2}-\frac{(d-1)}{d-2}\,\log2\right)\frac{R^{d-2}}{\d^{d-2}}- \frac{3d^2-13d+12}{2(d-2)(d-3)}\,\frac{R^{d-3}}{\d^{d-3}}   +\cdots \right]
\nn
&&\qquad +\frac{L^{d-1}}{4\pi^2 G_N}\,\frac{\O_{d-2}}{d-1}    \log \left(\frac{L}{\a\,\d } \right) \left[ \frac{R^{d-1}}{\d^{d-1}} - \frac{d-1 }{d-2}\,\frac{R^{d-2} }{\d^{d-2}}  -  \frac{d-1}{2}\,\frac{R^{d-3}}{\d^{d-3}} +\cdots     \right]  \,. \nonumber
\eea 
Note that we can recognize the leading divergences ($\propto1/\d^{d-1}$) as being proportional to the volume of the ball-shaped region, \ie $\mm{V}(B)=\Omega_{d-2}R^{d-1}/(d-1)$. Now we expect the coefficients of the subleading divergences are also proportional to various geometric factors. However, in this example, both the background curvature and the extrinsic curvature of the time slice vanish.
Hence, the next pair of divergences ($\propto1/\d^{d-2}$) must be proportional to the volume of the boundary, \ie
$\mm{V}(\pa B)=\Omega_{d-2}R^{d-2}$. Similarly the coefficients of the terms proportional to $1/\d^{d-3}$ involve an integral of $\bs_i\tilde{K}^i =(d-2)/R$ over $\pa B$, as in eq.~\reef{track}, while the higher order terms will involve higher powers of the boundary extrinsic curvature. One notable feature of eq.~\reef{trumpet} is that the coefficient $\beta$ has canceled out in the total action. In fact, it is straightforward to show that this cancellation extends to include the $R^{d-6}/\d^{d-6}$ terms and higher. We return to discuss this feature in section \ref{discuss}.

Now given the results of the above calculation and our previous experience, we expect the the CA duality produces the following general form for the divergences in subregion complexity:
\bea
\ca(A) &=&\frac{1}{\delta^{d-1}}\,\int_A d^{d-1}\sigma\,\sqrt{{h}}\,\left[ v_1(\mathcal{R},K)+\log\left(\frac{L}{\alpha\,\d}\right)\ v_2(\mathcal{R},K)\right]
\labell{cute1x}\\
&&\qquad+ \frac{1}{\delta^{d-2}}\int_{\pa A} d^{d-2}\tilde{\s}\,\sqrt{\tilde{h}}\,\left[ b_1(\mathcal{R},\tilde{K};{\bf s},\bt)+\log\left(\frac{L}{\tilde\alpha\,\d}\right)\ b_2(\mathcal{R},\tilde{K};{\bf s},\bt)\right]
\nonumber\\
&&\quad{\rm with}\quad\qquad
\ v_k(\mathcal{R},K)=\,\sum_{n=0}^{\lfloor{\frac{d-1}{2}}\rfloor}\,\sum_{i}\  c^{[k]}_{i,n}(d)\ \d^{2n}\ [\mathcal{R},K]^{2n}_{i} \,,
\labell{cute2x}\\
&&\qquad\qquad \ b_k(\mathcal{R},\tilde{K};{\bf n},\bt)=\ \sum_{n=0}^{d-1}\ \sum_{i}\  \tilde c^{[k]}_{i,n}(d)\,\delta^{n}\,[\mathcal{R},\tilde{K};{\bf s},\bt]^{n}_{i}\,.
\labell{cute3x}
\eea
with $d$ boundary dimensions. The expressions in eq.~\reef{cute2x} would be identical to those found in eq.~\reef{cute2} and the only difference here is that the corresponding integral in eq.~\reef{cute1x} is now restricted to the subregion $A$. 
As in the previous section, we also find additional divergences that are associated with the entangling surface $\partial A$.  The corresponding integrands $b_{1,2}$ involve: $\mathcal{R}$, the background curvatures of the AdS boundary; $\tilde{K}^i_{ab}$, the extrinsic curvatures of the codimension-two entangling surface; and also ${\bf s}^i$ and ${\bf t}^i$, a preferred basis in the space transverse to $\pa A$ --- see the description under eq.~\reef{eq:jdkkks}. The appearance of the coefficients $\a$ and $\tilde \a$ in the logarithmic factors introduces some additional ambiguity in the above expressions and we will return to discuss this point in section \ref{discuss}. 
%rcm new sentence
We note that the contribution associated with the entangling surface may involve divergences proportional to $\log^2\delta$ coming from the $b_2$ term, as discussed for the example above.

We note that invariance under time reversal results in only even powers of $\delta$ appearing in the integrands $v_{1,2}$, but this is not the case for $b_{1,2}$. There it only imposes the weaker constraint: $b_k(\mathcal{R},\tilde{K};{\bf s},\bt)=b_k(\mathcal{R},\tilde{K};{\bf s},-\bt)$ --- see discussion around eq.~\reef{goal}. Further, with CV duality, we were able to find further restrictions by considering the case of time symmetry, \eg in eq.~\reef{purple}, $\tilde c_{1,0}$ vanishes. However, the same considerations cannot be made here because
\beq
\ca(\S) \ne \ca(A)+\ca(\bar{A})  \,,
\labell{pint}
\eeq
even in the special case of time symmetry and in fact, we saw contributions  proportional to ${\cal V}(\pa A)$, the volume of the entangling surface, explicitly appear the example above in eq.~\reef{trumpet}.

% % % % % % % % % % % % % % % % % % % % % % % % % % % % % % % % % % % % % % % % % % % % % % % % % % % % % % % % % % % % % % % % % % % % % % % % % % % % % % % % % % % % % % % % % %

\section{Discussion \labell{discuss}}

In this paper, we studied the two conjectures for holographic complexity: the complexity=action (CA) conjecture and the complexity=volume (CV) conjecture. In particular, we examined the structure of UV divergences in the complexity following from these two conjectures. We found that both $\ca$ and $\cv$ contain a series of power law divergences and the coefficients of these divergences are determined by local integrals of various geometric invariants over the corresponding time slice $\S$, as shown in eqs.~\reef{eq:e1} and \reef{cute1}. These coefficients also contain dimensionless parameters characterizing the underlying CFT, \eg $C_T\sim L^{d-1}/G_N$ in the present case where the bulk is described by Einstein gravity.\footnote{If the boundary CFT is deformed by a relevant operator, it is straightforward to show that the corresponding (dimensionful) coupling will also appear in these coefficients \cite{new7}. In this case, the structure of the UV divergent terms is analogous to the results found for holographic entanglement entropy in \cite{Hung:2011ta}.} The leading divergence appearing with the CV duality is proportional to the volume of the boundary time slice, \ie $\cv(\S)\sim{\cal V}(\S)/\d^{d-1}$. A similar divergence appears with the CA duality, however, an extra factor proportional to $\log\d$ arises from the asymptotic joint contributions in the WDW action, \eg see eq.~\reef{pole}. Hence the leading divergence appearing with the CA duality takes the form
\beq
\ca(\S)\sim\log\!\left[{L}/({\a\,\d})\right]\, \frac{{\cal V}(\S)}{\d^{d-1}}\,,
\labell{leader}
\eeq
where $L$ is the AdS curvature scale and $\a$ is a (dimensionless) normalization constant --- see further discussion below. 

We note that the asymptotic joint contribution, which produces this divergence \reef{leader}, is essential for the consistency of our CA calculations in sections \ref{sec:action} and \ref{sec:subaction}. The bulk integral of the Einstein-Hilbert action \reef{racoon} contributes a divergence proportional to the boundary volume but the coefficient is negative, which follows from Einstein's equations and the negative cosmological constant. That is, we have $I_\mt{bulk}\sim-{\cal V}(\S)/\d^{d-1}$ but if this was the leading divergence, the resulting WDW action would be negative. Since by definition the complexity is positive, this would produce an inconsistency for CA duality.\footnote{Note that with regularization scheme illustrated in figure~\ref{fig:space}a, there is an additional GHY surface term which is positive and which will dominate over the bulk term, \eg compare eqs.~\reef{rounder} and \reef{poilu} for the example examined in appendix \ref{newA}. \label{footy}} However, the joint contribution is positive and contains an even stronger divergence with the extra $\log\d$ factor, \eg see eqs.~\reef{pole} and \reef{trumpet2}. Hence it is responsible for making the holographic complexity positive and ensuring the consistency of the CA duality.

As noted in the introduction, the locality of the coefficients in the holographic complexity suggests that these divergences should be associated with establishing local correlations down to the cutoff scale in the boundary CFT. On the gravity side, since the calculations in CV duality resemble those in holographic entanglement entropy so closely, it is not surprising that the coefficients of the power law divergences are determined by local integrals. Essentially, the initial terms in the FG expansion \reef{eq:poli} are expressed in terms of the boundary geometry and the equations determining the asymptotic shape of the extremal surface have a similar geometric interpretation \cite{Hung:2011ta}. On the other hand, it is not immediately clear that the CA duality should produce coefficients with a similar locality. Of course, our explicit calculations in section \ref{sec:action} demonstrate that this is the case, at least for the first few divergences. It would be interesting to thoroughly investigate if this locality extends to all of the UV divergences appearing in the holographic complexity, as assumed in the discussion at the end of section \ref{sec:action}. It is clear that locality continues to hold for all of the divergences in the joint contribution, since these only rely on evaluating the asymptotic expansion of the metric at $z=\d$. On the other hand, one needs a better understanding of the general geometry of the null boundaries of the WDW patch to determine if the bulk integral also produces coefficients which are always local. 

In passing, we note that in the context of holographic entanglement entropy, ref.~\cite{wrong} shows that this behaviour may fail in certain situations. Although the coefficients are still determined by local integrals, the integrand involves state dependent data in these cases. It would also be interesting to see if these results extend to holographic calculations of complexity.

We now turn to the factor of $\a$ appearing in the argument of the logarithm in eq.~\reef{leader}, or more generally in eq.~\reef{cute1} for the CA duality. As noted previously, this (dimensionless) coefficient is an arbitrary normalization constant for the null normals, \eg see eq.~\reef{noraX} or \reef{nnorm},\footnote{In general, we have two independent normalization constants for the normals on the future and past null boundaries, as in eq.~\reef{noraX}, in which case the factor of $\a$ is replaced by $\sqrt{\a_1\a_2}$, as in eq.~\reef{leader}.} which arises because of the freedom to rescale the affine parameter along the future and past null boundaries of the WDW patch \cite{Lehner:2016vdi}. In order to make a meaningful comparison of the gravitational action for different WDW patches, \eg in different spacetimes as in \cite{shira}, one must first fix this normalization constant in a consistent way. The suggestion of \cite{Lehner:2016vdi} was to impose a normalization condition on the null normals near the asymptotic AdS boundary. In particular, one can choose 
\beq
{\bf k}\cdot \hat t=\pm\a
\labell{hop}
\eeq
at the AdS boundary. Here, ${\bf k}$ is the normal to the future (+) or past (--) null boundary (written as an outward pointing one-form --- see appendix \ref{sec:manuel}); $\hat t=\partial_t$ is the timelike vector in the asymptotic AdS geometry which is normalized to describe the time flow in the boundary theory; and $\a$ is an arbitrary positive constant. 

One simple choice that was suggested in \cite{Lehner:2016vdi} is $\a=1$. However there is a puzzle here as follows: With $\a=1$, the result, \eg in eq.~\reef{leader} takes the form $\ca(\S)\sim\log({L}/\d)\,{\cal V}(\S)/\d^{d-1}$ and so the complexity explicitly depends on the AdS curvature scale $L$, which has no interpretation in the boundary theory. Hence it seems another choice is required to eliminate this dependence. Let us instead set $\a=L/\ell$ where $\ell$ is some scale in the boundary theory which is common to all states and geometries for which we might want to evaluate the complexity. One candidate would be $\ell=\d$, the short-distance cutoff, however, with this choice, the argument of the log reduces to one and the contribution in eq.~\reef{leader} vanishes. Unfortunately, as noted in the discussion above, this would leave us with a negative complexity\footnote{As emphasized in footnote \ref{footy}, we re-iterate that this is a feature of the regularization illustrated in figure \ref{fig:space}b. With the alternate regularization in figure \ref{fig:space}a, the GHY contribution on the regulator surface dominates over the bulk term to produce a positive action, without the joint contribution --- see appendix \ref{newA}.} and so this choice of $\ell$ appears inconsistent. Another choice would be the size of the boundary time slice, \ie $\ell = \ell_{\cal V}\sim {\cal V}^{1/(d-1)}$. However, with this choice, the complexity becomes superextensive, \ie the leading contribution grows faster than the volume of the time slice. As this contribution to the complexity seems most naturally related to the establishing very short-distance correlations in the boundary CFT, it seems that this contribution should only be proportional to ${\cal V}(\S)$ and should not be superextensive. Unfortunately, these two choices seem to be the only scales in the boundary theory that will naturally arise in any geometry and for any state, and they both yield undesirable results. However, it may then be that $\ell$ is a new scale defined by the precise microscopic rules used to define the complexity. For example, if we set $\ell=e^\sigma\d$ (\ie $\a=e^{-\sigma}L/\d$) where $\sigma$ is some numerical factor,\footnote{Of course, given the previous discussion, there must be a lower bound on $\sigma$. Examining eq.~\reef{trumpet2},  we find $\sigma>1/(d-1)$.} then eq.~\reef{leader} becomes $\ca(\S)\sim\sigma\,{\cal V}(\S)/\d^{d-1}$ and one can imagine that different choices of $\sigma$ correspond to different choices for the set of universal gates which are used to prepare states and define the complexity, \eg $\ell$ might be related to the maximum range over which the universal gates act in the CFT.
Note that with this prescription, there is no real distinction between the two families of divergent terms appearing in eq.~\reef{cute1} since $\d$ is eliminated from the argument of the logarithm. However, generically the  null boundaries for the WDW patch will end on joints deep in the interior of the bulk geometry, and the corresponding null joint terms will introduce new contributions where $\log\d$  is mixed with IR features, \eg if one considers complexity of the thermofield double state dual to a black hole. Further, this $\log\d$ may also `infect' quantities which might otherwise be expected to be finite, such as the rate of growth of the complexity in certain situations \cite{Brown:2015lvg,new8}.\footnote{Recently, ref.~\cite{ross16} divergences in the WDW action using a additional boundary term introduced in \cite{Lehner:2016vdi}, which renders the action invariant under reparameterizations of the null boundary coordinates. In particular, it was found that this boundary term also removes the $\log({L}/\d)\,{\cal V}(\S)/\d^{d-1}$ divergence.}
\begin{figure}[!h]
	\centering
	\includegraphics[scale=0.7]{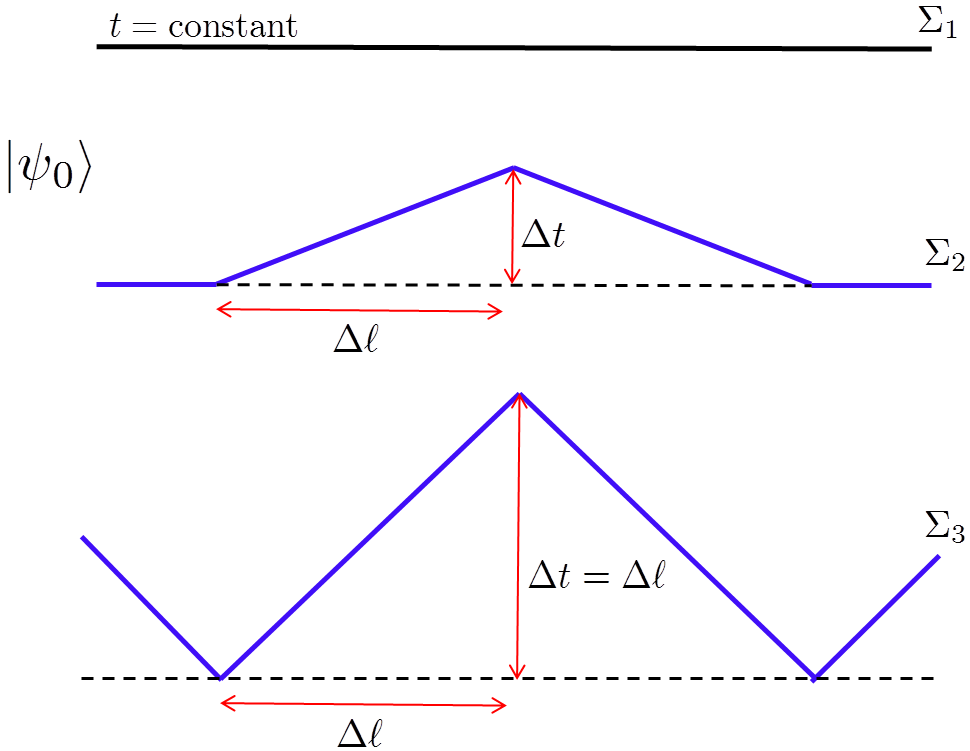}	
	\caption{We consider the ground state $|\psi_0\rangle$ of the boundary theory but evaluate the complexity on three different time slices, $\Sigma_1$, $\S_2$ and $\S_3$. Comparing the first two time slices, the complexity sees a large reduction of $\Sigma_2$ because the proper volume of this time slice is reduced by  $\Delta\calv=(\sqrt{\Delta\ell^2-\Delta t^2}-\Delta\ell)\,\calv_\mt{trans}$ where $\calv_\mt{trans}$ is the volume in the transverse directions. The third time slice $\S_3$ is composed of null segments and so the leading divergence in the complexity vanishes. \labell{bump2}}
\end{figure}

One of the key features which was observed here was the geometric nature of the coefficients in the various power law divergences appearing in the holographic complexity. While this feature is entirely expected given our experiences from holographic entanglement entropy \cite{Hung:2011xb,Hung:2011ta,ajay}, it means that the complexity has some unusual features. To illustrate this point, recall again that the leading term for both CA and CV duality is proportional to the volume of the time slice, \ie  $\cv(\S)\sim{\cal V}(\S)/\d^{d-1}$. Now consider the ground state of the boundary theory in flat space but let us evaluate the complexity on two different time slices, $\Sigma_1$ and $\S_2$, as illustrated in figure \ref{bump2}. In the second case, we have pushed $\Sigma_2$ forward in time over a portion of the time slice and hence, because of the Lorentzian signature of the boundary theory, the proper volume is reduced. Comparing these two time slices illustrated in the figure, $\Delta\calv=\calv(\S_2)-\calv(\S_1)=(\sqrt{\Delta\ell^2-\Delta t^2}-\Delta\ell)\,\calv_\mt{trans}<0$ where $\calv_\mt{trans}$ is the volume in the directions transverse to the page. Hence there is an enormous reduction in the corresponding complexity: $\Delta\calc\sim\Delta\calv/\d^{d-1}<0$. In fact, the leading divergence can be completely removed by evaluating the complexity on a time slice composed of a series of null segments, as illustrated by $\Sigma_3$ in figure \ref{bump2}. In this case, we expect that the complexity will still contain otherwise subleading divergences associated with the `folds' between the null segments in this case,\footnote{A logarithmic divergence appears with $d=2$ and  in this case, the holographic calculations for the CV duality would be closely related to those evaluating the cusp anomaly in holographic gauge theories  \cite{martink,Alday}.} \ie $\calc(\S_3)\sim\calv_\mt{trans}/\d^{d-2}$. 

However, it seems challenging to understand this behaviour from the usual perspective of circuit complexity. The latter involves using discrete gates to prepare a lattice approximation of the state in the boundary field theory. While this provides an intuitive picture for the complexity of states on a constant time slice, it seems ill-suited to discuss states (even ground states) that are defined on Cauchy surfaces which vary in time. Given a state defined on a discrete lattice for a constant time slice, one might consider evolving it to a time-varying slice with a differential application of the (local) Hamiltonian across the lattice. However, it is not at all clear why this process should significantly reduce the complexity of the state. Of course, similar issues arise if one considers entanglement entropy for discrete lattice models. However, in this case, we have a field theoretic approach where the entanglement entropy can be defined in terms of a path integral approach. Hence we naturally anticipate that the divergences in the entanglement entropy are defined in terms of covariant geometric quantities in a curved background or with a time-varying Cauchy surface. Of course, this discussion highlights the challenge of developing an analogous field theoretic approach to define complexity in a covariant manner. Perhaps, the techniques developed in \cite{Nielsen:2006:GAQ:2011686.2011688,Nielsen1133,Dowling:2008:GQC:2016985.2016986} or in \cite{beni} can provide better insight towards developing such a covariant approach.

\subsection*{Subregion Complexity}

In sections \ref{sec:subregions} and \ref{sec:subaction}, we also considered  generalizing the CV and CA conjectures to the complexity of the mixed state produced by reducing the boundary state to a specific subregion of the boundary time slice.
Our proposals were motivated by the idea that this mixed state should be encoded in the corresponding entanglement wedge in the bulk \cite{ewedge1,Headrick:2014cta}. While our suggestion for CV duality applies for general time-dependent situations, it reduces in a time-independent case to the proposal first studied in \cite{Alishahiha:2015rta} --- see also \cite{Ben-Ami:2016qex}. 

The original notion of circuit complexity that was introduced in holography, \eg \cite{Susskind:2014rva,
Susskind:2014moa,Brown:2015bva,Brown:2015lvg}, referred to pure states in the boundary theory. For a subregion complexity, we are instead considering the preparation of a mixed state, described by the density matrix $\rho_A$ which comes from reducing the global pure state to the region $A$. Since a pure reference state will never become mixed by the application of a unitary circuit, we should instead think in terms of preparing $\rho_A$ with a completely positive trace-preserving (CPTP) map acting on the reference state. From this perspective, the set of allowed universal gates would be extended to include `ancillary' and `erasure' gates, which add and remove additional degrees of freedom \cite{watrous2009quantum,Aharonov:1998zf}. However, the dilation theorems \cite{dilaton} imply that the most general CPTP maps acting on a system of qubits can be realized as unitary evolution of the system coupled to ancillary qubits \cite{watrous2009quantum}. That is, we may also think of subregion complexity as first extending the Hilbert space of $A$ with new ancillary degrees of freedom to purify the state $\rho_A$ and then determining the minimum number of universal gates needed to prepare the resulting pure state from a reference state in the extended Hilbert space. However, we expect in the present holographic context (or in quantum field theory, more generally) that the specific rules defining subregion complexity must restrict the allowed ancilla and how they are permitted to interact with the QFT degrees of freedom in the subregion. For example, locality of the QFT may suggest that the ancilla only interact with the degrees of freedom near the boundary of the subregion. Better insight into these restrictions may come from further studies of holographic subregion complexity.
	
Turning to the structure of the UV divergences revealed by our calculations in sections \ref{sec:subregions} and \ref{sec:subaction}, we found that these were more or less the same as found for the pure states. In particular, for a given region $A$, both $\ca(A)$ and $\cv(A)$ contained power law divergences and the coefficients of these divergences are again determined by local integrals of various geometric invariants, as in eqs.~\reef{eq:jdkkks} and \reef{cute1x}. However, there are now two types of integrals: The first were ($d$--1)-dimensional integrals over the entire region $A$ and the integrands were identical to those found in the previous calculations for a pure state on an entire time slice. The second were ($d$--2)-dimensional integrals over the boundary $\pa A$ and the integrands involved geometric invariants constructed on this geometry, as described schematically  in eqs.~\reef{purple} and \reef{cute3x}. In the discussion of complexity for pure states, we observed that it seems natural that the UV divergences multiplying `bulk' integrals over $A$ should be associated with the necessity of establishing correlations between the CFT degrees of freedom down to the arbitrarily short distance scales. Similarly then, the divergences multiplying the `boundary' integrals must be related to the UV structure of the portion of $\rho_A$ describing the degrees of freedom near the boundary $\pa A$. In particular, we note that these degrees of freedom behave as though they are nearly maximally mixed or strongly entangled with ancilla, \ie the near-boundary degrees of freedom appear as though they are in the Rindler vacuum with a local temperature that diverges at $\pa A$ --- see discussions in \cite{Arias:2016nip,Bianchi:2012ev}.

We note that there were no essential differences between the CV and CA duality for the divergences associated with the integrals over $A$. On the other hand, those associated with the boundary integrals seemed to show some more interesting differences. For example, the CA duality generally produces a divergence proportional to $\calv(\pa A)/\d^{d-2}$, while the analogous divergence never arises in the CV duality. Similarly, for a configuration which is time-symmetric about a time slice $\S=A\cup\bar{A}$,  we argued that $\cv(\S) = \cv(A)+\cv(\bar{A})$ in eq.~\reef{eq:jdkkks33}, while it is clear that $\ca(\S) \ne \ca(A)+\ca(\bar{A})$ even in the time-symmetric case. These differences must be related to differences in the implicit microscopic rules defining the subregion complexity for these two dualities. In particular, as discussed above, the different dualities may introduce different types of ancilla and allow for different types of interactions between the ancilla and the CFT degrees of freedom in the subregion. 
%\comment{might be interesting to study $\cv(\S) - \cv(A)-\cv(\bar{A})$ in general, \eg black holes.}

We close here with a technical observation  about our proposal for subregion complexity with the CA duality. One notable feature of eq.~\reef{trumpet} is that the coefficient $\beta$ does not appear in the final complexity. That is, while this coefficient which appears in the two joint contributions individually in eqs.~\reef{tiara} and \reef{stroll}, it cancels out in the total action. In fact, we will now argue that this cancellation is complete, rather than only holding to some high order in the expansion near the asymptotic boundary. The first observation is that $\beta$ dependence in eqs.~\reef{tiara} and \reef{stroll} is proportional to the volume of the corresponding surface, \ie
\beq
I^{(2)}\sim  - \frac{\log\b}{4\pi G_N}   \,\calv(C^+\cap C^-)\,,\quad
I^{(3)}\sim  + \frac{\log\b}{8\pi G_N}  \,\calv(C^+\cap S^+)\,,\quad
I^{(4)}\sim  + \frac{\log\b}{8\pi G_N}  \,\calv(C^-\cap S^-)\,.
\labell{ghost}
\eeq
Next, the key observation is that for the particular geometry which we are considering in the example in section \ref{sec:subaction} (\ie the bulk geometry is empty AdS and the boundary region is a ball on a constant time slice), the boundary \reef{ncone} of the entanglement wedge is actually a Killing horizon and the corresponding normals \reef{nnorm} are null Killing vectors, \eg \cite{Casini:2011kv,Faulkner2}. Hence,  $C^+\cap C^-$, which corresponds the a portion of the bifurcation surface, is mapped to either $C^+\cap S^+$ or $C^-\cap S^-$ by the Killing flow along the horizon -- see figure \ref{Wedge2}. Hence the `area' of these three cross-sections of the Killing horizon are identical, \ie $\calv(C^+\cap C^-)=\calv(C^+\cap S^+)=\calv(C^-\cap S^-)$, and hence the sum of the three expressions in eq.~\reef{ghost} exactly cancel, ensuring that the total action contains on $\beta$ dependence.

It was a fortunate coincidence that $\beta$ did not appear in the subregion complexity for the simple example considered in section \ref{sec:subaction}. For more generic situations, we would expect the subregion complexity to depend on the analogous normalization constant for the null generators of the boundary of entanglement wedge. Essentially, the same issues which were discussed above for $\a$, the normalization constant for the null generators on the boundaries of the WDW patch, will arise again here for $\beta$. It may seem natural to normalize the corresponding null normals near the asymptotic AdS boundary, in a manner similar to eq.~\reef{hop}. However, we point out that in the generic situation, the generators of the boundary of the entanglement wedge intersect the AdS boundary outside of the domain of dependence of the particular subregion of interest \cite{Headrick:2014cta}. As a result, we have the somewhat unsettling possibility that the subregion complexity may depend of features of the background geometry or of the global state which are casually disconnected from the subregion.
Certainly, our holographic proposals for evaluating subregion complexity should be studied further to make sure that they are consistent with the expectations which come from a quantum information perspective.\\

To conclude the discussion, we make a few comments on other possible future directions:  It would be interesting to study the rate of growth or time-dependence of subregion complexity. For example, if one considers the thermofield double state and a subregion that includes portions on both boundaries of the dual black hole, then we know that at some late time, there will be a transition to the RT surface which is disconnected and the holographic entanglement entropy saturates at some constant value \cite{tom33}. Similarly, we expect that the subregion complexity defined using either eq.~\reef{defineCVx} or eq.~\reef{trump} will also saturate at the same time. Another possibility would be to extend the present considerations to holographic complexity in other bulk geometries, such as Lifshitz black holes and D$p$-branes.

Recently, ref.~\cite{Freedman:2016zud} introduced the concept of `bit threads', which are flow lines emanating from a boundary region $A$ and threading the extremal RT surface, to define holographic entanglement in terms of the max flow-min cut principle. It would be interesting to understand if some property of the bit threads or the corresponding flows can be related to holographic complexity.
It might also be of interest to consider other covariantly defined geometric features of the bulk, \eg the spacetime volume of the entire entanglement wedge or the graviational action evaluated on this region \cite{Ben-Ami:2016qex}. It might also be interesting to extend the holographic complexity calculations to connect to the constructions appearing in \cite{Lashkari:2015hha,deBoer:2015kda,deBoer:2016pqk,Czech:2016xec}, where other fields are integrated over surfaces and regions in the bulk.

\section*{Acknowledgements}	
We thank Dorit Aharonov, Amin Faraji Astaneh, Omer Ben-Ami, Horacio Casini, Shira Chapman, Patrick Hayden, Michal Heller, Luis Lehner,  Aitor Lewkowycz, Hugo Marrochio, Eric Poisson, Sergey Solodukhin, Sotaro Sugishita, Brian Swingle, Guifre Vidal, Andrew Waldron, Run-Qiu Yang and Ying Zhao for useful comments and discussions. Research at Perimeter Institute is supported by the Government of Canada through the Department of Innovation, Science and Economic Development and by the Province of Ontario through the Ministry.  The work of DC is partially supported by the Israel Science Foundation (grant 1989/14), the US-Israel Binational  Science Foundation (grant 2012383) and the German-Israeli Foundation for Scientific Research and Development (grant I-244-303.7-2013). DC is also grateful for support from the Visiting Graduate Fellows program at the Perimeter Institute. PR is grateful to the Perimeter Scholars International program at Perimeter Institute. RCM is supported by funding from the Natural Sciences and Engineering Research Council of Canada, from the Canadian Institute for Advanced Research and from the Simons Foundation through the ``It from Qubit'' collaboration.

% % % % % % % % % % % % % % % % % % % % % % % % % % % % % % % % % % % % % % % % % % % % % % % % % % % % % % % % % % % % % % % % % % % % % % % % % % % % % %

\appendix

\section{Action User's Manual
\labell{sec:manuel}} 

The CA duality \cite{Brown:2015bva,Brown:2015lvg} requires evaluating the gravitational action for a bulk spacetime region with null boundaries. It is only very recently that a careful analysis was made of the boundary terms which must be added to the gravitational action for null boundary surfaces and for joints where such null boundaries intersect with other boundary surfaces \cite{Lehner:2016vdi} --- see also \cite{Pad1}. We review these results here but present them with a slightly different set of conventions. In particular, as discussed below, the normals to the boundary surfaces are always directed outward from the region of interest, and we do not make any special account for their orientation in time.

To begin, we write the gravitational action as
\beqa
&&\qquad\quad I= \frac1{16\pi G_N}\int_\calm d^{d+1}x \sqrt{-g} \left({R}-2\L
\right) +\frac1{8\pi G_N}\int_\calb d^{d}x \sqrt{|h|} \,K 
\nn
&&- \frac1{8\pi G_N}\int_{\calb'}d\lambda\, d^{d-1}\theta \sqrt{\gamma}\, \kappa+ \frac1{8\pi G_N}\int_\Sigma d^{d-1}x \sqrt{\sigma}\, \eta
+ \frac1{8\pi G_N}\int_{\Sigma'} d^{d-1}x \sqrt{\sigma}\, a\,.
\labell{actall}
\eeqa
In the first line, we have the standard Hilbert action, with a cosmological constant, and the Gibbons-Hawking boundary term \cite{York:1972sj,Gibbons:1977mu}. We have normalized the {\it negative} cosmological constant here such that $L$ is the curvature scale of the anti-de Sitter vacuum. 

Note that the Gibbons-Hawking boundary term is written in a way that it can be evaluated on either spacelike or timelike boundaries. However, to do so, we have a convention where the normal one-form is directed away from or out of the region of interest. That is, if a certain point of the boundary is determined by an equation $f(x)=0$, then the function $f(x)$ increases as we move out of the region of interest and hence, the form $df$ is directed outward. This may seem somewhat unconventional \cite{Lehner:2016vdi} since for a spacelike boundary, the (timelike) normal form ${\bf t} = t_\mu\,dx^\mu$ is outward directed but the normal vector $\vec t=t^\mu\,\partial_\mu$ is then inward directed. 

The first term in the second line of eq.~\reef{actall} is the corresponding surface term for null boundaries. The constant $\kappa$ is defined by the equation
\beq
k^\rho\,\nabla_{\!\rho}\, k_\mu =\kappa\,k_\mu
\labell{cape}
\eeq
where $\nkk=k_\mu\,dx^\mu$ is the outward directed null normal.
We can think that $\kappa$ measures the failure of $\lambda$ to be an affine parameter on the null generators of the null boundary. Of course, by choosing the normalization $\nkk$ appropriately then, we can always set $\kappa=0$.

\begin{figure}[b]
	\centering
	\includegraphics[scale=0.63]{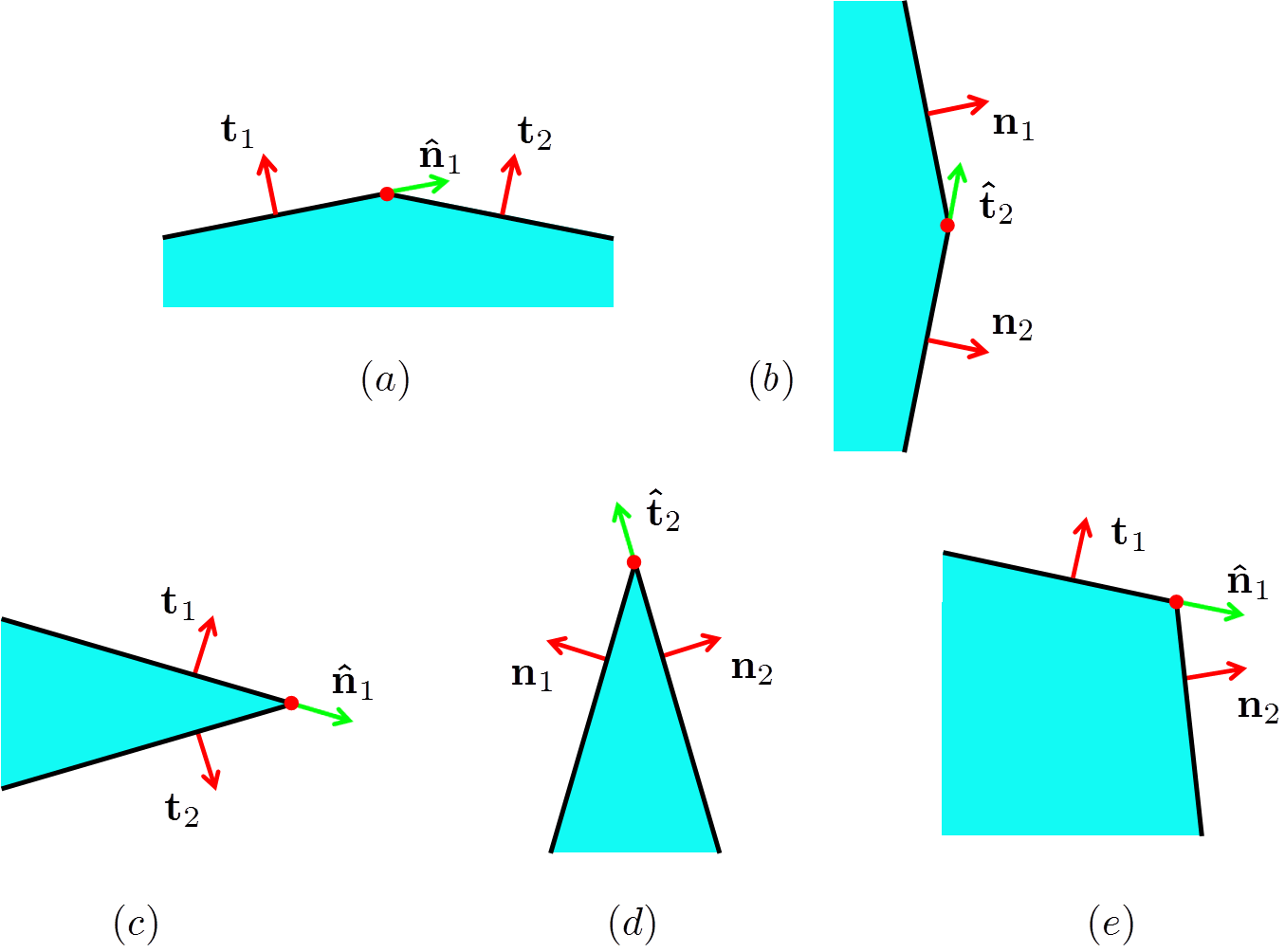} 
	\caption{Various joints or junctions considered by Hayward \cite{Hayward:1993my}. Class I junctions are shown in (a) and (b) while Class II junctions are given in (c), (d) and (e). \labell{joint}}
\end{figure}

For nonsmooth boundaries, we have the two joint terms appearing as the second and third terms in the second line of eq.~\reef{actall}. Here we are only considering the contributions required for spacelike joints. In particular, the first term is required for spacelike jointss of spacelike and timelike boundary regions \cite{Hayward:1993my}, as illustrated in figure \ref{joint} --- see also \cite{Brill:1994mb}. The integrand $\eta$ is given by:
\beqa
(a)\ \&\ (c)\ :&&\ \ \cosh\eta\equiv |\nt1\cdot\nt2|\quad\ {\rm with}\ \ sign(\eta)=-sign(\nt1\cdot\nt2)\,sign(\hn1\cdot\nt2) \labell{aax}\\
(b)\ \&\ (d)\ :&&\ \ \cosh\eta\equiv |\nlj1\cdot\nlj2| \quad{\rm with}\ \ sign(\eta)=-sign(\nlj1\cdot\nlj2)\,sign(\nlj1\cdot\hnt2)\labell{bbx}\\
(e)\ :&&\ \ \sinh\eta\equiv \eps\,\nt1\cdot\nlj2 \quad{\rm with}\ \ \eps=-sign(\nlj2\cdot\hn1)\labell{eex}
\eeqa
Our notation here distinguishes timelike and spacelike normals. In particular, timelike normals are denoted $\nt{i}$ with $\nt{i}\cdot\nt{i}=-1$, and spacelike normals are denoted $\nlj{i}$ with $\nlj{i}\cdot\nlj{i}=+1$. Implicitly again we are referring to the outward directed normal one forms. We have also introduced auxiliary unit vectors, $\hn{i}$ and $\hnt{i}$ --- {\it vectors}, not one-forms. These are defined as the unit vector that is in the tangent space of the  appropriate boundary region, orthogonal to the joint and pointing outward from the boundary region. We have chosen to normalize these auxiliary vectors as unit vectors but the signs in eqs.~(\ref{aax}--\ref{eex}) are independent of the normalization of these vectors. Further note that although the expression for the sign of $\eta$ is not symmetric in 1 and 2 (in the first two expressions), the result does not depend on which surfaces are labeled 1 or 2 since, \eg we have $sign(\hn1\cdot\nt2) = sign(\hn2\cdot\nt1)$ in eq.~\ref{eex}.

\begin{figure}[t]
	\centering
	\includegraphics[scale=0.5]{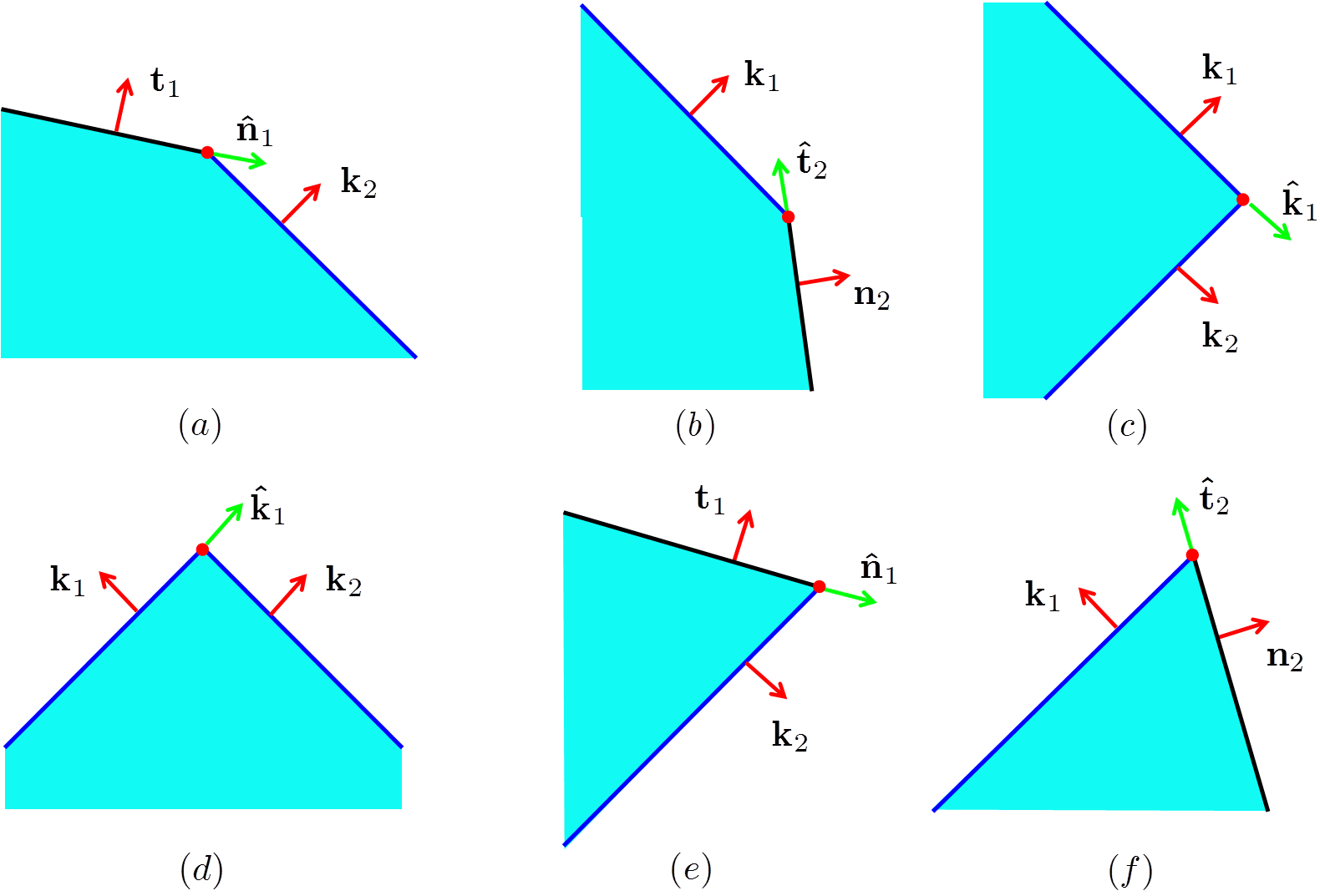} 
	\caption{Various different joints involving null boundaries.
		The null boundaries are indicated in blue.  \labell{full}}
\end{figure}

The last term in eq.~\reef{actall} is the appropriate boundary term for a spacelike joint involving one or two null boundary surfaces, as illustrated in figure \ref{full}.
Here, the integrand $a$ is given by:
\beqa
(a)\ \&\ (e)\ :&&\quad a\equiv \eps \,\log| \nt1\cdot\nk2|\quad\ \ \ \, {\rm with}\ \ \eps=-{\rm sign}(\nt1\cdot\nk2)\,{\rm sign}(\hn1\cdot\nk2)\,, \labell{aaeex}\\
(b)\ \&\ (f)\ :&&\quad a\equiv \eps \,\log| \nk1\cdot\nlj2|\quad\ \ \ {\rm with}\ \ \eps=-{\rm sign}(\nk1\cdot\nlj2)\,{\rm sign}(\nk1\cdot\hnt2)\,, \labell{bbff}\\
(c)\ \&\ (d)\ :&&\quad a\equiv \eps \,\log|\nk1\cdot\nk2/2|\quad{\rm with}\ \ \eps=-{\rm sign}(\nk1\cdot\nk2)\,{\rm sign}(\hnk1\cdot \nk2)\,.\labell{ccddx}
\eeqa
Implicitly again we are referring to outward directed normal null one forms with $\nk1$ or $\nk2$. Again, we also introduce auxiliary null vectors $\hnk{i}$ --- {\it vectors}, not one-forms. These are defined as the null vector that is in the tangent space of the  appropriate boundary region, orthogonal to the joint and pointing outward from the boundary region. Again, although the expression for the sign of $a$ is not symmetric in 1 and 2 (in the last expression), the result does not depend on which surfaces are labeled 1 or 2 since, \eg we have $sign(\hnk1\cdot\nk2) = sign(\hnk2\cdot\nk1)$ in eq.~\reef{ccddx}.

Further, we should recall from the discussion in \cite{Lehner:2016vdi}, the boundary terms in eq.~\reef{actall} associated with the null boundary surfaces  and null joints  are somewhat ambiguous. By construction, the variation of these boundary terms is well-defined and cancels the corresponding total derivative terms coming from the variation of the bulk action. However, when the gravitational action is evaluated on a particular spacetime geometry, it will generally yield different numerical values depending on different choices that can be made in constructing these boundary terms.   In particular, $\kappa$ depends on an arbitrary choice
for the parameterization for the null generators.  Further, for the null joints, $a$
depends on the arbitrary normalization of the null tangent $k^\alpha$ and in principle, we could add an additional function $a_0$ to $a$ in eqs.~(\ref{aaeex}--\ref{ccddx}), which remains fixed when the action is varied. 

Now as discussed \cite{Lehner:2016vdi}, there is a natural prescription to these ambiguities in the gravitational action.  As mentioned above, the $\kappa$
ambiguity is easily resolved by choosing the generators of the null boundary surfaces to be affinely parametrized, and then the corresponding boundary terms simply vanish. Further, eqs.~(\ref{aaeex}--\ref{ccddx}) make a particular choice for the functions $a_0$ at the null joints which guaranteed additivity for the gravitational
action. These choices leave only the freedom to rescale the affine parameter
along any of the null boundaries by a constant factor. However, this final ambiguity can be removed by imposing a normalization condition on the null normals near the asymptotic AdS boundary. One particularly appealing aspect of these choices is that they allow us to make a meaningful comparison of the action for different WDW patches, including in different bulk spacetimes.  For the most part, we simply adopt these choices formulated in \cite{Lehner:2016vdi} for our calculations of $I_\mt{WDW}$. However, we will not choose a fixed normalization condition for the null normals at the AdS boundary in sections \ref{sec:action} and \ref{sec:subaction}  --- see discussion in section \ref{discuss}.

% % % % % % % % % % % % % % % % % % % % % % % % % % % % % % % % % % % % % % % % % % % % % % % % % % % % % % % % % % % % % % % % % % % % % % % % % % % % % % % % % % % % % % % % % % % % % 

\section{Example: Extremal volume for a spherical boundary \labell{sec:extremal3}}

Let us consider an explicit example of a codimension-one slice of the boundary where both the intrinsic and extrinsic curvatures are non-vanishing. For simplicity, we will consider Euclidean AdS$_{d+1}$ in a foliation where the full boundary metric is simply $S^{d}$, with standard coordinates $\lbrace \theta,\phi_1,\dotsb,\phi_{d-1}\rbrace$ --- see figure \ref{fig:extremal}. Then we can consider a codimension-one slice with the geometry $S^{d-1}$ given by $\theta=\theta_0$, for which the extrinsic curvature is nonvanishing as long as $\theta_0 \neq \frac{\pi}{2}$. To obtain the corresponding extremal surface in the bulk, we could write the volume functional \reef{eq:lkopj} and attempt to solve the Euler-Lagrange equations \reef{eq:eom}. From spherical symmetry, we know that $\theta=\theta\left(\rho\right)$ where $\rho$ is the bulk radial direction which certainly simplifies the latter task. 
\begin{figure}[h]
	\center
	\includegraphics[scale=0.35]{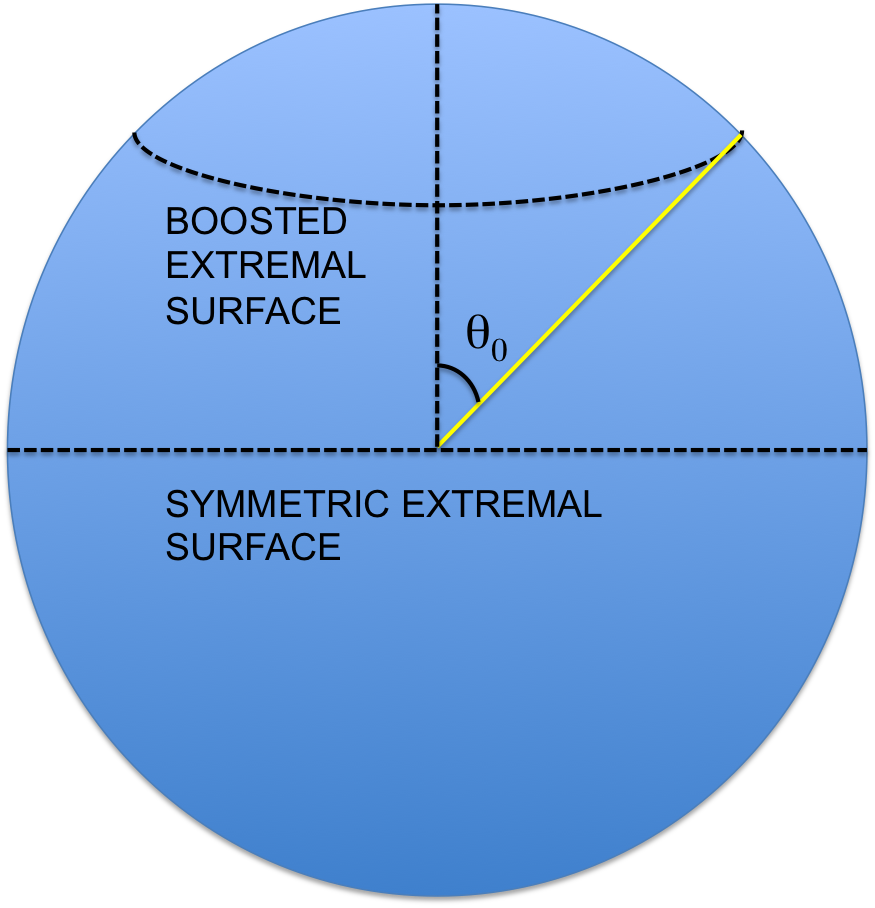}
	\caption{Extremal surface with $\theta_0 \neq \frac{\pi}{2}$ obtained by boosting the symmetric extremal surface at $\theta_0 =\frac{\pi}{2}$.\labell{fig:extremal}}
\end{figure}

However, we proceed with a useful trick following the discussion in \cite{Casini:2011kv}. That is, we use the fact that AdS can be embedded in flat space in one higher dimension
\begin{equation}
ds^2 = - dy_{-1}^2 + \sum_{i=0}^{d}\, dy_{i}^2\,.
\labell{eq:d2}
\end{equation}
Now $AdS_{d+1}$ is basically a hyperbolic slice of this, such that
\begin{equation}
 y_{-1}^2 - \sum_{i=0}^{d}\, y_{i}^2 = L^2
 \labell{eq:d3}
\end{equation}
Consider the foliation 
\bea
&&\qquad \qquad \qquad y_{-1}=L \cosh  u \ , \qquad \  y_0=L \sinh u \,\cos \theta
\nn
&&y_1=L\,\sinh u\,\sin \theta\,\cos \phi_1\, ,\quad  \cdots\,, \quad y_d=L\,\sinh u\,\sin\theta\cdots \sin\phi_{d-1}
\eea
which then yields the induced metric for the AdS geometry
\beq
ds^2 = L^2 \left[du^2 + \sinh^2\! u\ d\Omega^2_d \right]
\label{red}
\eeq
where $d\Omega^2_d$ is the usual line element on a unit $S^d$.
Let us look at the plane $y_0=0$ $\implies\,\theta=\frac{\pi}{2}$. By symmetry, this must be an extremal surface and its intersection with the boundary which lies at $u\rightarrow \infty$, is the equator of the boundary $S^d$. Hence we can see the bulk surface described by $y_0=0$ is a codimension-one extremal surface with an $S^{d-1}$ boundary. Now we simply perform a boost with boost parameter $\beta$ in the embedding space \reef{eq:d2}, which shifts the plane and hence its intersection surface \reef{eq:d3},
\begin{align*}
y_{-1}={}&(\cosh\,\beta)L\,\cosh\,u + (\sinh \,\beta)L\,\sinh\, u\,\cos\,\theta'\,,\\	
y_0={}&(\sinh\,\beta)L\,\cosh\,u + (\cosh \,\beta)L\,\sinh\, u\,\cos\,\theta'\,.
\end{align*}
where $\theta'$ is the angular coordinate for the symmetric extremal surface. Setting $\theta'=\frac{\pi}{2}$, we get an equation for $\theta (u)$
\beq
\cos\theta=\coth u\,\sinh\beta \labell{redder}
\eeq
where, as $u \rightarrow \infty$, $\theta \rightarrow \cos^{-1}(\,\sinh\,\beta)=\theta_0\neq\frac{\pi}{2}$. We can check that this bulk surface \reef{redder} satisfies the required Euler-Lagrange equation. Transforming the radial coordinate with $u=\log(2L/z)$, the AdS metric \reef{red} is put in FG form \reef{FGform}
\begin{equation}
ds^2=\frac{L^2}{z^2}\left[dz^2 + \left(1-\frac{z^2}{4L^2}\right)^2L^2\,d\Omega_{d}^2\right]
\end{equation}
with the extremal surface becoming $\theta(\rho)=\cos^{-1}\left(\left(\frac{1+\rho/4}{1-\rho/4}\right)\cos\,\theta_0\right)$. From this result, we can obtain the induced metric which, when substituted into the volume functional, yields the complexity
\begin{equation}
\cv = \frac{L^{d-1}}{G_N}\, \frac{\sin^{d-1}\theta_0\,\Omega_{d-1}}{d-1}\,\left(\frac{L^{d-1}}{\delta^{d-1}}-\frac{(d-1)^2L^{d-3}}{4(d-3)\,\delta^{d-3}}-\frac{(d-1)(d-2)\,\cot^2\theta_0\,L^{d-3}}{2(d-3)\,\delta^{d-3}}+\dotsb\right)\,,\labell{blue}
\end{equation}
where as in the main text, we have set the regulator surface at $z=\delta$. 

Now the boundary metric is $ds^2_\mt{bdy}=L^2\,d\Omega_{d}^2$. Hence we  recognize $L^{d-1}\sin^{d-1}\theta_0\,\Omega_{d-1}$ as the volume of the $S^{d-1}$ boundary slice. Further we can evaluate
\begin{equation}
\mathcal{R}=\frac{d(d-1)}{L^2},\quad \mathcal{R}^a_a=\frac{(d-1)^2}{L^2},\quad K=\frac{(d-1)\cot\,\theta_0}{L}
\end{equation}
Then we can explicitly confirm that the expansion of the complexity in eq.~\reef{blue} matches eq.~\reef{eq:divergence}, with integrals of boundary curvature invariants. There is one minor discrepancy in this comparison, namely, the $\cot^2\theta_0$ term above appears with a positive sign while the sign of the $K^2$ term in eq.~\reef{eq:divergence} is negative. This difference occurs because implicitly this sign is set by a factor of $\bn\cdot\bn$ and while the calculations in the section \ref{sec:extremal} use a Lorentzian signature, in this appendix, we work with a Euclidean signature.

% % % % % % % % % % % % % % % % % % % % % % % % % % % % % % % % % % % % % % % % % % % % % % % % % % % % % % % % % % % % % % % % % % % % % % % % % % % % % % % % % % % % % % % % % %

\section{Example: Wheeler-DeWitt action for global AdS \labell{newA}}

Using the rules prescribed in appendix \ref{sec:manuel}, here we study the divergence structure of the WDW action in the simple example of a constant time slice on the boundary of global AdS$_{d+1}$. We will also compare the results found using the two different regularization procedures illustrated in figure~\ref{fig:space}. In either case, we introduce a standard (timelike) regulator surface at a distance $\d$ from the boundary of AdS. Then in figure~\ref{fig:space}a, we discard the portion of  the WDW patch extending beyond this surface, \ie we only integrate the bulk action out to this maximum radius. However, the regulated WDW region then has a new timelike boundary segment and two null joints at this surface, which contribute to $I_\mt{WDW}$. In figure~\ref{fig:space}b, we instead regulate the calculation by simply shifting the edge of the WDW patch inwards to the regulator surface. We will show that the structure of the UV divergences in the corresponding complexity $\ca$ is the same for both procedures. 

The $AdS_{d+1}$ metric with boundary geometry $R \times S^{d-1}$ can be written in the following form: 
\begin{equation}
ds^2 = \frac{L^2}{\cos^2\theta}\left(-d\tau^2+d\theta^2+\sin^2\theta\,d\Omega_{d-1}^2\right)\,,
\labell{howto}
\end{equation}
where $L$ is the AdS radius, and the boundary is at $\theta=\pi/2.$
From the previous discussion of the geometries in figure \ref{fig:space}, we see that the WDW action may receive contributions from the Einstein-Hilbert bulk term, the Gibbons-Hawking-York boundary term, and the null joint terms in eq.~\reef{actall}:
\bea
I_\mt{WDW}={}&I_\mt{bulk}+ I_\mt{GHY}+I_\mt{jnt}\,.
\eea 
As discussed in appendix \ref{sec:manuel}, we are assuming that the generators on null boundaries are affinely parametrized so that we may ignore the null boundary terms, \ie $\kappa=0$. One may also examine the contributions to the gravitational action coming from the caustics at the tips of the WDW patch where all of the null generators meet. However, these contributions were examined in detail in \cite{shira} and were shown to vanish there.

Using the regularization of Figure~\ref{fig:space}a, the boundaries of the WDW patch are
\beqa
S^+\ :&\quad \theta= \frac{\pi}2-\tau  & \quad{\rm for}\ \ \ \ 
\ \ \frac{\pi}2\ge\tau\ge \d'\,, 
\nonumber\\
S^-\ :&\quad \theta= \frac{\pi}2+\tau & \quad{\rm for}\ \ -\frac{\pi}2\le\tau\le -\d'\,, 
\labell{bounding}\\
R\,\ :&\quad \theta= \frac{\pi}2-\delta' & \quad{\rm for}\ \ -\d'\le\tau\le \d' \,,
\nonumber
\eeqa
where $S^+$ and $S^-$ are the future and past null boundaries, while $R$ is the UV regulator surface. Of course, implicitly we have chosen the boundary time slice to be $\tau=0$. The bulk contribution to the WDW action then becomes
\begin{align}
\label{eq:poilu2}
I_\mt{bulk} 
&=-\frac{d\, L^{d-1}\Omega_{d-1}}{4\pi G_N}\int^{\frac{\pi}{2}}_{\d'} d\tth \ \tth \, \cot^{d-1}\!\tth \, \csc^2\!\tth
\end{align}
where $\tth=\frac\pi2-\theta$ and $\Omega_{d-1}$ is the area of a unit ($d$--1)-sphere. The integral above can be evaluated in terms of hypergeometric functions, but here we are only looking for  the leading behavior as $\d'\rightarrow0$, which can be extracted using a series expansion for small $\tth$, 
\beq
I_\mt{bulk}  =-\frac{d\, L^{d-1}\Omega_{d-1}}{4\pi G_N} \bigg[ \frac{1}{(d-1)\d'^{d-1}} - \frac{d-2}{30(d-3)\d'^{d-3}} +\ldots    \bigg]
\labell{rounder}
\eeq
 To evaluate the GHY and null joint terms, we must consider the normals to the boundary surfaces \reef{bounding}, 
\bea
&&S^+:\ \nk1=\a_1\,L\,(d\theta+d\tau)\,,\qquad 
 S^-:\ \nk2=\a_2\,L\,(d\theta-d\tau)\,,\qquad
R:\ \nnv=\frac{L}{\sin\d'}\,d\theta\,,
\labell{nnorm2}
\eea
where $\a_{1,2}$ are (dimensionless) normalization constants. For the regulator surface $R$, we have $K=\frac{1}{L}\left(\frac{d-1}{\cos\d'}+ \cos\d' \right)$ and hence the corresponding boundary term becomes
\bea
I_\mt{GHY} = \frac{L^{d-1}\Omega_{d-1} }{4 \pi G_N}\, \d'\,\frac{\cos^d \d'}{\sin^d \d'} \left(\frac{d-1}{\cos^2\d'}+ 1 \right)=\frac{L^{d-1}\Omega_{d-1} }{4 \pi G_N}\left( \frac{d}{\d^{d-1}}- \frac{d^2-3d+3}{3\d'^{d-3}}+\ldots\right)\,,
\labell{poilu}
\eea
 where we are considering the limit $\d'\to0$ in the final expression. For the null joint at $S^+\cap R$, eq.~\reef{bbff} yields $a_1=-\log\left(\nnv\cdot\nk1\right) = -\log(\a_1 \,\sin \d')$. Similarly for $S^-\cap R$,  $a_2= -\log(\a_2 \,\sin \d')$.
Hence combining the two joint contributions yields
\bea
I_\mt{jnt} &=&-\frac{L^{d-1}\Omega_{d-1}}{4 \pi G_N} \frac{\cos^{d-1} \d'}{\sin^{d-1} \d'}   \log\left(\sqrt{\a_1 \a_2 }\sin \,\d'\right)
\nn
&=& \frac{L^{d-1}\Omega_{d-1}}{4 \pi G_N}\bigg[ \log\left(\frac1{\sqrt{\a_1 \a_2 }\,\d'}\right)
\left(\frac{1}{\d'^{d-1}}- \frac{d-1}{3\d'^{d-3}}+\cdots\right) +  \left(\frac{1}{6\d'^{d-3}} - \frac{10d-11}{180\d'^{d-5}}   \right) \bigg]
\labell{round}
\eea
 Combining all of the above results, we see the divergence structure of the corresponding complexity \reef{none} emerges as
\bea
\ca&=&   \frac{L^{d-1}\Omega_{d-1}}{4 \pi^2 G_N}\left[ \frac{d(d-2)}{(d-1)\,\d'^{d-1}}-\frac{10d^3-61d^2+117d-75}{30(d-3)\d'^{d-3}}+\dotsb  \right]
\nn
&&\quad+\frac{L^{d-1}\Omega_{d-1}}{4 \pi^2 G_N}\,\log\left(\frac{1}{\sqrt{\a_1 \a_2 }\,\d'}\right)  \left[\frac{1}{\d'^{d-1}}-\frac{d-1}{3\,\d'^{d-3}}+\dotsb\right]  
\labell{putter}
\eea
Note that the leading divergence coming from the joint term is positive. Likewise, the leading power divergence is positive in this regularization.

This result is expressed in terms of the (dimensionless) boundary regulator $\d '$, which was convenient in the present coordinates \reef{howto}. To relate $\d'$ to the short-distance cutoff $\d$ appearing in the main text, we introduce the coordinate transformation
\begin{equation}
z=\frac{2L\,\cos\theta}{1+\sin\theta}\,.
\end{equation}
Thus, the two regulators are related by
\begin{align}
\d &= \frac{2L\,\sin\d'}{1+\cos\d'}\ \ \longrightarrow\ \
\d'=\frac{\d}{L} - \frac{ \d^3}{12\, L^3}+\cdots\,.
\labell{cutter}
\end{align} 
Then, in terms of $\d$, the complexity \reef{putter} becomes
\bea
\ca&=&   \frac{L^{d-1}\Omega_{d-1}}{4 \pi^2 G_N}\left[ \frac{d(d-2)}{d-1}\,\frac{L^{d-1}}{\d^{d-1}}-\frac{15d^3-97d^2+199d-135}{60(d-3)}\,\frac{L^{d-3}}{\d^{d-3}}+\dotsb  \right]
\labell{putter2}\\
&& \quad + \frac{L^{d-1}\Omega_{d-1}}{4 \pi^2 G_N}\,\log\left(\frac{L}{\sqrt{\a_1 \a_2 }\,\d}\right)   \left[\frac{L^{d-1}}{\d^{d-1}}-\frac{d-1}{4}\,\frac{L^{d-3}}{\d^{d-3}}+\dotsb\right] \,.
\nonumber
\eea

Alternatively, we could have used the second regularization illustrated Figure~\ref{fig:space}b. For this case there is no time-like boundary, and the null normals $\nk1$ and $\nk2$ are the same as in eq.~\reef{nnorm2}. The boundaries of the WDW patch are:
\beqa
S^+\ :&\quad \theta= \frac{\pi}2-\tau -\d'  & \quad{\rm for}\ \ \ \ 
\ \ \frac{\pi}2-\d' \ge\tau\ge 0\,, 
\nonumber\\
S^-\ :&\quad \theta= \frac{\pi}2+\tau -\d' & \quad{\rm for}\ \ -\frac{\pi}2 +\d '\le\tau\le 0\,, 
\labell{bounding22}
\eeqa 
 For this case there is no space-like boundary. The joint terms turn out to be the same in the two regularizations. Expressed in terms of the cutoff $\d$, eq.~\reef{round} becomes
 \bea
I_\mt{jnt} 
&=& \frac{L^{d-1}\Omega_{d-1}}{4 \pi G_N}\bigg[ \log\left(\frac{L}{\sqrt{\a_1 \a_2}\,\d}\right)
\left(\frac{L^{d-1}}{\d^{d-1}}- \frac{d-1}{4}\,\frac{L^{d-3}}{\d^{d-3}}+\cdots\right) +  \left(\frac{L^{d-3}}{4\d^{d-3}} + \cdots   \right) \bigg]\,.
\labell{round2}
\eea 
However,  the bulk term is slightly modified:
\begin{align}
\label{eq:poilu3}
I_{bulk}&= \frac{2}{16 \pi G_N}\int_{0}^{\frac{\pi}{2}-\d'} d\theta \int^{\frac{\pi}{2}-\theta - \d'}_{0}\!\!\! d\tau \int d\Omega_{d-1} \frac{L^{d+1}\, \sin^{d-1}\theta}{\cos^{d+1}\theta} \left(\frac{-2d}{L^2}\right)
\nn
&= -\frac{d\, L^{d-1}\Omega_{d-1}}{4\pi G_N}\int^{\frac{\pi}{2}}_{\d'} dx (x-\d') \cot^{d-1}x \, \csc^2\,x
\nn
&=-\frac{L^{d-1}\Omega_{d-1}}{4\pi G_N}\left[ \frac{1}{(d-1)\d'^{d-1}}
%\frac{1}{} 
-\frac{d}{3(d-3)}\frac{1}{\d'^{d-3}} +\ldots \right]\nn
&=-\frac{L^{d-1}\Omega_{d-1}}{4\pi G_N} \left[\frac{L^{d-1}}{(d-1)\d^{d-1}}-\frac{(d+1)}{4(d-3)}\,\frac{L^{d-3}}{\d^{d-3}}+\cdots\right]
\end{align}
Combining these two contributions for the action, we find
\bea
\ca&=&   -\frac{L^{d-1}\Omega_{d-1}}{4\pi^2 G_N}\left[\frac{L^{d-1}}{(d-1)\d^{d-1}}-\frac{d-1}{2(d-3)}\,\frac{L^{d-3}}{\d^{d-3}}+\cdots \right]\labell{gromit1}\\
&&\quad+ \frac{L^{d-1}\Omega_{d-1}}{4 \pi^2 G_N}\,\log\left(\frac{L}{\sqrt{\a_1 \a_2 }\,\d}\right)   \left[\frac{L^{d-1}}{\d^{d-1}}-\frac{d-1}{4}\,\frac{L^{d-3}}{\d^{d-3}}+\dotsb\right] \nonumber
\eea
This result can also be compared with the general geometric expression in eq.~\reef{trumpet2}. The required curvature invariants for the present example are
\beq
\mathcal{R}=\mathcal{R}^a_a=\frac{(d-1)(d-2)}{L^2}\,,\qquad
%\mathcal{R}_{ij}\,n^i\, n^j = 
K^2= K_{ab}K^{ab}=0\,. \labell{gromit}
\eeq
Now substituting these expressions into eq.~\reef{trumpet2} reproduces precisely the divergences given above in eq.~\reef{gromit1}.

Comparing eqs.~\reef{putter2} and \reef{gromit1}, we see that the form of divergences remains the same between the two regularizations, however, the coefficients are typically different. Generally, these coefficients are not universal and so these differences are not at all surprising. However, it is of interest to compare the logarithmic contribution appearing with the two regulators since the coefficient of this term is usually regarded as universal. However, unfortunately a careful analysis shows that in general the two regularizations produce different coefficients for this term as well. We expect that this difference is related to the ambiguity in the choice of the normalization constants, $\a_1$ and $\a_2$, discussed in section \ref{discuss}.

% % % % % % % % % % % % % % % % % % % % % % % % % % % % % % % % % % % % % % % % % % % % % % % % % % % % % % % % % % % % % % % % % % % % % % % % % % % % % % % % % % % % % % % % % % % %
% % % % % % % % % % % % % % % % % % % % % % % % % % % % % % % % % % % % % % % % % % % % % % % % % % % % % % % % % % % % % % % % % % % % % % % % % % % % % % % % % % % % % % % % % % % %
% % % % % % % % % % % % % % % % % % % % % % % % % % % % % % % % % % % % % % % % % % % % % % % % % % % % % % % % % % % % % % % % % % % % % % % % % % % % % % % % % % % % % % % % % % % %

\section{Geometric details for CA duality calculation
	\labell{sec:actionXXX}}

Two coefficients, $q_0^{(2)}$ and $q_2^{(0)}$, appear in eq.~\reef{big22} and here we will derive eq.~\reef{rumpt} which provides a geometric translation for these factors. Recall that these coefficients were defined by the double expansion of the measure $\sqrt\g=\sqrt{{\rm det} [g_{ab}(x,z)]}$ in eq.~\reef{pout}, which we reproduce here for the reader's convenience: 
\bea 
\sqrt{\g}= \sqrt{h (\s)}\Big([1+q_0^{(2)}(\s^a)z^2+\ldots ]+[q_1^{(0)}(\s^a)+q_1^{(2)}(\s^a)z^2 +\ldots]t+ [q_2^{(0)}(\s^a)+q_2^{(2)}(\s^a)z^2 +\ldots]t^2  +\dotsb \Big)\,.
\labell{pott}
\eea
Expanding the determinant using FG expansion \reef{eq:poli}, as well as eq.~\reef{eq:g1}, then yields
\bea
\sqrt{\g}=  \sqrt{h}\,\Bigg(1+\frac{z^2}{2}\,\overset{(0)}{g}{}^{a b}\,\overset{(1)}{g_{a b}}+\dotsb \Bigg)=
 \sqrt{h}\, \Bigg(1-\frac{z^2}{2(d-2)} \left(\mathcal{R}^{a}_{a}-\frac{1}{2}\,\mathcal{R}\right)+\dotsb \Bigg)
 \labell{pott2}
\eea
where  $\mathcal{R}^a_{a}=h^{ab}\,\mathcal{R}_{ab}$ --- recall that the boundary metric takes the form given in eq.~\reef{eqLpolikmnb} and we are considering the time slice $t=0$. Hence comparing eqs.~\reef{pott} and \reef{pott2}, we see
\bea
	q_0^{(2)}(\s^a) =-\frac{1}{2(d-2)}\left.\left(\mathcal{R}^{a}_{a}-\frac{1}{2}\,\mathcal{R}\right)\right|_{t=0} \,. 
\eea 

To evaluate $q_1^{(0)}$ and $q_2^{(0)}$, we first set 
$z=0$ to reduce eq.~\reef{pott} to
\bea
\sqrt{\g}\big|_{z=0} =\sqrt{h}\,\Big(1 +q_1^{(0)}(\s^a)t +q_2^{(0)}(\s^a)t^2  +\dotsb \Big)\,.
\eea
Now differentiating with respect to time, we find
\bea
\frac{1}{\sqrt{h}}\,\pa_t \sqrt{\g}\big|_{ t=0}  =   q_1^{(0)}(\s^a)    \ , \qquad\qquad  \frac{1}{\sqrt{h}}\, \pa^2_t \sqrt{\g}\big|_{ t=0}  = 2  q_2^{(0)}(\s^a)   \,.
\label{slash}
\eea
Now we may write the trace of the extrinsic curvature as
\begin{equation}
\label{eq:bla1}
K(t,\s)= \nabla_\m n^\m= \frac{1}{\sqrt{-g}}\pa_\m (n^\m\sqrt{-g}) =\frac{1}{\sqrt{-g}}\pa_t (n^t\sqrt{-g}) =\frac{1}{\sqrt{-g}}\,\partial_{t}\!\left(\frac{\sqrt{-g}}{\sqrt{-g_{tt}}}\right) = \frac{1}{\sqrt{\g}}\,\partial_{t}\sqrt{\g}\,,
\end{equation}
since we have fixed the boundary metric as in eq.~\reef{eqLpolikmnb}.
Hence comparing the last two equations, we see
\begin{equation}
q_1^{(0)}(\s^a) =	K(t,\s) \big|_{t=0}\,.  
\end{equation}

Now turning on to $q_2^{(0)}$, eq.~\ref{eq:bla1} gives
\begin{equation}
\sqrt{\g}\, K(t,\s)= \partial_{t} \sqrt{\g}\,. \labell{combo}
\end{equation}
By differentiating this result, we can rewrite the expression for $q_2^{(0)}$ in eq.~\reef{slash} as
\bea
\label{eq:opnssd}
q_2^{(0)}(\s^a) &=& \frac{1}{2\sqrt{h }}\, \partial^2_{t} \sqrt{\g}\big|_{t=0} =  \frac{1}{2\sqrt{h}}\,\pa_t \big[\sqrt{\g}\, K(t,\s) \big]\big|_{t=0}
%\nn
%&=&\frac{1}{2\sqrt{\gamma }}\, \big[ \pa_t (\sqrt{\s}) K(t,x) + \sqrt{\s} \pa_t K\big]\big|_{t=0} 
=  \frac{1}{2}\, \big[  K^2 +  \pa_t K\big]\big|_{t=0}
\,,
\eea 
where the final result uses eq.~\reef{combo} again.
To find a covariant replacement for $\pa_t K$, we can use the following identity (\eg see \cite{Gourgoulhon:2007ue})
\bea
\frac{1}{N}\Big[  \mm{L}_m K +   D_a D^a N \Big]=  K_{ab}K^{ab}+  n^i n^j \mathcal{R}_{ij}
\eea 
where $N$ is the lapse function.
In our case, $\mm{L}_m = \pa_t$ and $N=1$, and hence we may write
\bea
\pa_t K =  K_{ab}K^{ab}+  n^i n^j \mathcal{R}_{ij}  =  K_{ab}K^{ab}+ \mathcal{R}^{a}_{a}- \mathcal{R}
\eea
where we used $n^i n^j \mathcal{R}_{ij}  =  \mathcal{R}^{a}_{a}- \,\mathcal{R} $.
Thus from eq.\reef{eq:opnssd}, we have
\bea
q_2^{(0)}(\s^a)   
%= \frac{1}{2} [  K^2 +  \pa_t K]\Big|_{t=0} 
= \frac{1}{2}\left. \Big(  K^2 +   K_{ab}K^{ab}+ \mathcal{R}^{a}_{a}- \,\mathcal{R}  \Big)\right|_{t=0}\,.
\eea

% % % % % % % % % % % % % % % % % % % % % % % % % % % % % % % % % % % % % % % % % % % % % % % % % % % % % % % % % % % % % % % % % % % % % % % % % % % % % % % % % % % % % % % % % % % %
% % % % % % % % % % % % % % % % % % % % % % % % % % % % % % % % % % % % % % % % % % % % % % % % % % % % % % % % % % % % % % % % % % % % % % % % % % % % % % % % % % % % % % % % % % % %
% % % % % % % % % % % % % % % % % % % % % % % % % % % % % % % % % % % % % % % % % % % % % % % % % % % % % % % % % % % % % % % % % % % % % % % % % % % % % % % % % % % % % % % % % % % %

% % % % % % % % % % % % % % % % % % % % % % % % % % % % % % % % % % % % % % % % % % % % % % % % % % % % % % % % % % % % % % % % % % % % % % % % % % % % % % % % % % % % % % % % % %

% % % % % % % % % % % % % % % % % % % % % % % % % % % % % % % % % % % % % % % % % % % % % % % % % % % % % % % % % % % % % % % % % % % % % % % % % % % % % % % % % % % % % % % % %

\bibliographystyle{utphys}

\bibliography{New_Draft}

\providecommand{\href}[2]{#2}\begingroup\raggedright\begin{thebibliography}{10}

\bibitem{watrous2009quantum}
J.~Watrous, ``Quantum computational complexity,'' {\em {\rm in} Encyclopedia of
  Complexity and Systems Science, ed., R. A. Meyers} (2009) 7174--7201,
  \href{http://arxiv.org/abs/0804.3401}{{\ttfamily arXiv:0804.3401
  [quant-ph]}}.

\bibitem{2014arXiv1401.3916G}
S.~Gharibian, Y.~Huang, Z.~Landau, and S.~W. Shin, ``Quantum hamiltonian
  complexity,'' \href{http://dx.doi.org/10.1561/0400000066}{{\em Foundations
  and Trends in Theoretical Computer Science} {\bfseries 10} (2015) 159--282},
  \href{http://arxiv.org/abs/1401.3916}{{\ttfamily arXiv:1401.3916
  [quant-ph]}}.

\bibitem{0034-4885-75-2-022001}
T.~J. Osborne, ``Hamiltonian complexity,'' {\em Reports on Progress in Physics}
  {\bfseries 75} (2012) 022001,
  \href{http://arxiv.org/abs/1106.5875}{{\ttfamily arXiv:1106.5875
  [quant-ph]}}.

\bibitem{Susskind:2014rva}
L.~Susskind, ``{Computational Complexity and Black Hole Horizons},''
  \href{http://dx.doi.org/10.1002/prop.201500092}{{\em Fortsch. Phys.}
  {\bfseries 64} (2016) 24--43},
\href{http://arxiv.org/abs/1402.5674}{{\ttfamily arXiv:1402.5674 [hep-th]}}.
%%CITATION = ARXIV:1402.5674;%%.

\bibitem{2014arXiv1403.5695S}
L.~{Susskind}, ``{Addendum to: Computational Complexity and Black Hole
  Horizons},'' {\em Fortsch. Phys.} {\bfseries 64} (2016) 44,
\href{http://arxiv.org/abs/1403.5695}{{\ttfamily arXiv:1403.5695 [hep-th]}}.
%%CITATION = doi:10.1002/prop.201500093;%%.

\bibitem{Stanford:2014jda}
D.~Stanford and L.~Susskind, ``{Complexity and Shock Wave Geometries},''
  \href{http://dx.doi.org/10.1103/PhysRevD.90.126007}{{\em Phys. Rev.}
  {\bfseries D90} (2014) 126007},
\href{http://arxiv.org/abs/1406.2678}{{\ttfamily arXiv:1406.2678 [hep-th]}}.
%%CITATION = ARXIV:1406.2678;%%.

\bibitem{Susskind:2014jwa}
L.~Susskind and Y.~Zhao, ``{Switchbacks and the Bridge to Nowhere},''
\href{http://arxiv.org/abs/1408.2823}{{\ttfamily arXiv:1408.2823 [hep-th]}}.
%%CITATION = ARXIV:1408.2823;%%.

\bibitem{Susskind:2014moa}
L.~Susskind, ``{Entanglement is not enough},''
  \href{http://dx.doi.org/10.1002/prop.201500095}{{\em Fortsch. Phys.}
  {\bfseries 64} (2016) 49--71},
\href{http://arxiv.org/abs/1411.0690}{{\ttfamily arXiv:1411.0690 [hep-th]}}.
%%CITATION = ARXIV:1411.0690;%%.

\bibitem{Brown:2015bva}
A.~R. Brown, D.~A. Roberts, L.~Susskind, B.~Swingle, and Y.~Zhao,
  ``{Holographic Complexity Equals Bulk Action?},''
  \href{http://dx.doi.org/10.1103/PhysRevLett.116.191301}{{\em Phys. Rev.
  Lett.} {\bfseries 116} (2016) 191301},
\href{http://arxiv.org/abs/1509.07876}{{\ttfamily arXiv:1509.07876 [hep-th]}}.
%%CITATION = ARXIV:1509.07876;%%.

\bibitem{Brown:2015lvg}
A.~R. Brown, D.~A. Roberts, L.~Susskind, B.~Swingle, and Y.~Zhao,
  ``{Complexity, action, and black holes},''
  \href{http://dx.doi.org/10.1103/PhysRevD.93.086006}{{\em Phys. Rev.}
  {\bfseries D93} (2016) 086006},
\href{http://arxiv.org/abs/1512.04993}{{\ttfamily arXiv:1512.04993 [hep-th]}}.
%%CITATION = ARXIV:1512.04993;%%.

\bibitem{Couch:2016exn}
J.~Couch, W.~Fischler, and P.~H. Nguyen, ``{Noether charge, black hole volume
  and complexity},''
\href{http://arxiv.org/abs/1610.02038}{{\ttfamily arXiv:1610.02038 [hep-th]}}.
%%CITATION = ARXIV:1610.02038;%%.

\bibitem{Christodoulou:2014yia}
M.~Christodoulou and C.~Rovelli, ``{How big is a black hole?},''
  \href{http://dx.doi.org/10.1103/PhysRevD.91.064046}{{\em Phys. Rev.}
  {\bfseries D91} (2015) 064046},
\href{http://arxiv.org/abs/1411.2854}{{\ttfamily arXiv:1411.2854 [gr-qc]}}.
%%CITATION = ARXIV:1411.2854;%%.

\bibitem{Christodoulou:2016tuu}
M.~Christodoulou and T.~De~Lorenzo, ``{Volume inside old black holes},''
  \href{http://dx.doi.org/10.1103/PhysRevD.94.104002}{{\em Phys. Rev.}
  {\bfseries D94} (2016) 104002},
\href{http://arxiv.org/abs/1604.07222}{{\ttfamily arXiv:1604.07222 [gr-qc]}}.
%%CITATION = ARXIV:1604.07222;%%.

\bibitem{Lehner:2016vdi}
L.~Lehner, R.~C. Myers, E.~Poisson, and R.~D. Sorkin, ``{Gravitational action
  with null boundaries},''
  \href{http://dx.doi.org/10.1103/PhysRevD.94.084046}{{\em Phys. Rev.}
  {\bfseries D94} (2016) 084046},
\href{http://arxiv.org/abs/1609.00207}{{\ttfamily arXiv:1609.00207 [hep-th]}}.
%%CITATION = ARXIV:1609.00207;%%.

\bibitem{Ryu:2006bv}
S.~Ryu and T.~Takayanagi, ``{Holographic derivation of entanglement entropy
  from AdS/CFT},'' \href{http://dx.doi.org/10.1103/PhysRevLett.96.181602}{{\em
  Phys. Rev. Lett.} {\bfseries 96} (2006) 181602},
\href{http://arxiv.org/abs/hep-th/0603001}{{\ttfamily arXiv:hep-th/0603001
  [hep-th]}}.
%%CITATION = HEP-TH/0603001;%%.

\bibitem{Ryu2}
S.~Ryu and T.~Takayanagi, ``{Aspects of Holographic Entanglement Entropy},''
  \href{http://dx.doi.org/10.1088/1126-6708/2006/08/045}{{\em JHEP} {\bfseries
  08} (2006) 045},
\href{http://arxiv.org/abs/hep-th/0605073}{{\ttfamily arXiv:hep-th/0605073
  [hep-th]}}.
%%CITATION = HEP-TH/0605073;%%.

\bibitem{Hung:2011xb}
L.-Y. Hung, R.~C. Myers, and M.~Smolkin, ``{On Holographic Entanglement Entropy
  and Higher Curvature Gravity},''
  \href{http://dx.doi.org/10.1007/JHEP04(2011)025}{{\em JHEP} {\bfseries 04}
  (2011) 025},
\href{http://arxiv.org/abs/1101.5813}{{\ttfamily arXiv:1101.5813 [hep-th]}}.
%%CITATION = ARXIV:1101.5813;%%.

\bibitem{Hung:2011ta}
L.-Y. Hung, R.~C. Myers, and M.~Smolkin, ``{Some Calculable Contributions to
  Holographic Entanglement Entropy},''
  \href{http://dx.doi.org/10.1007/JHEP08(2011)039}{{\em JHEP} {\bfseries 08}
  (2011) 039},
\href{http://arxiv.org/abs/1105.6055}{{\ttfamily arXiv:1105.6055 [hep-th]}}.
%%CITATION = ARXIV:1105.6055;%%.

\bibitem{Bombelli:1986rw}
L.~Bombelli, R.~K. Koul, J.~Lee, and R.~D. Sorkin, ``{A Quantum Source of
  Entropy for Black Holes},''
\href{http://dx.doi.org/10.1103/PhysRevD.34.373}{{\em Phys. Rev.} {\bfseries
  D34} (1986) 373--383}.
%%CITATION = PHRVA,D34,373;%%.

\bibitem{Srednicki:1993im}
M.~Srednicki, ``{Entropy and area},''
  \href{http://dx.doi.org/10.1103/PhysRevLett.71.666}{{\em Phys. Rev. Lett.}
  {\bfseries 71} (1993) 666--669},
\href{http://arxiv.org/abs/hep-th/9303048}{{\ttfamily arXiv:hep-th/9303048
  [hep-th]}}.
%%CITATION = HEP-TH/9303048;%%.

\bibitem{Alishahiha:2015rta}
M.~Alishahiha, ``{Holographic Complexity},''
  \href{http://dx.doi.org/10.1103/PhysRevD.92.126009}{{\em Phys. Rev.}
  {\bfseries D92} (2015) 126009},
\href{http://arxiv.org/abs/1509.06614}{{\ttfamily arXiv:1509.06614 [hep-th]}}.
%%CITATION = ARXIV:1509.06614;%%.

\bibitem{Ben-Ami:2016qex}
O.~Ben-Ami and D.~Carmi, ``{On Volumes of Subregions in Holography and
  Complexity},'' \href{http://dx.doi.org/10.1007/JHEP11(2016)129}{{\em JHEP}
  {\bfseries 11} (2016) 129},
\href{http://arxiv.org/abs/1609.02514}{{\ttfamily arXiv:1609.02514 [hep-th]}}.
%%CITATION = ARXIV:1609.02514;%%.

\bibitem{ewedge1}
B.~Czech, J.~L. Karczmarek, F.~Nogueira, and M.~Van~Raamsdonk, ``{The Gravity
  Dual of a Density Matrix},''
  \href{http://dx.doi.org/10.1088/0264-9381/29/15/155009}{{\em Class. Quant.
  Grav.} {\bfseries 29} (2012) 155009},
\href{http://arxiv.org/abs/1204.1330}{{\ttfamily arXiv:1204.1330 [hep-th]}}.
%%CITATION = ARXIV:1204.1330;%%.

\bibitem{Headrick:2014cta}
M.~Headrick, V.~E. Hubeny, A.~Lawrence, and M.~Rangamani, ``{Causality \&
  holographic entanglement entropy},''
  \href{http://dx.doi.org/10.1007/JHEP12(2014)162}{{\em JHEP} {\bfseries 12}
  (2014) 162},
\href{http://arxiv.org/abs/1408.6300}{{\ttfamily arXiv:1408.6300 [hep-th]}}.
%%CITATION = ARXIV:1408.6300;%%.

\bibitem{Emparan:1999pm}
R.~Emparan, C.~V. Johnson, and R.~C. Myers, ``{Surface terms as counterterms in
  the AdS/CFT correspondence},''
  \href{http://dx.doi.org/10.1103/PhysRevD.60.104001}{{\em Phys. Rev.}
  {\bfseries D60} (1999) 104001},
\href{http://arxiv.org/abs/hep-th/9903238}{{\ttfamily arXiv:hep-th/9903238
  [hep-th]}}.
%%CITATION = HEP-TH/9903238;%%.

\bibitem{sken1}
S.~de~Haro, S.~N. Solodukhin, and K.~Skenderis, ``{Holographic reconstruction
  of space-time and renormalization in the AdS/CFT correspondence},''
  \href{http://dx.doi.org/10.1007/s002200100381}{{\em Commun. Math. Phys.}
  {\bfseries 217} (2001) 595--622},
\href{http://arxiv.org/abs/hep-th/0002230}{{\ttfamily arXiv:hep-th/0002230
  [hep-th]}}.
%%CITATION = HEP-TH/0002230;%%.

\bibitem{sken2}
K.~Skenderis, ``{Lecture notes on holographic renormalization},''
  \href{http://dx.doi.org/10.1088/0264-9381/19/22/306}{{\em Class. Quant.
  Grav.} {\bfseries 19} (2002) 5849--5876},
\href{http://arxiv.org/abs/hep-th/0209067}{{\ttfamily arXiv:hep-th/0209067
  [hep-th]}}.
%%CITATION = HEP-TH/0209067;%%.

\bibitem{Schwimmer:2008yh}
A.~Schwimmer and S.~Theisen, ``{Entanglement Entropy, Trace Anomalies and
  Holography},'' \href{http://dx.doi.org/10.1016/j.nuclphysb.2008.04.015}{{\em
  Nucl. Phys.} {\bfseries B801} (2008) 1--24},
\href{http://arxiv.org/abs/0802.1017}{{\ttfamily arXiv:0802.1017 [hep-th]}}.
%%CITATION = ARXIV:0802.1017;%%.

\bibitem{Fefferman:2007rka}
C.~Fefferman and C.~R. Graham, ``{The ambient metric},''
\href{http://arxiv.org/abs/0710.0919}{{\ttfamily arXiv:0710.0919 [math.DG]}}.
%%CITATION = ARXIV:0710.0919;%%.

\bibitem{Graham:1999pm}
C.~R. Graham and E.~Witten, ``{Conformal anomaly of submanifold observables in
  AdS/CFT correspondence},''
  \href{http://dx.doi.org/10.1016/S0550-3213(99)00055-3}{{\em Nucl. Phys.}
  {\bfseries B546} (1999) 52--64},
\href{http://arxiv.org/abs/hep-th/9901021}{{\ttfamily arXiv:hep-th/9901021
  [hep-th]}}.
%%CITATION = HEP-TH/9901021;%%.

\bibitem{GraWit}
C.~R. Graham and E.~Witten, ``{Conformal anomaly of submanifold observables in
  AdS / CFT correspondence},''
  \href{http://dx.doi.org/10.1016/S0550-3213(99)00055-3}{{\em Nucl. Phys.}
  {\bfseries B546} (1999) 52--64},
\href{http://arxiv.org/abs/hep-th/9901021}{{\ttfamily arXiv:hep-th/9901021
  [hep-th]}}.
%%CITATION = HEP-TH/9901021;%%.

\bibitem{gb2}
A.~Buchel, J.~Escobedo, R.~C. Myers, M.~F. Paulos, A.~Sinha, and M.~Smolkin,
  ``{Holographic GB gravity in arbitrary dimensions},''
  \href{http://dx.doi.org/10.1007/JHEP03(2010)111}{{\em JHEP} {\bfseries 03}
  (2010) 111},
\href{http://arxiv.org/abs/0911.4257}{{\ttfamily arXiv:0911.4257 [hep-th]}}.
%%CITATION = ARXIV:0911.4257;%%.

\bibitem{Myers:2010jv}
R.~C. Myers, M.~F. Paulos, and A.~Sinha, ``{Holographic studies of
  quasi-topological gravity},''
  \href{http://dx.doi.org/10.1007/JHEP08(2010)035}{{\em JHEP} {\bfseries 08}
  (2010) 035},
\href{http://arxiv.org/abs/1004.2055}{{\ttfamily arXiv:1004.2055 [hep-th]}}.
%%CITATION = ARXIV:1004.2055;%%.

\bibitem{corner}
P.~Bueno and R.~C. Myers, ``{Corner contributions to holographic entanglement
  entropy},'' \href{http://dx.doi.org/10.1007/JHEP08(2015)068}{{\em JHEP}
  {\bfseries 08} (2015) 068},
\href{http://arxiv.org/abs/1505.07842}{{\ttfamily arXiv:1505.07842 [hep-th]}}.
%%CITATION = ARXIV:1505.07842;%%.

\bibitem{Solodukhin:2008dh}
S.~N. Solodukhin, ``{Entanglement entropy, conformal invariance and extrinsic
  geometry},'' \href{http://dx.doi.org/10.1016/j.physletb.2008.05.071}{{\em
  Phys. Lett.} {\bfseries B665} (2008) 305--309},
\href{http://arxiv.org/abs/0802.3117}{{\ttfamily arXiv:0802.3117 [hep-th]}}.
%%CITATION = ARXIV:0802.3117;%%.

\bibitem{ajay}
R.~C. Myers and A.~Singh, ``{Entanglement Entropy for Singular Surfaces},''
  \href{http://dx.doi.org/10.1007/JHEP09(2012)013}{{\em JHEP} {\bfseries 09}
  (2012) 013},
\href{http://arxiv.org/abs/1206.5225}{{\ttfamily arXiv:1206.5225 [hep-th]}}.
%%CITATION = ARXIV:1206.5225;%%.

\bibitem{Hubeny:2007xt}
V.~E. Hubeny, M.~Rangamani, and T.~Takayanagi, ``{A Covariant holographic
  entanglement entropy proposal},''
  \href{http://dx.doi.org/10.1088/1126-6708/2007/07/062}{{\em JHEP} {\bfseries
  07} (2007) 062},
\href{http://arxiv.org/abs/0705.0016}{{\ttfamily arXiv:0705.0016 [hep-th]}}.
%%CITATION = ARXIV:0705.0016;%%.

\bibitem{ha}
X.~Dong, A.~Lewkowycz, and M.~Rangamani, ``{Deriving covariant holographic
  entanglement},'' \href{http://dx.doi.org/10.1007/JHEP11(2016)028}{{\em JHEP}
  {\bfseries 11} (2016) 028},
\href{http://arxiv.org/abs/1607.07506}{{\ttfamily arXiv:1607.07506 [hep-th]}}.
%%CITATION = ARXIV:1607.07506;%%.

\bibitem{shira}
S.~Chapman, H.~Marrochio, and R.~C. Myers, ``{Complexity of Formation in
  Holography},''
\href{http://arxiv.org/abs/1610.08063}{{\ttfamily arXiv:1610.08063 [hep-th]}}.
%%CITATION = ARXIV:1610.08063;%%.

\bibitem{new7}
D.~Carmi, ``{More on Holographic Volumes}.'' In preparation.

\bibitem{wrong}
D.~Marolf and A.~C. Wall, ``{State-Dependent Divergences in the Entanglement
  Entropy},'' \href{http://dx.doi.org/10.1007/JHEP10(2016)109}{{\em JHEP}
  {\bfseries 10} (2016) 109},
\href{http://arxiv.org/abs/1607.01246}{{\ttfamily arXiv:1607.01246 [hep-th]}}.
%%CITATION = ARXIV:1607.01246;%%.

\bibitem{new8}
D.~Carmi, S.~Chapman, H.~Marrochio, R.~C. Myers, and S.~Sugishita, ``{On the
  Time Dependence of Holographic Complexity}.'' In preparation.

\bibitem{ross16}
A.~Reynolds and S.~F. Ross, ``{Divergences in Holographic Complexity},''
\href{http://arxiv.org/abs/1612.05439}{{\ttfamily arXiv:1612.05439 [hep-th]}}.
%%CITATION = ARXIV:1612.05439;%%.

\bibitem{martink}
M.~Kruczenski, ``{A Note on twist two operators in N=4 SYM and Wilson loops in
  Minkowski signature},''
  \href{http://dx.doi.org/10.1088/1126-6708/2002/12/024}{{\em JHEP} {\bfseries
  12} (2002) 024},
\href{http://arxiv.org/abs/hep-th/0210115}{{\ttfamily arXiv:hep-th/0210115
  [hep-th]}}.
%%CITATION = HEP-TH/0210115;%%.

\bibitem{Alday}
L.~F. Alday and J.~M. Maldacena, ``{Gluon scattering amplitudes at strong
  coupling},'' \href{http://dx.doi.org/10.1088/1126-6708/2007/06/064}{{\em
  JHEP} {\bfseries 06} (2007) 064},
\href{http://arxiv.org/abs/0705.0303}{{\ttfamily arXiv:0705.0303 [hep-th]}}.
%%CITATION = ARXIV:0705.0303;%%.

\bibitem{Nielsen:2006:GAQ:2011686.2011688}
M.~A. Nielsen, ``A geometric approach to quantum circuit lower bounds,'' {\em
  Quantum Info. Comput.} {\bfseries 6} no.~3, (May, 2006) 213--262,
  \href{http://arxiv.org/abs/quant-ph/0502070}{{\ttfamily
  arXiv:quant-ph/0502070 [quant-ph]}}.

\bibitem{Nielsen1133}
M.~A. Nielsen, M.~R. Dowling, M.~Gu, and A.~C. Doherty, ``Quantum computation
  as geometry,'' \href{http://dx.doi.org/10.1126/science.1121541}{{\em Science}
  {\bfseries 311} no.~5764, (2006) 1133--1135},
  \href{http://arxiv.org/abs/quant-ph/0603161}{{\ttfamily
  arXiv:quant-ph/0603161 [quant-ph]}}.

\bibitem{Dowling:2008:GQC:2016985.2016986}
M.~R. Dowling and M.~A. Nielsen, ``The geometry of quantum computation,'' {\em
  Quantum Info. Comput.} {\bfseries 8} no.~10, (Nov., 2008) 861--899,
  \href{http://arxiv.org/abs/quant-ph/0701004}{{\ttfamily
  arXiv:quant-ph/0701004 [quant-ph]}}.

\bibitem{beni}
D.~A. Roberts and B.~Yoshida, ``{Chaos and complexity by design},''
\href{http://arxiv.org/abs/1610.04903}{{\ttfamily arXiv:1610.04903
  [quant-ph]}}.
%%CITATION = ARXIV:1610.04903;%%.

\bibitem{Aharonov:1998zf}
D.~Aharonov, A.~Kitaev, and N.~Nisan, ``{Quantum circuits with mixed states},''
  {\em {\rm in} Proceedings of the Thirtieth Annual ACM Symposium on Theory of
  Computing} (1998) 20--30,
\href{http://arxiv.org/abs/quant-ph/9806029}{{\ttfamily arXiv:quant-ph/9806029
  [quant-ph]}}.
%%CITATION = QUANT-PH/9806029;%%.

\bibitem{dilaton}
W.~F. Stinespring, ``Positive functions on c*-algebras,'' {\em Proceedings of
  the American Mathematical Society} {\bfseries 6} no.~2, (1955) 211--216.

\bibitem{Arias:2016nip}
R.~Arias, D.~Blanco, H.~Casini, and M.~Huerta, ``{Local temperatures and local
  terms in modular Hamiltonians},''
\href{http://arxiv.org/abs/1611.08517}{{\ttfamily arXiv:1611.08517 [hep-th]}}.
%%CITATION = ARXIV:1611.08517;%%.

\bibitem{Bianchi:2012ev}
E.~Bianchi and R.~C. Myers, ``{On the Architecture of Spacetime Geometry},''
  \href{http://dx.doi.org/10.1088/0264-9381/31/21/214002}{{\em Class. Quant.
  Grav.} {\bfseries 31} (2014) 214002},
\href{http://arxiv.org/abs/1212.5183}{{\ttfamily arXiv:1212.5183 [hep-th]}}.
%%CITATION = ARXIV:1212.5183;%%.

\bibitem{Casini:2011kv}
H.~Casini, M.~Huerta, and R.~C. Myers, ``{Towards a derivation of holographic
  entanglement entropy},''
  \href{http://dx.doi.org/10.1007/JHEP05(2011)036}{{\em JHEP} {\bfseries 05}
  (2011) 036},
\href{http://arxiv.org/abs/1102.0440}{{\ttfamily arXiv:1102.0440 [hep-th]}}.
%%CITATION = ARXIV:1102.0440;%%.

\bibitem{Faulkner2}
T.~Faulkner, M.~Guica, T.~Hartman, R.~C. Myers, and M.~Van~Raamsdonk,
  ``{Gravitation from Entanglement in Holographic CFTs},''
  \href{http://dx.doi.org/10.1007/JHEP03(2014)051}{{\em JHEP} {\bfseries 03}
  (2014) 051},
\href{http://arxiv.org/abs/1312.7856}{{\ttfamily arXiv:1312.7856 [hep-th]}}.
%%CITATION = ARXIV:1312.7856;%%.

\bibitem{tom33}
T.~Hartman and J.~Maldacena, ``{Time Evolution of Entanglement Entropy from
  Black Hole Interiors},''
  \href{http://dx.doi.org/10.1007/JHEP05(2013)014}{{\em JHEP} {\bfseries 05}
  (2013) 014},
\href{http://arxiv.org/abs/1303.1080}{{\ttfamily arXiv:1303.1080 [hep-th]}}.
%%CITATION = ARXIV:1303.1080;%%.

\bibitem{Freedman:2016zud}
M.~Freedman and M.~Headrick, ``{Bit threads and holographic entanglement},''
\href{http://arxiv.org/abs/1604.00354}{{\ttfamily arXiv:1604.00354 [hep-th]}}.
%%CITATION = ARXIV:1604.00354;%%.

\bibitem{Lashkari:2015hha}
N.~Lashkari and M.~Van~Raamsdonk, ``{Canonical Energy is Quantum Fisher
  Information},'' \href{http://dx.doi.org/10.1007/JHEP04(2016)153}{{\em JHEP}
  {\bfseries 04} (2016) 153},
\href{http://arxiv.org/abs/1508.00897}{{\ttfamily arXiv:1508.00897 [hep-th]}}.
%%CITATION = ARXIV:1508.00897;%%.

\bibitem{deBoer:2015kda}
J.~de~Boer, M.~P. Heller, R.~C. Myers, and Y.~Neiman, ``{Holographic de Sitter
  Geometry from Entanglement in Conformal Field Theory},''
  \href{http://dx.doi.org/10.1103/PhysRevLett.116.061602}{{\em Phys. Rev.
  Lett.} {\bfseries 116} (2016) 061602},
\href{http://arxiv.org/abs/1509.00113}{{\ttfamily arXiv:1509.00113 [hep-th]}}.
%%CITATION = ARXIV:1509.00113;%%.

\bibitem{deBoer:2016pqk}
J.~de~Boer, F.~M. Haehl, M.~P. Heller, and R.~C. Myers, ``{Entanglement,
  holography and causal diamonds},''
  \href{http://dx.doi.org/10.1007/JHEP08(2016)162}{{\em JHEP} {\bfseries 08}
  (2016) 162},
\href{http://arxiv.org/abs/1606.03307}{{\ttfamily arXiv:1606.03307 [hep-th]}}.
%%CITATION = ARXIV:1606.03307;%%.

\bibitem{Czech:2016xec}
B.~Czech, L.~Lamprou, S.~McCandlish, B.~Mosk, and J.~Sully, ``{A Stereoscopic
  Look into the Bulk},'' \href{http://dx.doi.org/10.1007/JHEP07(2016)129}{{\em
  JHEP} {\bfseries 07} (2016) 129},
\href{http://arxiv.org/abs/1604.03110}{{\ttfamily arXiv:1604.03110 [hep-th]}}.
%%CITATION = ARXIV:1604.03110;%%.

\bibitem{Pad1}
K.~Parattu, S.~Chakraborty, B.~R. Majhi, and T.~Padmanabhan, ``{A Boundary Term
  for the Gravitational Action with Null Boundaries},''
  \href{http://dx.doi.org/10.1007/s10714-016-2093-7}{{\em Gen. Rel. Grav.}
  {\bfseries 48} (2016) 94},
\href{http://arxiv.org/abs/1501.01053}{{\ttfamily arXiv:1501.01053 [gr-qc]}}.
%%CITATION = ARXIV:1501.01053;%%.

\bibitem{York:1972sj}
J.~W. York, Jr., ``{Role of conformal three geometry in the dynamics of
  gravitation},''
\href{http://dx.doi.org/10.1103/PhysRevLett.28.1082}{{\em Phys. Rev. Lett.}
  {\bfseries 28} (1972) 1082--1085}.
%%CITATION = PRLTA,28,1082;%%.

\bibitem{Gibbons:1977mu}
G.~W. Gibbons and S.~W. Hawking, ``{Cosmological Event Horizons,
  Thermodynamics, and Particle Creation},''
\href{http://dx.doi.org/10.1103/PhysRevD.15.2738}{{\em Phys. Rev.} {\bfseries
  D15} (1977) 2738--2751}.
%%CITATION = PHRVA,D15,2738;%%.

\bibitem{Hayward:1993my}
G.~Hayward, ``{Gravitational action for space-times with nonsmooth
  boundaries},''
\href{http://dx.doi.org/10.1103/PhysRevD.47.3275}{{\em Phys. Rev.} {\bfseries
  D47} (1993) 3275--3280}.
%%CITATION = PHRVA,D47,3275;%%.

\bibitem{Brill:1994mb}
D.~Brill and G.~Hayward, ``{Is the gravitational action additive?},''
  \href{http://dx.doi.org/10.1103/PhysRevD.50.4914}{{\em Phys. Rev.} {\bfseries
  D50} (1994) 4914--4919},
\href{http://arxiv.org/abs/gr-qc/9403018}{{\ttfamily arXiv:gr-qc/9403018
  [gr-qc]}}.
%%CITATION = GR-QC/9403018;%%.

\bibitem{Gourgoulhon:2007ue}
E.~Gourgoulhon, ``{3+1 formalism and bases of numerical relativity},''
\href{http://arxiv.org/abs/gr-qc/0703035}{{\ttfamily arXiv:gr-qc/0703035
  [gr-qc]}}.
%%CITATION = GR-QC/0703035;%%.

\end{thebibliography}\endgroup
\end{document}